\documentclass[a4paper,11pt]{article}
\usepackage[dvips]{graphicx}
\usepackage{amsmath}
\usepackage{amssymb}
\usepackage{cite}
\usepackage{cancel}
\usepackage{fancybox}
\usepackage{ulem}
\usepackage{pifont}
\usepackage{colortbl}
\usepackage{multicol}

\usepackage{colortbl}
\usepackage{multicol}
\usepackage{longtable}

\makeatletter
 
 \@addtoreset{equation}{section}
\makeatother

\begin{document}

\begin{titlepage}

\begin{center}

\vspace{1cm}
{\Large\textbf{
Single Higgs Boson Production at Electron-Positron Colliders\\ 
in Gauge-Higgs Unification
}}
\vspace{1cm}

\renewcommand{\thefootnote}{\fnsymbol{footnote}}
Shuichiro Funatsu${}^{1}$,
Hisaki Hatanaka${}^{2}$,
Yuta Orikasa${}^{3}$,
and
Naoki Yamatsu${}^{4}$\footnote[1]{yamatsu@phys.ntu.edu.tw}
\vspace{5mm}

\textit{
$^1${Ushiku, Ibaraki 300-1234, Japan } \\
$^2${Osaka, Osaka 536-0014, Japan} \\
$^3${Institute of Experimental and Applied Physics, Czech Technical
 University in Prague, \\ 
Husova 240/5, 110 00 Prague 1, Czech Republic}\\
$^4${Department of Physics, National Taiwan University, Taipei, Taiwan 10617, R.O.C.}
}

\date{\today}

\abstract{
 We examine contributions to single Higgs boson production
 processes via $Z'$ and $W'$ bosons in the $SU(3)_C\times SO(5)_W\times
 U(1)_X$  gauge-Higgs unification (GHU) model.
 In particular, we analyze the cross sections of three single Higgs boson
 production processes $e^-e^+\to Zh$, $e^-e^+\to \nu\bar{\nu}h$, and
 $e^-e^+\to e^-e^+h$ in the SM and the GHU model.
 For the Higgs strahlung process $e^-e^+\to Zh$,  we show that for a
 parameter region satisfying the current experimental  constraints, a
 maximum deviation from the SM is about up to 6\% for  the
 center-of-mass energy of the initial electron and positron 
 $\sqrt{s}=250$\,GeV and that from the SM is about up to 20\%  for
 $\sqrt{s}=500$\,GeV, depending on the initial polarization of electron
 and positron. 
 The deviation from the SM is monotonically increasing with respect to
 $\sqrt{s}$ for $\sqrt{s}\lesssim 1$\,TeV.
 By using the Higgs strahlung process,  it is possible to explore up to
 the region of tens of TeV in terms of the Kaluza-Klein (KK) mass.
 We also show that the sign of the bulk mass of a lepton multiplet in
 the GHU  model can be determined by examining the deviation of the
 left-right 
 symmetry of the $e^-e^+\to Zh$ process from the SM. 
 The contributions to the
 cross sections of the $e^-e^+\to \nu\bar{\nu}h$ and $e^-e^+\to e^-e^+h$
 processes
 via the $Z'$ and $W'$ bosons are relatively small compared to that for
 $e^-e^+\to Zh$ at least for $\sqrt{s}\leq 1$\,TeV. 
 As is the same as in the SM, the $e^-e^+\to\nu\bar{\nu}h$ process gives
 a main contribution to the single Higgs boson production processes
 for $\sqrt{s}\gtrsim 500$\,GeV, but large deviations from the SM  are
 only observed on energy scales close to the masses of the first KK
 gauge bosons or higher. 
}

\end{center}
\end{titlepage}

\section{Introduction}
\label{Sec:Introduction}

The Standard Model (SM) in particle physics is well established at low
energies. However, it is not yet clear whether the properties of the
observed Higgs boson are exactly the same as those of the Higgs boson in 
the SM. Current and future experiments 
such as 
the Large Hadron Collider (LHC) \cite{ATLAS:2008xda,CMS:2008xjf}, 
the High-Luminosity Large Hadron Collider (HL-LHC)
\cite{Cepeda:2019klc},
the International Linear Collider (ILC)
\cite{Behnke:2013xla,Baer:2013cma,Adolphsen:2013jya,Adolphsen:2013kya,Behnke:2013lya,Fujii:2019zll,ILCInternationalDevelopmentTeam:2022izu},
the Compact Linear Collider (CLIC)\cite{Linssen:2012hp},
the Future Circular Collider
(FCC-hh, FCC-ee)\cite{FCC:2018vvp,Blondel:2021ema},
the Cool Copper Collider (C${}^3$)
\cite{Dasu:2022nux},
and Circular Electron Positron Collider 
(CEPC)\cite{CEPCStudyGroup:2018ghi}
are needed to more precisely determine the
couplings of the Higgs boson to quarks, leptons, and SM gauge bosons, as
well as the self-coupling of the Higgs boson.

The SM Higgs boson sector remains unsatisfactory: the dynamics of the SM
gauge bosons, the photon, the $W$ and $Z$ bosons, and the gluons, are
governed by 
the gauge principle, but the dynamics of the SM Higgs boson is not
governed by the gauge principle. The Higgs boson coupling of quarks and
leptons and the self-coupling of the Higgs boson are also not governed by
the principle. At the quantum level, there are large corrections to the
mass of the Higgs boson.
To reproduce the observed mass of the Higgs boson
$m_h=125.25\pm 0.17$\,GeV \cite{ParticleDataGroup:2022pth}, 
an extremely strong cancellation between the bare mass and quantum effects
must be required. One known way to stabilize the mass of the Higgs boson
against quantum corrections is to identify the Higgs boson as the zero
mode of the five-dimensional component of the gauge potential. This
scenario is called gauge-Higgs unification (GHU) 
\cite{Hosotani:1983xw,Hosotani:1988bm,Davies:1987ei,Davies:1988wt,Hatanaka:1998yp,Hatanaka:1999sx}.

In GHU models, the Higgs boson appears as a fluctuating mode in the
5-dimensional (5D) Aharonov-Bohm (AB) phase $\theta_H$; 
$SU(3)_C \times SO(5)_W\times U(1)_X$ GHU models in the Randall-Sundrum
(RS) warped space are proposed in
Refs.~\cite{Agashe:2004rs,Medina:2007hz,Hosotani:2008tx,Funatsu:2019xwr}.
The phenomena of the GHU models below the electroweak (EW) scale are
very close to those of the SM
\cite{Funatsu:2019xwr,Funatsu:2013ni,Funatsu:2014fda,Funatsu:2015xba,Funatsu:2016uvi,Funatsu:2019fry,Funatsu:2020znj,Funatsu:2020haj,Funatsu:2021gnh,Funatsu:2021yrh,Funatsu:2022spb}. 
The gauge coupling constants of quarks and leptons to the $Z$ and $W$
bosons in the GHU model 
differ by less than 0.1\% from the coupling constants in the SM
for the parameter region that satisfies the current experiments, i.e.,
the Kaluza-Klein (KK) mass $m_{\rm KK}\gtrsim13$\,TeV and the AB phase
$\theta_H \lesssim 0.1$ \cite{Funatsu:2021yrh}.
The gauge coupling constants of the Higgs boson to the $W$ and $Z$ bosons 
are approximately equal to the corresponding coupling constants in the
SM multiplied by $\cos\theta_H$.
The coupling constants of the Higgs boson to the quarks and the leptons
are approximately equal to those in the SM multiplied by $\cos\theta_H$
or $\cos^2(\theta_H/2)$ depending on the $SO(5)_W$ representations of
the quarks and the leptons.  
The deviation of the coupling constants of the Higgs boson to the $W$
and $Z$ bosons, the quarks, and the leptons from the SM in the GHU model
is less than 1\% for $\theta_H\leq 0.1$ \cite{Funatsu:2020znj}. 
The decay of the Higgs boson to two photons in the GHU model is finite
and very small, even when the contribution of KK excitation modes is
taken into account\cite{Funatsu:2013ni}. 
The electroweak phase transition (EWPT) in the GHU model is a weakly first
order phase transition \cite{Funatsu:2021gnh}, which is very similar to
the EWPT in the SM \cite{Senaha:2020mop}.

The GHU models predict massive vector bosons $Z'$ and $W'$ bosons.
The $Z'$ bosons are mixed vector bosons of $U(1)_X$, 
$U(1)_L(\subset SU(2)_L)$, and $U(1)_R(\subset SU(2)_R)$, where
$SU(2)_L\times SU(2)_R \subset SO(5)$.
The $W'$ bosons are mixed vector bosons of 
$SO(5)_W/(U(1)_L\times U(1)_R)$. 
The $SU(3)_C\times SO(5)_W\times U(1)_X$ GHU models can be roughly
classified into two types of models depending on whether the $SO(5)_W$
representations of the quarks and the leptons are vector or spinor
representations.
The GHU model whose fermions belong to the spinor representation of
$SO(5)_W$
\cite{Funatsu:2019xwr,Funatsu:2019fry,Funatsu:2020znj,Funatsu:2020haj,Funatsu:2021gnh,Funatsu:2021yrh,Funatsu:2022spb}
can be regarded as a low-energy effective description of the GHU model
based on the $SO(11)$ grand unified gauge symmetry 
\cite{Hosotani:2015hoa,Yamatsu:2015rge,Furui:2016owe,Hosotani:2017edv,Hosotani:2017ghg,Englert:2019xhz,Englert:2020eep},
where the SM gauge symmetry and field content are incorporated into
grand unified theory (GUT) 
\cite{Georgi:1974sy,Inoue:1977qd,Fritzsch:1974nn,Gursey:1975ki,Slansky:1981yr,Yamatsu:2015gut}
in higher dimensional framework
\cite{Kawamura:1999nj,Kawamura:2000ir,Kawamura:2000ev,Hall:2001pg,Hall:2001zb,Burdman:2002se,Lim:2007jv,Kojima:2011ad,Kojima:2016fvv,Kojima:2017qbt,Yamatsu:2017sgu,Yamatsu:2017ssg,Yamatsu:2018fsg,Maru:2019lit,Kawamura:2022ecd}.
We focus on the spinor type model in this paper.
In the GHU model, $\theta_H\lesssim0.10$ and $m_{\rm KK}\gtrsim 13$TeV
are obtained as parameter constraints of the GHU model by using the
$Z'$ and $W'$ boson search results for the $pp\to\ell\nu$ and
$pp\to\ell^-\ell^+$ processes at the LHC at $\sqrt{s}=13$\,TeV with up
to 140\,fb${}^{-1}$ data
\cite{Aad:2019fac,ATLAS:2020yat,ATLAS:2019lsy,ATLAS:2019fgd,CMS:2019gwf,CMS:2021ctt}.

The $e^-e^+$ collider experiment is capable of searching for 
signals in the GHU model up to the KK mass scale $m_{\rm KK}$ of tens of
times the center-of-mass energy $\sqrt{s}$ of $e^-e^+$
\cite{Funatsu:2017nfm,Yoon:2018xud,Yoon:2018vsc,Funatsu:2019ujy,Funatsu:2020haj,Funatsu:2022spb,Bilokin:2017lco,Richard:2018zhl,Irles:2019xny,Irles:2020gjh,Fujii:2017vwa,Aihara:2019gcq,Bambade:2019fyw,Poschl:2022ecb}.
Large parity violation appears in the couplings of quarks and leptons to
KK gauge bosons, especially to the first KK modes.
The sign of the bulk mass parameter of the fermions in the GHU model
determines whether the coupling constants of the $Z'$ and $W'$ bosons to
right- or left-handed fermions are larger.
In Ref.~\cite{Funatsu:2020haj}, the SM and the GHU model are compared in
detail for observables such as cross sections,  forward-backward
asymmetries \cite{Schrempp:1987zy,Kennedy:1988rt}, 
left-right asymmetries
\cite{Schrempp:1987zy,Kennedy:1988rt,Abe:1994wx,Abe:1996nj}, 
and left-right forward-backward asymmetries
\cite{Blondel:1987gp,Kennedy:1988rt,Abe:1994bj,Abe:1994bm,Abe:1995yh}
in the processes $e^-e^+\to f\bar{f}$
($f\bar{f}= \mu^-\mu^+, c\bar{c}, b\bar{b}, t\bar{t})$.
For example, 
the total cross section of the $e^-e^+\to \mu^-\mu^+$ process
at $\sqrt{s}=250$\,GeV in the GHU model whose $m_{\rm KK}\simeq13$\,TeV
and $\theta_H=0.10$ is different at most about 3\% compared to that in
the SM, where it strongly depends on the initial polarizations of 
the electron and the positron.
Due to the very large cross section of the fermion pair production
processes, we can clearly observe deviations from the SM in the early
stage of the ILC experiment $(\sqrt{s}=250$\,GeV, integral luminosity 
$L_{\rm int}=250$\,fb$^{-1}$) 
even when $m_{\rm KK}$ is slightly larger than the current experimental
constraint $m_{\rm KK}\simeq13$\,TeV.
The cross section of the Bhabha scattering $e^-e^+\to e^-e^+$ process 
\cite{Abe:1994sh,MoortgatPick:2005cw,Schael:2013ita,Bardin:2017mdd,Borodulin:2017pwh}
is also very large, and  the deviation from the SM in the GHU model can
be also observed in the early-stage of the ILC experiment 
when $m_{\rm KK}$ is slightly larger than the current experimental
constraint $m_{\rm KK}=13$\,TeV \cite{Richard:2018zhl,Funatsu:2022spb}.
Furthermore, since the cross sections of the fermion pair production
processes are very sensitive to the initial polarizations of the
electron and the positron, the sign of the corresponding bulk mass
of each fermion in each final state in the GHU model can also be
determined by analyzing the polarization dependence.

To clarify the nature of the Higgs boson in the SM and the GHU model, it
is essential to understand Higgs boson production processes. Since the
mass of the SM Higgs boson is about $125$\,GeV, a Higgs strahlung process
$e^-e^+\to Zh$ is the main single Higgs boson production process for
$\sqrt{s}\lesssim500$\,GeV, and the $W^-W^+$ gauge boson fusion process 
$e^-e^+\to \nu\bar{\nu}W^-{}^*W^+{}^*\to \nu\bar{\nu}h$ is the main single
Higgs boson production process for $\sqrt{s}\gtrsim500$\,GeV
\cite{Djouadi:2005gi,Yan:2016xyx,ILCInternationalDevelopmentTeam:2022izu},
where the contributions depend on the initial polarizations of the
electron and the positron. 
There are the $Z'$ and $W'$ bosons in the GHU model, so these additional
gauge bosons affect the Higgs boson production processes. 
The Higgs boson production processes in the GHU model have not been
analyzed in detail. In particular, the dependence of the initial
polarizations of the electron and the positron has not been discussed at
all. In this paper, we analyze the cross sections and the left-right
asymmetries of the $e^-e^+\to Zh$ and 
$e^-e^+\to f\bar{f}h$ $(f\bar{f}=\nu\bar{\nu},e^-e^+)$
in the SM and the GHU model within the Born
approximation. The Higgs boson production processes in the SM have been
analyzed not only in the Born approximation but also including quantum
corrections 
\cite{Belanger:2002ik,Boudjema:2004eb,Denner:2003yg,Denner:2003iy}.
The Higgs boson production processes in other models beyond the
SM have been also discussed in e.g.,
Refs.~\cite{Yue:2003yk,Yue:2005av,Wang:2006ui,Han:2015orc}.

In this paper, we consider three single Higgs boson production processes 
$e^-e^+\to Zh$, $e^-e^+\to \nu\bar{\nu}h$, and $e^-e^+\to e^-e^+h$
to clarify the difference between the predictions of
the cross sections in the SM and the GHU model.
For the Higgs strahlung process $e^-e^+\to Zh$,
we show that for a parameter region satisfying the current experimental
constraints, 
a maximum deviation from the SM is about up to 6\% for
$\sqrt{s}=250$\,GeV and that from the SM is about up to 20\% for
$\sqrt{s}=500$\,GeV,
depending on the initial polarization of the electron and the positron.
The deviation from the SM is monotonically increasing
when $\sqrt{s}$ is increasing for $\sqrt{s}\lesssim 1$\,TeV.
By using the Higgs strahlung process, it is possible to explore up to
the region of tens of TeV in terms of the KK mass.
We also show that the sign of the bulk mass of a lepton multiplet in the
GHU model can be determined by examining the deviation of the left-right
symmetry of the $e^-e^+\to Zh$ process from the SM. 
The deviation of the left-right asymmetry of the $e^-e^+\to Zh$ process
from the SM in the GHU model can be large even when $\sqrt{s}$ is not so
large. We show that the deviations of the $e^-e^+\to \nu\bar{\nu}h$ and
$e^-e^+\to e^-e^+h$ processes from the SM in the GHU model 
are relatively small compared to the deviations of the $e^-e^+\to Zh$ 
process at least for $\sqrt{s}\leq 1$\,TeV.

The paper is organized as follows.
In Sec.~\ref{Sec:Model}, the $SU(3)_C\times SO(5)_W\times U(1)_X$ GHU
model is introduced. 
In Sec.~\ref{Sec:Parameter-sets},
we give some parameter sets of the GHU model.
In Sec.~\ref{Sec:Results},
we present numerical results for the cross sections
of the three Higgs boson production processes
$e^-e^+\to Zh$, $e^-e^+\to \nu\bar{\nu}h$, and $e^-e^+\to e^-e^+h$.
Section~\ref{Sec:Summary} is devoted to summary and discussions.
In Appendix~\ref{Sec:Cross-section}, 
we give the formulas for the cross sections of the 
$e^-e^+\to Zh$ and $e^-e^+\to f\bar{f}h$ 
$(f\bar{f}=\nu_e\bar{\nu}_e,e^-e^+)$ processes, 
involving the $Z'$ and $W'$ bosons as well as the $Z$ and $W$ bosons.

\section{Model}
\label{Sec:Model}

In this paper, we focus on observables related with the  EW gauge bosons
and leptons at tree level. The $SU(3)_C$ gauge bosons and fermions 
except leptons are not directly involved, so we omit them. For the full
field content in the GHU model, see Ref.~\cite{Funatsu:2019xwr}, 
in which the $SU(3)_C \times SO(5)_W\times U(1)_X$ GHU model was
originally proposed.

The GHU model is defined in the RS warped space with the following
\cite{Randall:1999ee}:
\begin{align}
 ds^2= g_{MN} dx^M dx^N =e^{-2\sigma(y)} \eta_{\mu\nu}dx^\mu dx^\nu+dy^2,
\end{align} 
where $M,N=0,1,2,3,5$, $\mu,\nu=0,1,2,3$, $y=x^5$,
$\eta_{\mu\nu}=\mbox{diag}(-1,+1,+1,+1)$,
$\sigma(y)=\sigma(y+ 2L)=\sigma(-y)$,
and $\sigma(y)=ky$ for $0 \le y \le L$.
By using the conformal coordinate $z=e^{ky}$
($1\leq z\leq z_L=e^{kL}$) in the region $0 \leq y \leq L$,
the metric is rewritten by 
\begin{align}
ds^2= \frac{1}{z^2}
\bigg(\eta_{\mu\nu}dx^{\mu} dx^{\nu} + \frac{dz^2}{k^2}\bigg).
\end{align} 
The bulk region $0<y<L$ ($1<z<z_L$) is anti-de Sitter (AdS) spacetime 
with a cosmological constant $\Lambda=-6k^2$, which is sandwiched by the
UV brane at $y=0$ ($z=1$) and the IR brane at $y=L$ ($z=z_L$).  
The KK mass scale is $m_{\rm KK}=\pi k/(z_L-1)$.

\subsection{Field content}

The EW symmetry $SU(2)_L\times U(1)_Y$ is embedded into 
$SO(5)_W\times U(1)_X$ symmetry. The associated gauge fields of
$SO(5)_W$ and $U(1)_X$ are denoted by $A_M^{SO(5)_W}$ and $A_M^{U(1)_X}$,
respectively. The orbifold boundary conditions (BCs) $P_j(j=0,1)$ 
of the gauge fields on the UV brane $(y=0)$ and the IR brane $(y=L)$ are
given by 
\begin{align}
&\begin{pmatrix} A_\mu \cr  A_{y} \end{pmatrix} (x,y_j-y) =
P_{j} \begin{pmatrix} A_\mu \cr  - A_{y} \end{pmatrix} (x,y_j+y)P_{j}^{-1}
\label{Eq:BC-gauge}
\end{align}
for each gauge field, where $(y_0, y_1) = (0, L)$. 
For the $U(1)_X$ gauge boson $A_M^{U(1)_X}$, $P_0=P_1=1$.
For the $SO(5)_W$ gauge boson $A_M^{SO(5)_W}$, 
$P_0=P_1=P_{\bf 5}^{SO(5)_W}$,
where $P_{\bf 5}^{SO(5)_W}=\mbox{diag}\left(I_{4},-I_{1}\right)$.
The orbifold BCs of the $SO(5)_W$ symmetry break
$SO(5)_W$ to $SO(4)_W\simeq SU(2)_L \times SU(2)_R$.
$W$, $Z$ bosons and $\gamma$ (photon) are zero modes in the 
$SO(5)_W\times U(1)_X$ of 4 dimensional (4D) gauge bosons, whereas the
4D Higgs boson is a zero mode in the $SO(5)_W/SO(4)_W$ part of the 5th
dimensional gauge boson. 
In the GHU model, extra neutral gauge bosons $Z'$ correspond to
the KK photons $\gamma^{(n)}$, the  KK $Z$ bosons $Z^{(n)}$,
and the KK $Z_R$ bosons $Z_R^{(n)}$ ($n \ge 1$),
where the $\gamma$, and $Z$, $Z_R$ bosons are the mass eigen states of
the electro-magnetic (EM) $U(1)_{\rm EM}$ neutral gauge bosons of
$SU(2)_L$, $SU(2)_R$, and $U(1)_X$.
Extra charged gauge bosons $W'$ correspond to
the KK $W$ boson $W^{\pm(n)}$ ($n \ge 1$) and  the KK $W_R$ bosons
$W_R^{\pm(n)}$ ($n \ge 1$).

Leptons are introduced both in the 5D bulk and on the UV
brane. They are listed in Table~\ref{Tab:matter}.
The SM lepton multiples are identified with the zero modes of the lepton
multiplets $\Psi_{({\bf 1,4})}^{\alpha}$ $(\alpha=1,2,3)$.
The bulk fields $\Psi_{({\bf 1,4})}^{\alpha}$ obey the following BCs:
\begin{align}
\Psi_{({\bf 1,4})}^{\alpha} (x, y_j - y) = 
 - P_{\bf 4}^{SO(5)_W} \gamma^5 \Psi_{({\bf 1,4})}^{\alpha} (x, y_j + y),
\label{leptonBC1}
\end{align}
where $P_{\bf 4}^{SO(5)_W}=\mbox{diag}\left(I_{2},-I_{2}\right)$.
With the BCs in Eq.~(\ref{leptonBC1}), the parity assignment of the
leptons is summarized in Table~\ref{Tab:parity}.
Neutral fermions are introduced as the brane fermions 
$\chi^\alpha$ $(\alpha=1,2,3)$ on the UV brane,
which are responsible for reproducing tiny neutrino masses
via the seesaw mechanism in the GHU model \cite{Hosotani:2017ghg}.

\begin{table}[tbh]
\begin{center}
\begin{tabular}{|c|c|}
\hline
 \rowcolor[gray]{0.9}
&$(SU(3)_C\times SO(5)_W)_X$\\
\hline
Lepton
 &$\strut ({\bf 1}, {\bf 4})_{-\frac{1}{2}}$ \\
\hline
Brane fermion
 &$({\bf 1}, {\bf 1})_{0} $ \\
\hline
Brane scalar &$({\bf 1}, {\bf 4})_{\frac{1}{2}} $ \\
\hline
\end{tabular}
\caption{\small
The field content for the lepton sector in the GHU model is shown.
}
\label{Tab:matter}
\end{center}
\end{table}

\begin{table}[tbh]
\begin{center}
\begin{tabular}{|c|c|c|c|c|c|}
\hline
 \rowcolor[gray]{0.9}
 Field & $(SU(3)_C\times SO(5)_W)_{X}$&$G_{22}$ &Left-handed &Right-handed &Name\\
\hline
$\Psi_{({\bf 1,4})}^{\alpha}$ &$({\bf 1,4})_{-\frac{1}{2}}$&$[{\bf 2,1}]$ 
&$(+,+)$ &$(-,-)$ &$\begin{matrix} \nu_e  & \nu_\mu & \nu_\tau \cr e & \mu & \tau \end{matrix}$\\
\cline{3-6}
&&$[{\bf 1,2}]$
&$(-,-)$ &$(+,+)$ &$\begin{matrix} \nu_e'  & \nu_\mu' & \nu_\tau' \cr e' & \mu' & \tau' \end{matrix}$\\
\hline
\end{tabular}
\caption{\small
 Parity assignment $(P_0^{SO(5)_W}, P_1^{SO(5)_W})$ of the lepton
 multiplets in the  bulk is shown. $G_{22}$ stands for 
 $SU(2)_L\times SU(2)_R(\subset SO(5)_W)$.
}
\label{Tab:parity}
\end{center}
\end{table}

The brane scalar field $\Phi_{({\bf 1}, {\bf 4})}(x)$ in
Table~\ref{Tab:matter} is responsible for breaking $SO(5)_W\times U(1)_X$
to $SU(2)_L\times U(1)_Y$. 
A spinor {\bf 4} of $SO(5)_W$ is decomposed into
$[{\bf 2}, {\bf 1}] \oplus [{\bf 1}, {\bf 2}]$ of
$SO(4)_W \simeq SU(2)_L \times SU(2)_R$.
We assume that the brane scalar $\Phi_{({\bf 1}, {\bf 4})}$ develops a
nonvanishing vacuum expectation value (VEV):
\begin{align}
\Phi_{({\bf 1,4})} =
\begin{pmatrix} \Phi_{[{\bf 2,1}]} \cr \Phi_{[{\bf 1,2}]} \end{pmatrix},
 \ \ 
\langle  \Phi_{[{\bf 1,2}]} \rangle = \begin{pmatrix} 0 \cr w \end{pmatrix},
\label{scalarVEV}
\end{align}
which reduces the symmetry
$SO(4)_W \times U(1)_X$ to 
the EW gauge symmetry  $SU(2)_L\times U(1)_Y$.
It is assumed that $w \gg m_{\rm KK}$, which ensures that the orbifold
BCs for the 4D components of the $SU(2)_R \times U(1)_X/U(1)_Y$ gauge
fields become effectively Dirichlet conditions at the UV brane
\cite{Furui:2016owe}. 
Accordingly the mass of the neutral physical mode of
$\Phi_{({\bf 1,4})}$ is much larger  than $m_{\rm KK}$.

The $U(1)_Y$ gauge boson is a mixed state of $U(1)_R(\subset SU(2)_R)$
and $U(1)_X$ gauge bosons. The $U(1)_Y$ gauge field $B_M^Y$ is
given in terms of the $SU(2)_R$ gauge fields $A_M^{a_R}$
$(a_R=1_R,2_R,3_R)$ and the $U(1)_X$ gauge field $B_M$  by 
\begin{align}
&B_M^Y = \sin\phi A_M^{3_R} + \cos\phi  B_M ~.
\end{align}
Here the mixing angle $\phi$ between $U(1)_R$ and $U(1)_X$ is given by 
$\cos \phi= {g_A}/{\sqrt{g_A^2+g_B^2}}$ and
$\sin \phi= {g_B}/{\sqrt{g_A^2+g_B^2}}$, where
$g_A$ and $g_B$ are gauge couplings in $SO(5)_W$ and $U(1)_X$,
respectively. 
The 4D $SU(2)_L$ gauge coupling is given by $g_w = g_A/\sqrt{L}$.  
The 5D gauge coupling constant $g_Y^{\rm 5D}$ of $U(1)_{Y}$ and the 4D
bare Weinberg angle at the tree level, $\theta_W^0$, are given by
\begin{align}
&g_Y^{\rm 5D} =\frac{g_Ag_B}{\sqrt{g_A^2+g_B^2}}, \ \ \
\sin \theta_W^0 = \frac{\sin\phi}{\sqrt{{1 +\sin^2\phi}}}.
\label{Eq:gY-sW}
\end{align}

The 4D Higgs boson $\phi_H(x)$ is the zero mode contained in the 
$A_z = (kz)^{-1} A_y$ component:
\begin{align}
A_z^{(j5)}(x,z)= \frac{1}{\sqrt{k}} \, \phi_j (x) u_H (z) + \cdots,\
u_H (z) = \sqrt{ \frac{2}{z_L^2 -1} } \, z,\ \ 
\phi_H(x) = \frac{1}{\sqrt{2}} \begin{pmatrix} \phi_2 + i \phi_1 \cr \phi_4 - i\phi_3 \end{pmatrix} .
\end{align}
Without loss of generality, we assume that
$\langle \phi_1 \rangle , \langle \phi_2 \rangle , 
\langle \phi_3 \rangle  =0$ and  
$\langle \phi_4 \rangle \not= 0$, 
which is related to the AB phase $\theta_H$ in the fifth dimension by
$\langle \phi_4 \rangle  = \theta_H f_H$, where
$f_H  = 2g_w^{-1}k^{1/2}L^{-1/2}(z_L^2 -1)^{-1/2}$.

\subsection{Action}

The bulk part of the action for the EW gauge and lepton sectors is given
by 
\begin{align}
S_{\rm bulk}^{\rm EW+lepton}&=
S_{\rm bulk}^{\rm EW\, gauge}+S_{\rm bulk}^{\rm lepton},
\label{Eq:Action-bulk}
\end{align}
where $S_{\rm bulk}^{\rm EW\, gauge}$ and $S_{\rm bulk}^{\rm lepton}$ are
bulk actions of the EW gauge bosons and the leptons, respectively.
The action of each gauge field, $A_M^{SO(5)_W}$ or $A_M^{U(1)_X}$,
is given in the form 
\begin{align}
S_{\rm bulk}^{\rm EW\, gauge}&=
\int d^5x\sqrt{-\det G}\, \bigg[-\mbox{tr}\left(
\frac{1}{4}F_{}^{MN}F_{MN}
+\frac{1}{2\xi}(f_{\rm gf})^2+{\cal L}_{\rm gh}\right)\bigg],
\label{Eq:Action-bulk-gauge}
\end{align}
where $\sqrt{-\det G}=1/k z^5$, $z=e^{ky}$,
$\mbox{tr}$ is a trace over all group generators for each group,
and 
$F_{MN}$ is a field strength defined by  
\begin{align}
F_{MN}&:=
\partial_MA_N-\partial_NA_M-i g[A_M,A_N]
\end{align}
with each 5D gauge coupling constant $g$.
The second and third terms in Eq.~(\ref{Eq:Action-bulk-gauge})
are the gauge fixing term and the ghost term given in
Ref.~\cite{Funatsu:2019xwr},
respectively.

Each lepton multiplet $\Psi_{({\bf 1,4})}^{\alpha}(x,y)$ in the bulk has
its own bulk-mass parameter $c_L^\alpha$ $(\alpha=1,2,3)$. The covariant
derivative is given by 
\begin{align}
&{\cal D}(c_L^\alpha)= \gamma^A {e_A}^M
\bigg( D_M+\frac{1}{8}\omega_{MBC}[\gamma^B,\gamma^C]  \bigg) 
-c_L^\alpha\sigma'(y),
\nonumber\\
&D_M =  \partial_M -i g_A A_M^{SO(5)_W} +i \frac{1}{2}g_B A_M ^{U(1)_X}.
\label{covariantD}
\end{align}
Here $\sigma'(y):=d\sigma(y)/dy$ and $\sigma'(y) =k$ for $0< y < L$. 
Then the action for the lepton sector in the bulk is given by
\begin{align}
&S_{\rm bulk}^{\rm lepton} =  \int d^5x\sqrt{-\det G} \,
\sum_{\alpha=1}^3
\overline{\Psi_{({\bf 1,4})}^{\alpha}}  {\cal D} (c_{L}^{\alpha}) 
\Psi_{({\bf 1,4})}^{\alpha},
\label{fermionAction1}
\end{align} 
where $\overline{\Psi_{({\bf 1,4})}^{\alpha}}  =
 i \Psi_{({\bf 1,4})}^{\alpha}{}^\dag\gamma^0$.
By using  $\check{\Psi}_{({\bf 1,4})}^{\alpha}
:=\Psi_{({\bf 1,4})}^{\alpha}/z^2$,
the bulk part of the fermion action is rewritten by
\begin{align}
&S_{\rm bulk}^{\rm lepton}=
\int d^4x\int_1^{z_L}\frac{dz}{k}
\sum_{\alpha=1}^3 \overline{\check{\Psi}_{({\bf 1,4})}^{\alpha}} 
\Big[ \gamma^\mu D_\mu + 
k \Big( \gamma^5 D_z - \frac{c_L^\alpha}{z}\Big) \Big] 
 \check{\Psi}_{({\bf 1,4})}^{\alpha}.
\label{fermionAction2}
\end{align}

The action for the brane scalar field $\Phi_{({\bf 1,4})}(x)$
is given by
\begin{align}
S_{\rm brane}^{\Phi} & = 
\int d^5x\sqrt{-\det G} \,  \delta(y) 
\left\{
-(D_\mu\Phi_{({\bf 1,4})})^{\dag}D^\mu\Phi_{({\bf 1,4})}
-\lambda_{\Phi_{({\bf 1,4})}}
\big(\Phi_{({\bf 1,4})}^\dag\Phi_{({\bf 1,4})} - |w|^2  \big)^2
\right\},
\label{Eq:Action-brane-scalar}
\end{align}
where 
\begin{align}
D_\mu\Phi_{({\bf 1,4})}&=
\bigg\{\partial_\mu- ig_A 
 \sum_{\alpha=1}^{10}   A_{\mu}^{\alpha} T^{\alpha}  
 -i\frac{1}{2}g_B B_\mu  \bigg\}\Phi_{({\bf 1,4})} ~.
\end{align}
Here $SO(5)_W$ generators $\{ T^\alpha \}$ consist of
$SU(2)_L$ and  $SU(2)_R$ generators $\{ T^{a_L}, T^{a_R} \}$ ($a=1,2,3$)
and $SO(5)_W/SO(4)_W$ generators $\{ T^{\hat p} = T^{p5}/\sqrt{2} \}$
($p=1,2,3,4$).
The corresponding canonically normalized gauge fields are given by
\begin{align}
A_{M}^{a_L} = \frac{1}{\sqrt{2}}
\Big(\frac{1}{2} \epsilon^{abc}A_M^{bc}+A_M^{a4}\Big),\ \
A_{M}^{a_R} = \frac{1}{\sqrt{2}}
\Big(\frac{1}{2} \epsilon^{abc}A_M^{bc} -A_M^{a4}\Big),\ \ 
A_M^{\hat p} = A_M^{p5}.
\end{align}
$B_M$ represents the $U(1)_X$ gauge field.

The action for the gauge-singlet brane fermion $\chi^\alpha (x)$
is given by
\begin{align}
S_{\rm brane}^\chi = &
\int d^5x\sqrt{-\det G} \, \delta(y) \bigg\{  
\frac{1}{2}\overline{\chi}^\alpha \gamma^\mu\partial_\mu \chi^\alpha
 - \frac{1}{2} M^{\alpha \beta}  \overline{\chi}^\alpha \chi^{\beta}
 \bigg\},
\label{Eq:Action-brane-fermion-chi}
\end{align}
where $\chi^\alpha (x)$ satisfies the Majorana condition
$\chi^c=\chi$;
\begin{align}
\chi = \begin{pmatrix} \xi \cr \eta \end{pmatrix} , ~~
\chi^c = \begin{pmatrix} + \eta^c \cr - \xi^c \end{pmatrix} 
=e^{i\delta_C} \begin{pmatrix} + \sigma^2 \eta^* \cr - \sigma^2 \xi^* \end{pmatrix} .
\label{Majorana1}
\end{align}

On the UV brane, the brane interaction terms among the bulk leptons,
the brane fermions, and the brane scalar are given by
\begin{align}
&S_{\rm brane}^{\rm int}=
-\int d^5x\sqrt{-\det G} \, \delta(y)
\Big\{ 
\widetilde{\kappa}_{\bf 1}^{\alpha \beta} \,
\overline{\chi}^\beta 
\widetilde{\Phi}_{({\bf 1,4})}^\dag \Psi_{({\bf 1,4})}^{\alpha}   + {\rm h.c.} \Big\} ~, 
\label{Eq:Action-brane-fermion}
\end{align}
where $\widetilde{\kappa}_{\bf 1}^{\alpha\beta}$ is a coupling constant.

The nonvanishing VEV $\langle\Phi_{({\bf 1,4})}\rangle\not=0$ generates
brane mass terms on the UV brane from the interaction term in
Eq.~(\ref{Eq:Action-brane-fermion}). Together with the Majorana
mass term in Eq.~(\ref{Eq:Action-brane-fermion-chi})
the brane fermion mass terms are given by
\begin{align}
&S_{\rm brane\ mass}^{\rm fermion}=
\int d^5x\sqrt{-\det G} \, \delta(y)
\left[
-\frac{m_B^{\alpha \beta}}{\sqrt{k}} \, 
 ( \overline{\chi}^\beta  \check{\nu}_{R}^{\prime\alpha}
 +\overline{\check{\nu}} {}_{R}^{\prime\alpha}  \chi^\beta )
-\frac{1}{2}  M^{\alpha \beta}\overline{\chi}^\alpha \chi^{\beta}
\right],
\label{braneFmass1}
\end{align}
where
$m_B^{\alpha\beta}=\widetilde{\kappa}_{\bf 1}^{\alpha\beta}w\sqrt{k}$.

The VEV of the brane scalar field 
$\langle\Phi_{({\bf 1,4})}\rangle\not=0$ also generates additional brane
mass terms for  the 4D components of the $SO(5)_W\times U(1)_X$ gauge
fields. It follows from Eq.~(\ref{Eq:Action-brane-scalar}) that 
\begin{align}
&S_{\rm brane}^{\rm gauge}=
\int d^5x\sqrt{-\det G} \, \delta(y) 
 \bigg\{ -\frac{g_A^2|w|^2}{4}
 \big( A_{\mu}^{1_R} A^{1_R \mu}+A_{\mu}^{2_R} A^{2_R \mu} \big)
 -\frac{(g_A^2+g_B^2)|w|^2}{4}
 A_{\mu}^{3_R'} A^{3_R' \mu} \bigg\}.
 \label{gaugeBranemass1}
\end{align}

\section{Parameter sets}
\label{Sec:Parameter-sets}

To evaluate cross sections and other observable quantities in single
Higgs boson production processes $e^-e^+\to Zh$, 
$e^-e^+\to \nu\bar{\nu}h$, 
$e^-e^+\to e^-e^+h$ at tree level in the GHU model, we need to know the
masses, decay widths, and coupling constants of the gauge bosons, the
Higgs boson, and the leptons. 
Parameters of the model are determined in the steps described in 
Refs.~\cite{Funatsu:2020haj}.

We present several parameter sets of the coupling constants of the
leptons, which are necessary for the present analysis.
In Sec.~\ref{Sec:Model}, we gave only the 5D Lagrangian of the GHU model,
but in the analysis, we use a 4D effective theory
with KK mode expansion of the 5th dimension.
By solving the equations of motions derived from the BCs of each 5D
multiplet, we can obtain the mass spectra of 4D modes for each 5D
field. Once mass spectra of a field are known, wave functions of the
zero mode and the KK modes of the field can be determined by substituting 
mass spectra into the mode function of the field.
Furthermore, coupling constants can be obtained by performing
overlap integrals of the wave functions of the corresponding fields.
For more details, see e.g., Ref.~\cite{Funatsu:2019xwr} for gauge boson
and fermion mass formulas and wave functions and
Ref.~\cite{Funatsu:2020znj} for the Higgs boson mass and coupling
formulas.

We will describe the steps to fix parameter sets in the GHU model,
where we will show parameter sets for leptons, the gauge bosons, and the
Higgs boson.
\begin{enumerate}
 \item We pick the values of $\theta_H$ and
       $m_{\rm KK}=  \pi k (z_L-1)^{-1}$.
       From the constraints on $\theta_H$ and $m_{\rm KK}$ from 
       the LHC-Run 2 results in the GHU model \cite{Funatsu:2021yrh}, we
       only consider parameters satisfying
       $\theta_H\leq 0.10$ and $m_{\rm KK}\geq 13$\,TeV.

       (Note that possible values of $m_{\rm KK}$ are restricted with
       given $\theta_H$ \cite{Funatsu:2020znj}. For example, for
       $\theta_H=0.10$ 
       $z_L\geq 10^{8.1}$ to reproduce the top quark mass, 
       and $z_L\leq 10^{15.5}$ to realize the EW symmetry breaking.
       They lead to $m_{\rm KK}\simeq[11,15] \,$TeV.)

 \item $k$ is determined in order for the $Z$ boson mass $m_Z$ to be
       reproduced, which fixes the warped factor $z_L$ as well.
       (For the mass formula of $Z$ boson, see
       Ref.~\cite{Funatsu:2019xwr}.) 

 \item The bare Weinberg angle $\theta_W^0$ in the GHU model is given 
       in Eq.~(\ref{Eq:gY-sW}).
       For each value of  $\theta_H$, the value of $\theta_W^0$ 
       is determined self-consistently to fit the observed 
       forward-backward asymmetry 
       $A_{\rm FB}(e^-e^+\to\mu^-\mu^+)=0.0169\pm0.0013$
       at $\sqrt{s}=m_Z$\cite{ALEPH:2005ema,ParticleDataGroup:2022pth},
       after evaluating the coupling constants of the muon to the $Z$
       boson with the procedure described below. We have checked that
       each self-consistent value of
       $\theta_W^0$ is found after a couple of iterations of this
       process.

\begin{table}[htb]
{
\begin{center}
\begin{tabular}{c|cc|cc|c}
\hline
 \rowcolor[gray]{0.9}
 &&&&&\\[-0.75em]
\rowcolor[gray]{0.9}
 Name&$\theta_H$&$m_{\rm KK}$&$z_L$&$k$
 &$\sin^2\theta_W^0$\\
\rowcolor[gray]{0.9}
 &\mbox{[rad.]}&[TeV]&&[GeV]&\\ 
\hline 
 &&&&\\[-0.75em]
 A$_{-}$&0.10&13.00&
 3.865$\times10^{11}$&1.599$\times10^{15}$&0.2306\\
 A$_{+}$&0.10&13.00&
 4.029$\times10^{11}$&1.667$\times10^{15}$&0.2318\\
 B$_{-}$&0.07&19.00&
 1.420$\times10^{12}$&8.589$\times10^{15}$&0.2309\\
 B$_{+}$&0.07&19.00&
 1.452$\times10^{12}$&8.779$\times10^{15}$&0.2315\\
 C$_{-}$&0.05&25.00&
 5.546$\times10^{10}$&4.413$\times10^{14}$&0.2310\\
 C$_{+}$&0.05&25.00&
 5.600$\times10^{10}$&4.456$\times10^{14}$&0.2313\\
\hline
\end{tabular}
 \caption{\small
 The name of the parameter set and the corresponding $z_L$, $k$, and
 $\sin^2\theta_W^0$ for each $\theta_H$ and $m_{\rm KK}$
 are listed.
 In the SM, $\sin^2\theta_W(\overline{\mbox{MS}})=0.23122\pm0.00004$
\cite{ParticleDataGroup:2022pth}.
 The column ``Name'' denotes each parameter set.
 }
\label{Table:Parameter-sets}
\end{center}
}
\end{table}

       The parameter sets of $(\theta_H,m_{\rm KK})$, named
       $A_{\pm}$,$B_{\pm}$ and $C_{\pm}$, used in this
       analysis are summarized in Table~\ref{Table:Parameter-sets},
       where the subscripts denote the sign of the  bulk masses of the
       leptons. For example, $A_{+}$ denotes the case where the bulk mass
       of the lepton is positive and $A_{-}$ denotes the case
       where the bulk mass of the lepton is negative.

 \item With given $\sin \theta_W^0$, wave functions of the gauge bosons
       are fixed. 
       Masses and widths of $Z'$, $W$, $W'$ bosons are listed for each
       parameter set in Table~\ref{Table:Mass-Width-Vector-Bosons}.
       Although information on parameters of the quark sector is also
       needed when we determine the decay widths of the gauge bosons,
       the specific values of the parameters of the quark sector are
       omitted   since they 
       are not directly  used in this analysis. 
       The parameters of the quark sector are obtained by the same
       procedure as for the lepton sector. 
       (For more detail, see Ref.~\cite{Funatsu:2019xwr}.)

\begin{table}[htb]
{
\begin{center}
\begin{tabular}{c|ccccccccccc}
\hline
 \rowcolor[gray]{0.9}
 &&&&&&&&&&\\[-0.75em]
\rowcolor[gray]{0.9}
 Name
 &$m_{Z^{(1)}}$&$\Gamma_{Z^{(1)}}$
 &$m_{Z_R^{(1)}}$&$\Gamma_{Z_R^{(1)}}$
 &$m_{W}$&$\Gamma_{W}$
 &$m_{W^{(1)}}$&$\Gamma_{W^{(1)}}$
 &$m_{W_R^{(1)}}$&$\Gamma_{W_R^{(1)}}$\\
\rowcolor[gray]{0.9}
 &[TeV]&[TeV]&[TeV]&[TeV]&[GeV]&[GeV]&[TeV]&[TeV]&[TeV]&[TeV]
 \\ 
\hline 
 &&&&&&&\\[-0.75em]
 A$_{-}$
 &10.196&7.840
 & 9.951&0.816
 & 79.98&2.02
 &10.196&9.802
 & 9.951&0.359
 \\
 A$_{+}$
 &10.196&5.982
 & 9.951&0.775
 & 79.92&2.01
 &10.196&7.163
 & 9.951&0.358
 \\
 B$_{-}$
 &14.886&11.962
 &14.544&1.231
 & 79.97&2.02
 &14.886&14.970
 &14.544&0.549
 \\
 B$_{+}$
 &14.886&9.162
 &14.544&1.168
 &79.94&2.02
 &14.886&10.979
 &14.544&0.548
 \\
 C$_{-}$
 &19.648&14.006
 &19.137&1.505
 &79.96&2.03
 &19.648&17.502
 &19.137&0.639
 \\
 C$_{+}$
 &19.648&10.700
 &19.137&1.432
 &79.95&2.02
 &19.648&12.784
 &19.137&0.638
 \\
\hline
\end{tabular}
 \caption{\small
 Masses and widths of $Z'$, $W$, $W'$ bosons are listed for each parameter
 set listed in Table~\ref{Table:Parameter-sets}.
 In the SM, $\alpha_{\rm EM}=1/128$ at the mass scale of $Z$ boson,
 $m_{Z}=91.1876$\,GeV, $\Gamma_{Z}=2.4952$\,GeV,
 $m_W=80.377\pm 0.012$\,GeV, $\Gamma_{W}=2.085\pm0.042$\,GeV,
 \cite{ParticleDataGroup:2022pth}.
 The column ``Name''   denotes each parameter set 
 in Table~\ref{Table:Parameter-sets}.
 $m_V$ and $\Gamma_V$ $(V=Z^{(1)},Z_R^{(1)},W,W^{(1)},W_R^{(1)})$ are
 the mass and decay width of each $V$ boson, respectively.
 Decay widths are calculated by using the formulas in
 Ref.~\cite{Funatsu:2020haj}, where all possible two-body final states
 are taken into account.
 }
\label{Table:Mass-Width-Vector-Bosons}
\end{center}
}
\end{table}

 \item The bulk masses of the leptons $c_L^\alpha$ in
       Eq.~(\ref{fermionAction2})
       are determined so as to reproduce the masses of charged leptons,
       where we denote $c_L^1,c_L^2,c_L^3$ as $c_e,c_\mu,c_\tau$.
       We use the masses of the electron, the muon, and the tau lepton
       given by $m_e=0.48657\,$MeV, $m_\mu=102.72\,$MeV, and
       $m_\tau=1.7462\,$GeV
       at $\mu=m_Z$ \cite{Xing:2007fb}.
       The parameters of the Majorana mass terms and brane interactions
       in the neutrino sector in Eq.~(\ref{braneFmass1}),
       $M^{\alpha\beta}$ and $m_B^{\alpha\beta}$,
       are determined so as to reproduce the neutrino masses.
       The neutrino masses cannot be completely fixed because only two
       mass-squared differences are known from observations
       \cite{ParticleDataGroup:2022pth}.
       In our analysis, we take
       $m_{\nu_e}=m_{\nu_\mu}=m_{\nu_\tau}=10^{-12}$\,GeV
       for reference.
       For simplicity we assume that $M^{\alpha\beta}$ and
       $m_B^{\alpha\beta}$ are diagonal to the flavor.
       We take $M^{\alpha\beta}=M_\ell\delta_{\alpha\beta}$
       $(\alpha,\beta=1,2,3)$,
       $m_B^{11}=m_{B_{e}}$, $m_B^{22}=m_{B_{\mu}}$, 
       $m_B^{33}=m_{B_{\tau}}$, and
       $m_B^{\alpha\beta}=0$ for $\alpha\not=\beta$.
       The neutrino masses are so small that a difference of one to two
       orders of magnitude in their masses will not affect the results
       of this analysis. Neutrino mixings are ignored, but they do not
       affect the present analysis.
       The bulk masses and the brane interaction
       parameters of the leptons are listed in
       Table~\ref{Table:Bulk-and-brane-prameters}.

\begin{table}[htb]
{
\begin{center}
\begin{tabular}{c|cccc|ccc}
\hline
 \rowcolor[gray]{0.9}
 &&&&&&&\\[-0.75em]
\rowcolor[gray]{0.9}
 Name&$c_e$&$c_\mu$&$c_\tau$
 &$M_\ell$
 &$m_{Be}$&$m_{B\mu}$&$m_{B\tau}$\\
\rowcolor[gray]{0.9}
 &&&&[GeV]&[GeV]&[GeV]&[GeV]\\ 
\hline 
 &&&&&&&\\[-0.75em]
 A$_{-}$
 &$-$1.0067&$-$0.7929&$-$0.6753
 &$10^6$
 &4.834$\times10^5$&1.342$\times10^8$&2.949$\times10^9$
 \\
 A$_{+}$
 &$+$1.0058&$+$0.7924&$+$0.6750
 &$10^6$
 &8.382$\times10^{22}$&6.373$\times10^{22}$&4.931$\times10^{22}$
 \\
 B$_{-}$
 &$-$0.9828&$-$0.7791&$-$0.6670
 &$10^6$
 &4.952$\times10^5$&1.375$\times10^8$&3.021$\times10^9$
 \\
 B$_{+}$
 &$+$0.9824&$+$0.7788&$+$0.6669
 &$10^6$
 &3.017$\times10^{23}$&2.294$\times10^{23}$&1.775$\times10^{23}$
 \\
 C$_{-}$
 &$-$1.0469&$-$0.8162&$-$0.6893
 &$10^6$
 &4.652$\times10^5$&1.292$\times10^8$&2.838$\times10^9$
 \\
 C$_{+}$
 &$+$1.0467&$+$0.8161&$+$0.6982
 &$10^6$
 &1.165$\times10^{22}$&8.857$\times10^{21}$&6.853$\times10^{21}$
 \\
\hline
\end{tabular}
 \caption{\small
 The bulk masses of electron, muon, and tau lepton, $c_e, c_\mu, c_\tau$ 
 and the brane interaction parameters of leptons 
 $M_\ell, m_{B_e}, m_{B_mu}, m_{B_\tau}$ are listed.
 In the SM, $m_e=0.48657\,$MeV, $m_\mu=102.72\,$MeV, 
 and $m_\tau=1.7462\,$GeV at $\mu=m_Z$ \cite{Xing:2007fb}.
 We take the neutrino masses:
 $m_{\nu_e}=m_{\nu_\mu}=m_{\nu_\tau}=10^{-12}$\,GeV.
The column ``Name'' denotes each parameter set
 in Table~\ref{Table:Parameter-sets}.
 }
\label{Table:Bulk-and-brane-prameters}
\end{center}
}
\end{table}

       In Table~\ref{Table:Bulk-and-brane-prameters},
       it may be problematic that the mass parameters
       $m_{B\ell}(\ell=e,\mu,\tau)$ 
       of the brane fermions are larger than the Planck mass for the
       cases     of positive bulk masses of the leptons. 
       Furthermore, it has been pointed out in
       Ref.~\cite{Funatsu:2019xwr} that the appearance of extra
       vector-like neutrinos with MeV-scale masses is for the case of
       the positive bulk masses of the leptons.
       The sign of the bulk masses causes a qualitative difference
       in the behavior of the coupling constants of fermions to the KK
       gauge bosons, so in this paper we consider the parameter sets of
       the positive bulk masses of the leptons as well as
       the negative bulk masses of the leptons for reference.

 \item With given the bulk masses and the brane interaction parameters,
       wave functions of fermions are fixed. 

 \item The mass and self-coupling constant of the Higgs boson can be
      obtained from the effective potential 
       of the Higgs boson
       \cite{Funatsu:2020znj}.
       The mass of the The Higgs boson is determined by adjusting the
       bulk mass of the dark fermions so that the mass of the Higgs
       boson is  $m_h=125.25\pm 0.17$\,GeV
       \cite{ParticleDataGroup:2022pth}.

\end{enumerate}

To evaluate cross sections for single Higgs boson production processes
such as $e^-e^+\to Zh$, $e^-e^+\to \nu\bar{\nu}h$, $e^-e^+\to e^-e^+h$, we
need to know not only the masses and the decay widths but also coupling
constants of the gauge bosons, the Higgs boson, and the leptons. They 
are obtained from the five-dimensional gauge interaction terms by
substituting the wave functions of gauge bosons and fermions or the 
Higgs boson and integrating over the fifth-dimensional
coordinate\cite{Funatsu:2014fda,Funatsu:2015xba,Funatsu:2016uvi}.
\begin{itemize}
\item  The coupling constants of the gauge bosons to the leptons are
       obtained by performing overlap integrals of the wave functions in
       the fifth dimension of gauge bosons and leptons.
       Coupling constants of gauge bosons to leptons
       are listed in Tables~\ref{Table:Gauge-Fermion-Couplings-W-lepton},
       \ref{Table:Gauge-Neutrino-Couplings}, and
       \ref{Table:Gauge-Charged-Lepton-Couplings}.

\begin{table}[htb]
\begin{center}
\begin{tabular}{c|cccccc}
\hline
 \rowcolor[gray]{0.9}
 &&&&&&\\[-0.75em]
\rowcolor[gray]{0.9}
 Name
 &$g_{We\nu_e}^L$&$g_{We\nu_e}^R$
 &$g_{W^{(1)}e\nu_e}^L$&$g_{W^{(1)}e\nu_e}^R$
 &$g_{W_R^{(1)}e\nu_e}^L$&$g_{W_R^{(1)}e\nu_e}^R$
 \\
 \hline
 A$_{-}$
 &$+$0.9976&0
 &$+$5.7451&0
 &$+$0.0145&0
 \\
 A$_{+}$
 &$+$0.9988&0
 &$-$0.2220&0
 &0&0
 \\
 B$_{-}$
 &$+$0.9988&0
 &$+$5.8583&0
 &$+$0.0073&0
 \\
 B$_{+}$
 &$+$0.9994&0
 &$-$0.2171&0
 &0&0
 \\
 C$_{-}$
 &$+$0.9994&0
 &$+$5.5706&0
 &$+$0.0035&0
 \\
 C$_{+}$
 &$+$0.9997&0
 &$-$0.2316&0
 &0&0
 \\
\hline
\end{tabular}\\[0.5em]
\begin{tabular}{c|cccccc}
\hline
 \rowcolor[gray]{0.9}
 &&&&&&\\[-0.75em]
\rowcolor[gray]{0.9}
 Name
 &$g_{W\mu\nu_\mu}^L$&$g_{W\mu\nu_\mu}^R$
 &$g_{W^{(1)}\mu\nu_\mu}^L$&$g_{W^{(1)}\mu\nu_\mu}^R$
 &$g_{W_R^{(1)}\mu\nu_\mu}^L$&$g_{W_R^{(1)}\mu\nu_\mu}^R$
 \\
 \hline
 A$_{-}$
 &$+$0.9976&0
 &$+$5.4705&0
 &$+$0.0139&0
 \\
 A$_{+}$
 &$+$0.9988&0
 &$-$0.2222&0
 &0&0
 \\
 B$_{-}$
 &$+$0.9988&0
 &$+$5.5850&0
 &$+$0.0069&0
 \\
 B$_{+}$
 &$+$0.9994&0
 &$-$0.2171&0
 &0&0
 \\
 C$_{-}$
 &$+$0.9994&0
 &$+$5.2941&0
 &$+$0.0034&0
 \\
 C$_{+}$
 &$+$0.9997&0
 &$-$0.2316&0
 &0&0
 \\
\hline
\end{tabular}\\[0.5em]
\begin{tabular}{c|cccccc}
\hline
 \rowcolor[gray]{0.9}
 &&&&&&\\[-0.75em]
\rowcolor[gray]{0.9}
 Name
 &$g_{W\tau\nu_\tau}^L$&$g_{W\tau\nu_\tau}^R$
 &$g_{W^{(1)}\tau\nu_\tau}^L$&$g_{W^{(1)}\tau\nu_\tau}^R$
 &$g_{W_R^{(1)}\tau\nu_\tau}^L$&$g_{W_R^{(1)}\tau\nu_\tau}^R$
 \\
 \hline
 A$_{-}$
 &$+$0.9976&0
 &$+$5.2878&0
 &$+$0.0134&0
 \\
 A$_{+}$
 &$+$0.9988&0
 &$-$0.2217&0
 &0&0
 \\
 B$_{-}$
 &$+$0.9988&0
 &$+$5.4046&0
 &$+$0.0067&0
 \\
 B$_{+}$
 &$+$0.9994&0
 &$-$0.2169&0
 &0&0
 \\
 C$_{-}$
 &$+$0.9994&0
 &$+$5.1077&0
 &$+$0.0032&0
 \\
 C$_{+}$
 &$+$0.9997&0
 &$-$0.2314&0
 &0&0
 \\
\hline
\end{tabular}\\[0.5em]
 \caption{\small
 Coupling constants of 
 charged vector bosons, $W,W'(=W^{(1)},W_R^{(1)})$ bosons, to
 leptons in units of $g_w/\sqrt{2}$
 are listed.
 The fine structure constant is the same as in the SM: 
 $\alpha(M_Z)^{-1}=127.951\pm 0.009$
 \cite{ParticleDataGroup:2022pth}.
 In the SM, $g_w=e/\sin\theta_W$, while
 in the GHU
 $g_w=e/\sin\theta_W^0$.
 The coupling constants of $W$ boson to leptons in the SM  are 
 $g_{We\nu_e}^L=1$.
 When the value is less than $10^{-4}$, we write $0$.
 }
\label{Table:Gauge-Fermion-Couplings-W-lepton}
\end{center}
\end{table}

\begin{table}[htb]
\begin{center}
\begin{tabular}{c|cccccc}
\hline
 \rowcolor[gray]{0.9}
 &&&&&&\\[-0.75em]
\rowcolor[gray]{0.9}
 Name
 &$g_{Z\nu_e}^L$&$g_{Z\nu_e}^R$
 &$g_{Z^{(1)}\nu_e}^L$&$g_{Z^{(1)}\nu_e}^R$
 &$g_{Z_R^{(1)}\nu_e}^L$&$g_{Z_R^{(1)}\nu_e}^R$
 \\
 \hline
 A$_{-}$
 &$+$0.5687&0
 &$+$3.2774&0
 &$-$1.0322&0
 \\
 A$_{+}$
 &$+$0.5691&0
 &$-$0.1264&0
 &0&0
 \\
 B$_{-}$
 &$+$0.5695&0
 &$+$3.3413&0
 &$-$1.0586&0
 \\
 B$_{+}$
 &$+$0.5697&0
 &$-$0.1237&0
 &0&0
 \\
 C$_{-}$
 &$+$0.5698&0
 &$+$3.1768&0
 &$-$1.0099&0
 \\
 C$_{+}$
 &$+$0.5699&0
 &$-$0.1320&0
 &0&0
 \\
\hline
\end{tabular}\\[0.5em]
\begin{tabular}{c|cccccc}
\hline
 \rowcolor[gray]{0.9}
 &&&&&&\\[-0.75em]
\rowcolor[gray]{0.9}
 Name
 &$g_{Z\nu_\mu}^L$&$g_{Z\nu_\mu}^R$
 &$g_{Z^{(1)}\nu_\mu}^L$&$g_{Z^{(1)}\nu_\mu}^R$
 &$g_{Z_R^{(1)}\nu_\mu}^L$&$g_{Z_R^{(1)}\nu_\mu}^R$
 \\
 \hline
 A$_{-}$
 &$+$0.5687&0
 &$+$3.1207&0
 &$-$0.9852&0
 \\
 A$_{+}$
 &$+$0.5691&0
 &$-$0.1264&0
 &0&0
 \\
 B$_{-}$
 &$+$0.5695&0
 &$+$3.1854&0
 &$-$1.0114&0
 \\
 B$_{+}$
 &$+$0.5697&0
 &$-$0.1237&0
 &0&0
 \\
 C$_{-}$
 &$+$0.5698&0
 &$+$3.0191&0
 &$-$0.9623&0
 \\
 C$_{+}$
 &$+$0.5699&0
 &$-$0.1320&0
 &0&0
 \\
\hline
\end{tabular}\\[0.5em]
\begin{tabular}{c|cccccc}
\hline
 \rowcolor[gray]{0.9}
 &&&&&&\\[-0.75em]
\rowcolor[gray]{0.9}
 Name
 &$g_{Z\nu_\tau}^L$&$g_{Z\nu_\tau}^R$
 &$g_{Z^{(1)}\nu_\tau}^L$&$g_{Z^{(1)}\nu_\tau}^R$
 &$g_{Z_R^{(1)}\nu_\tau}^L$&$g_{Z_R^{(1)}\nu_\tau}^R$
 \\
 \hline
 A$_{-}$
 &$+$0.5691&0
 &$+$3.0165&0
 &$-$0.9539&0
 \\
 A$_{+}$
 &$+$0.5687&0
 &$-$0.1262&0
 &0&0
 \\
 B$_{-}$
 &$+$0.5695&0
 &$+$3.0825&0
 &$-$0.9803&0
 \\
 B$_{+}$
 &$+$0.5697&0
 &$-$0.1236&0
 &0&0
 \\
 C$_{-}$
 &$+$0.5698&0
 &$+$2.9129&0
 &$-$0.9302&0
 \\
 C$_{+}$
 &$+$0.5699&0
 &$-$0.1319&0
 &0&0
 \\
\hline
\end{tabular}
 \caption{\small
 Coupling constants of neutral vector bosons, $Z,Z'(=Z^{(1)},Z_R^{(1)})$
 bosons, to neutrinos in units of $g_w=e/\sin\theta_W^0$
 are listed.
 The coupling constants of $Z$ boson to neutrinos in the
 SM are  $(g_{Z_{\nu}}^L,g_{Z_{\nu}}^R)=(0.5703,0)$
 in units of $g_w$.
 Other information is the same as in 
 Table~\ref{Table:Gauge-Fermion-Couplings-W-lepton}.
 }
\label{Table:Gauge-Neutrino-Couplings}
\end{center}
\end{table}

\begin{table}[htb]
\begin{center}
\begin{tabular}{c|cccccc}
\hline
 \rowcolor[gray]{0.9}
 &&&&&&\\[-0.75em]
\rowcolor[gray]{0.9}
 Name
 &$g_{Ze}^L$&$g_{Ze}^R$
 &$g_{Z^{(1)}e}^L$&$g_{Z^{(1)}e}^R$
 &$g_{Z_R^{(1)}e}^L$&$g_{Z_R^{(1)}e}^R$
 \\
 \hline
 A$_{-}$
 &$-$0.3058&$+$0.2629
 &$-$1.7621&$-$0.0584
 &$-$1.0444&0
 \\
 A$_{+}$
 &$-$0.3060&$+$0.2631
 &$+$0.0680&$+$1.5171
 &0&$+$1.3826
 \\
 B$_{-}$
 &$-$0.3062&$+$0.2633
 &$-$1.7964&$-$0.0572
 &$-$1.0646&0
 \\
 B$_{+}$
 &$-$0.3063&$+$0.2634
 &$+$0.0665&$+$1.5459
 &0&$+$1.4102
 \\
 C$_{-}$
 &$-$0.3064&$+$0.2634
 &$-$1.7082&$-$0.0610
 &$-$1.0128&0
 \\
 C$_{+}$
 &$-$0.3065&$+$0.2635
 &$+$0.0710&$+$1.4690
 &0&$+$1.3428
 \\
\hline
\end{tabular}\\[0.5em]
\begin{tabular}{c|cccccc}
\hline
 \rowcolor[gray]{0.9}
 &&&&&&\\[-0.75em]
\rowcolor[gray]{0.9}
 Name
 &$g_{Z\mu}^L$&$g_{Z\mu}^R$
 &$g_{Z^{(1)}\mu}^L$&$g_{Z^{(1)}\mu}^R$
 &$g_{Z_R^{(1)}\mu}^L$&$g_{Z_R^{(1)}\mu}^R$
 \\
 \hline
 A$_{-}$
 &$-$0.3058&$+$0.2629
 &$-$1.6778&$-$0.0584
 &$-$0.9969&0
 \\
 A$_{+}$
 &$-$0.3060&$+$0.2631
 &$+$0.0680&$+$1.4447
 &0&$+$1.3197
 \\
 B$_{-}$
 &$-$0.3062&$+$0.2633
 &$-$1.7126&$-$0.0572
 &$-$1.0173&0
 \\
 B$_{+}$
 &$-$0.3063&$+$0.2634
 &$+$0.0665&$+$1.4738
 &0&$+$1.3474
 \\
 C$_{-}$
 &$-$0.3064&$+$0.2634
 &$-$1.6234&$-$0.0610
 &$-$0.9651&0
 \\
 C$_{+}$
 &$-$0.3065&$+$0.2635
 &$+$0.0710&$+$1.3961
 &0&$+$1.2796
 \\
\hline
\end{tabular}\\[0.5em]
\begin{tabular}{c|cccccc}
\hline
 \rowcolor[gray]{0.9}
 &&&&&&\\[-0.75em]
\rowcolor[gray]{0.9}
 Name
 &$g_{Z\tau}^L$&$g_{Z\tau}^R$
 &$g_{Z^{(1)}\tau}^L$&$g_{Z^{(1)}\tau}^R$
 &$g_{Z_R^{(1)}\tau}^L$&$g_{Z_R^{(1)}\tau}^R$
 \\
 \hline
 A$_{-}$
 &$-$0.3058&$+$0.2629
 &$-$1.6218&$-$0.0584
 &$-$0.9652&0
 \\
 A$_{+}$
 &$-$0.3060&$+$0.2631
 &$+$0.0679&$+$1.3965
 &0&$+$1.2778
 \\
 B$_{-}$
 &$-$0.3062&$+$0.2633
 &$-$1.6473&$-$0.0571
 &$-$0.9859&0
 \\
 B$_{+}$
 &$-$0.3063&$+$0.2634
 &$+$0.0664&$+$1.4262
 &0&$+$1.3060
 \\
 C$_{-}$
 &$-$0.3064&$+$0.2634
 &$-$1.5663&$-$0.0610
 &$-$0.9330&0
 \\
 C$_{+}$
 &$-$0.3065&$+$0.2635
 &$+$0.0710&$+$1.3470
 &0&$+$1.2370
 \\
\hline
\end{tabular}
 \caption{\small
 Coupling constants of neutral vector bosons, $Z,Z'$ bosons, to
 charged leptons in units of $g_w=e/\sin\theta_W^0$
 are listed.
 The coupling constants of $Z$ boson to charged leptons in the SM are
 $(g_{Z_{e}}^L,g_{Z_{e}}^R)=(-0.3065,0.2638)$ 
 in units of $g_w$.
 Other information is the same as in 
 Table~\ref{Table:Gauge-Fermion-Couplings-W-lepton}.
 }
\label{Table:Gauge-Charged-Lepton-Couplings}
\end{center}
\end{table}

       As can be seen from Tables
       \ref{Table:Gauge-Fermion-Couplings-W-lepton},
       \ref{Table:Gauge-Neutrino-Couplings}, and 
       \ref{Table:Gauge-Charged-Lepton-Couplings}
       the coupling constants of the zero modes and the 1st KK modes of
       the $Z$ and $W$ bosons to the zero modes of the leptons in
       the GHU are very close to those in the SM. More specifically, the
       deviations from the SM are less than 1\%. There is almost no
       difference in the coupling constants of the $Z$ and $W$ bosons to
       the leptons for the positive and negative bulk masses of the
       leptons, while there is a large difference in the coupling
       constants of the KK gauge bosons, $Z'(=Z^{(1)},Z_R^{(1)})$ and 
       $W'(=W^{\pm(1)},W_R^{\pm(1)})$ bosons, to the zero modes of the
       leptons.
       When the bulk masses of leptons are negative, the coupling
       constants of the 1st KK gauge bosons to the zero modes of the
       left-handed leptons are large, and those of the 1st KK gauge
       bosons to the zero modes of the right-handed leptons are small. 
       On the other hand, when the bulk masses of the leptons are
       positive, the coupling constants of the 1st KK gauge bosons 
       to the zero modes of the right-handed leptons are large and those
       of the 1st KK gauge bosons to the zero modes of the left-handed
       leptons are small.

\item The coupling constants of gauge bosons to the Higgs boson are
      obtained by performing overlap integrals of the wave functions in
      the fifth dimension of gauge bosons and the Higgs boson.
      Coupling constants of gauge bosons to Higgs boson are listed
      Table~\ref{Table:Higgs-Gauge-Couplings}.

\begin{table}[htb]
\begin{center}
\begin{tabular}{c|cccccc}
\hline
 \rowcolor[gray]{0.9}
 &&&&&&\\[-0.75em]
\rowcolor[gray]{0.9}
 Name
 &$g_{WWh}$
 &$g_{WW^{(1)}h}$
 &$g_{W^{(1)}W^{(1)}h}$
 &$g_{WW_R^{(1)}h}$
 &$g_{W_R^{(1)}W_R^{(1)}h}$
 &$g_{W^{(1)}W_R^{(1)}h}$
 \\
\rowcolor[gray]{0.9}
 &[GeV]
 &[GeV]
 &[GeV]
 &[GeV]
 &[GeV]
 &[GeV]
 \\
 \hline
 A$_{-}$
 &$+$79.60&$+$395.80
 &$-$315.90&$+$405.14
 &0&$-$156.05
 \\
 A$_{+}$
 &$+$79.54&$+$395.81
 &$-$315.62&$+$405.14
 &0&$-$155.92
 \\
 B$_{-}$
 &$+$79.78&$+$406.12
 &$-$315.67&$+$414.29
 &0&$-$155.67
 \\
 B$_{+}$
 &$+$79.75&$+$406.12
 &$-$315.53&$+$414.29
 &0&$-$155.60
 \\
 C$_{-}$
 &$+$79.87&$+$381.16
 &$-$318.16&$+$389.29
 &0&$-$156.69
 \\
 C$_{+}$
 &$+$79.85&$+$381.17
 &$-$318.80&$+$389.29
 &0&$-$156.65
 \\
\hline
\end{tabular}\\[0.5em]
\begin{tabular}{c|cccccc}
\hline
 \rowcolor[gray]{0.9}
 &&&&&&\\[-0.75em]
\rowcolor[gray]{0.9}
 Name
 &$g_{ZZh}$
 &$g_{ZZ^{(1)}h}$
 &$g_{Z^{(1)}Z^{(1)}h}$
 &$g_{ZZ_R^{(1)}h}$
 &$g_{Z_R^{(1)}Z_R^{(1)}h}$
 &$g_{Z^{(1)}Z_R^{(1)}h}$
 \\
\rowcolor[gray]{0.9}
 &[GeV]
 &[GeV]
 &[GeV]
 &[GeV]
 &[GeV]
 &[GeV]
 \\
 \hline
 A$_{-}$
 &$+$103.47&$+$514.84
 &$-$410.55&$+$386.81
 &0&$-$148.87
 \\
 A$_{+}$
 &$+$103.55&$+$515.66
 &$-$410.82&$+$386.56
 &0&$-$148.64
 \\
 B$_{-}$
 &$+$103.74&$+$528.25
 &$-$410.43&$+$395.33
 &0&$-$148.48
 \\
 B$_{+}$
 &$+$103.78&$+$528.67
 &$-$410.57&$+$395.20
 &0&$-$148.37
 \\
 C$_{-}$
 &$+$103.86&$+$495.76
 &$-$414.64&$+$371.38
 &0&$-$149.45
 \\
 C$_{+}$
 &$+$103.88&$+$495.96
 &$-$414.71&$+$371.32
 &0&$-$149.39
 \\
\hline
\end{tabular}
 \caption{\small
 Coupling constants of gauge bosons to Higgs boson in units of $g_w$
 are listed.
 In the SM, $g_{WWh}/g_w=m_W=80.38$\,GeV and 
 $g_{ZZh}/g_w=m_Z /\cos\theta_W=104.00$\,GeV.
 When the value is less than $0.1$, we write $0$.
 }
\label{Table:Higgs-Gauge-Couplings}
\end{center}
\end{table}

      From Table~\ref{Table:Higgs-Gauge-Couplings},
      the coupling constants of the $Z$ and $W$ bosons to the Higgs
      boson in the GHU model are very close to those in the SM. 
      More specifically, the deviations from the SM are less than 1\%.

\end{itemize}

{

We comment on the expected improvement in the accuracy of
the Weinberg angle in the ILC experiment.
At the moment we use $A_{\rm FB}(e^-e^+\to\mu^-\mu^+)$ to
determine the value of the Weinberg angle $\theta_W^0$.
If the uncertainty of $A_{\rm FB}(e^-e^+\to\mu^-\mu^+)$ is reduced
by future experiments, we can use exactly the same method to
determine the parameters. 
If the uncertainty of $A_{\rm FB}(e^-e^+\to\mu^-\mu^+)$ is reduced by a
factor of 1/10, the uncertainty of $\sin^2\theta_W^0$ is also reduced by
a factor of about 1/10.
However, the (unpolarized) forward-backward asymmetry 
$A_{\rm FB}$ is not a direct observable in the ILC experiment
with polarized electrons and positrons. 
Although it is possible to obtain $A_{\rm FB}$ by combining
observables, it is known that the accuracy of Weinberg angle
determination can be improved by using more direct observables,
polarization asymmetry parameters 
$A_f(f=e,\mu,...)$ 
\cite{ILCInternationalDevelopmentTeam:2022izu,ParticleDataGroup:2022pth}.
The relations between the Weinberg angle and 
the asymmetry parameters are approximately
given as 
\begin{align}
A_f\simeq
8\left(\frac{1}{4}-|Q_f|\sin^2\theta_{\rm eff}^{f}\right).
\end{align}
From this relationship, the accuracy of
$\sin^2\theta_{\rm eff}^f$ is estimated to be 1/8 times the
decision accuracy of $A_f$. This is also true for
$\sin^2\theta_W^0$ in the GHU model.
At present, the accuracy of $\sin^2\theta_W^0$ is $O(0.1)\%$ by
using the currently forward-backward asymmetry 
$A_{\rm FB}(e^-e^+\to\mu^-\mu^+)$.
We expect that the decision accuracy of $\sin^2\theta_W^0$ will be
$O(0.01)\%$ by using the expected accuracy of the asymmetry parameter
$A_e$ at the ILC experiment with $\sqrt{s}=250$\,GeV and 
$L_{\rm int}=2$\,ab$^{-1}$
\cite{ILCInternationalDevelopmentTeam:2022izu};
$O(0.001)\%$ 
by using the expected accuracy of the asymmetry parameter
$A_e$ at the ILC experiment with $\sqrt{s}=m_Z\simeq 91.2$\,GeV and 
$L_{\rm int}=100$\,fb$^{-1}$
\cite{ILCInternationalDevelopmentTeam:2022izu}.

}

\section{Numerical analysis}
\label{Sec:Results}

We calculate cross sections of single Higgs boson production
processes.
More specifically, we analyze three processes $e^-e^+\to Zh$, 
$e^-e^+\to \nu\bar{\nu}h$, $e^-e^+\to e^-e^+h$ for 
the initial states of unpolarized and polarized electrons and positrons,
where we use the formula of the cross sections given in
Appendix~\ref{Sec:Cross-section}. For the values of the initial
polarizations, we mainly use $(P_{e^-},P_{e^+})=(\mp0.8,\pm0.3)$
with the ILC experiment in mind. 

\subsection{$e^-e^+\to Zh$}

Here we evaluate observables of the $e^-e^+\to Zh$ process in the SM and
the GHU model at tree level. We use the
parameter sets A$_{\pm}$, B$_{\pm}$, C$_{\pm}$ listed in
Tables~\ref{Table:Parameter-sets}, 
\ref{Table:Mass-Width-Vector-Bosons}, 
\ref{Table:Bulk-and-brane-prameters},
\ref{Table:Gauge-Fermion-Couplings-W-lepton},
\ref{Table:Gauge-Neutrino-Couplings},
\ref{Table:Gauge-Charged-Lepton-Couplings}, and
\ref{Table:Higgs-Gauge-Couplings}.

\begin{figure}[htb]
\begin{center}
\includegraphics[bb=0 0 415 290,height=5.5cm]{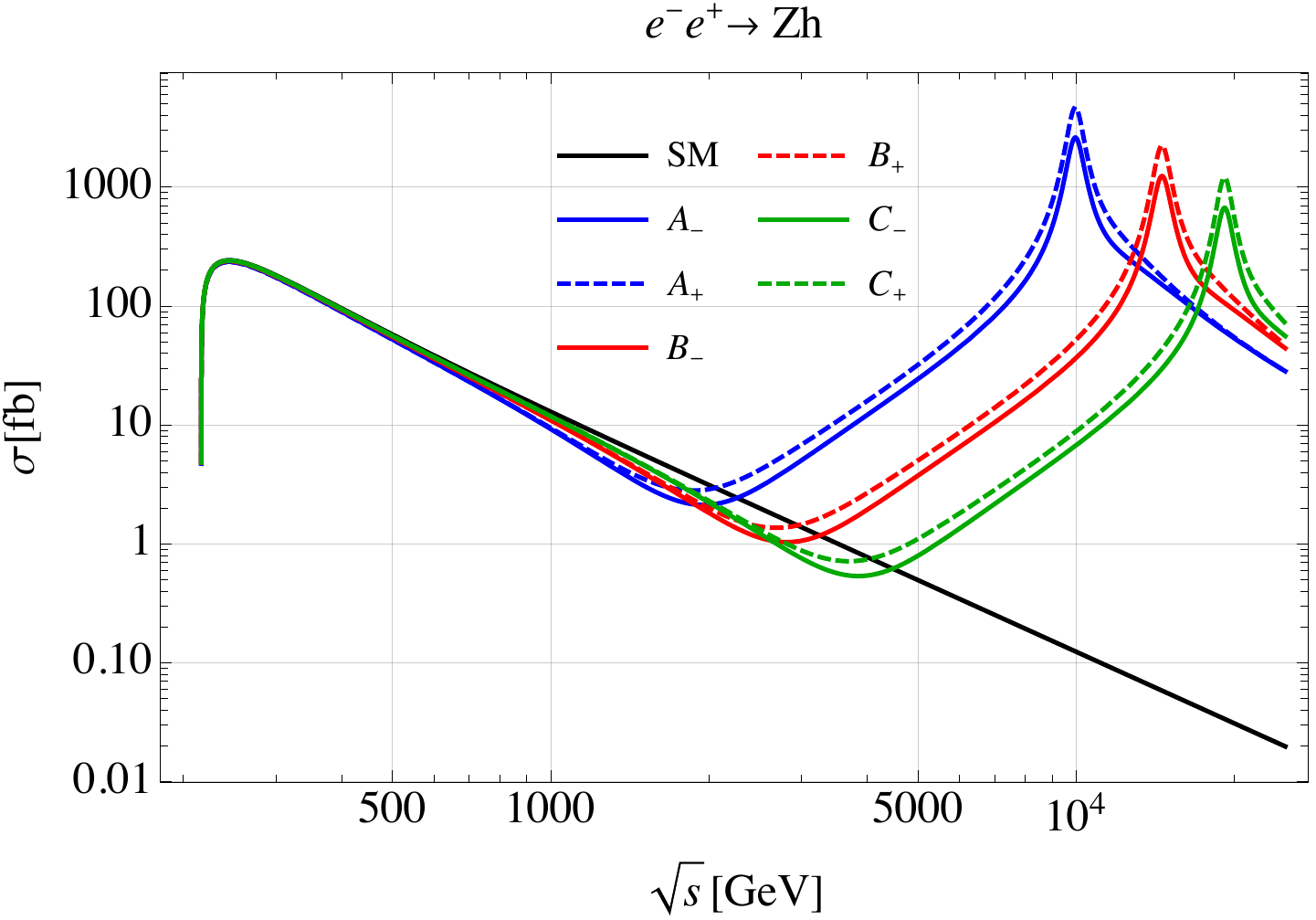}
\ \ 
\includegraphics[bb=0 0 504 357,height=5.5cm]{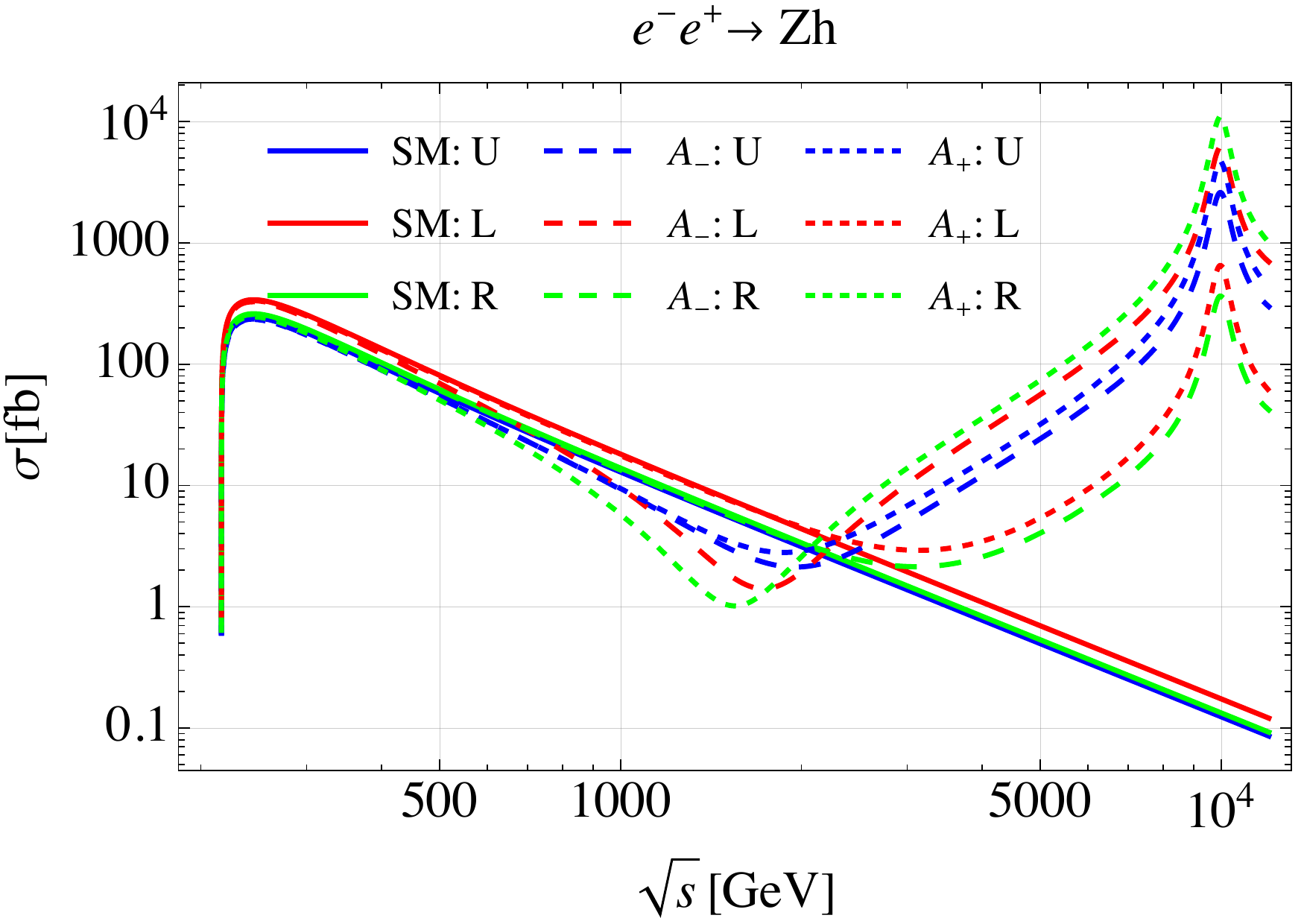}
 \caption{\small
 The total cross sections of the $e^-e^+\to Zh$ process
 in the SM and the GHU model are shown in wider range of $\sqrt{s}$.
 The left figure shows
 the $\sqrt{s}$ dependence of 
 $\sigma^{Zh}(P_{e^-}=0,P_{e^+}=0)$ in the SM and the GHU model
 with unpolarized electron and positron beams, where $A_\pm$, $B_\pm$,
 $C_\pm$ are 
 the names of the parameter sets listed in  
 Table~\ref{Table:Parameter-sets}.
 The right figure shows 
 the $\sqrt{s}$ dependence of 
 $\sigma^{Zh}(P_{e^-},P_{e^+})$ in the SM and the GHU model whose
 parameter sets are $A_\pm$ 
 with the three different polarizations U, L, R, where 
 U, L, R stand for $(P_{e^-},P_{e^+})=(0,0),(-0.8,+0.3),(+0.8,-0.3)$,
 respectively.
 }
 \label{Figure:sigma}
\end{center}
\end{figure}

In Figure~\ref{Figure:sigma}, we show the $\sqrt{s}$ dependence of 
the total cross sections of the $e^-e^+\to Zh$ process in wider range of
$\sqrt{s}$ 
in the SM and the GHU model with unpolarized $e^\pm$ beams. The
calculation incorporates contributions up to the 1st KK mode.
Above the first KK mass scale, 
the contributions from higher KK modes cannot be neglected, and
the region where the calculation is reliable and valid for this analysis
is well below the first KK mass scale.
The left figure in Figure~\ref{Figure:sigma} shows 
$\sigma^{Zh}(P_{e^-},P_{e^+})$ given in 
Eq.~(\ref{Eq:sigma-total-Zh})
in the SM and the GHU model with $(P_{e^-},P_{e^+})=(0,0)$, where
$A_\pm$, $B_\pm$, $C_\pm$ are the names of the parameter sets listed in  
Table~\ref{Table:Parameter-sets}.
The right figure shows $\sigma^{Zh}(P_{e^-},P_{e^+})$ inf the SM and GHU
models with the initial states of polarized and unpolarized electron and
positron. 
U, L, R stand $(P_{e^-},P_{e^+})=(0,0),(-0.8,+0.3),(+0.8,-0.3)$,
respectively.

From Figure~\ref{Figure:sigma}, in the $\sqrt{s}\lesssim O(1)$\,TeV
region, the cross sections in the GHU model are smaller than the cross
sections in the SM, independent of the parameter sets and initial
polarizations. From the right figure, we find that
the deviations from the SM for each initial polarization of $e^\pm$ beams
depend on the sign of the bulk masses.
The deviation from the SM is monotonically increasing with respect to
$\sqrt{s}$ for $\sqrt{s}\lesssim 1$\,TeV.
The cross sections in the GHU model are smaller than those
in the SM because of the interference effects between the $Z$ boson and
the KK gauge bosons, and the cross sections in the GHU begin to increase
in the energy scale region where the contributions to the cross sections
from the KK gauge bosons are larger than the contribution to the cross
section from the interference effects, indicating a turning point
located at $\sqrt{s}\simeq \sqrt{m_Zm_{\rm KK}}$, where 
$m_{\rm KK}=13,19,25$\,TeV for the parameter sets $A_\pm$, $B_\pm$,
$C_\pm$, respectively. 

\begin{figure}[htb]
\begin{center}
\includegraphics[bb=0 0 363 241,height=5.25cm]{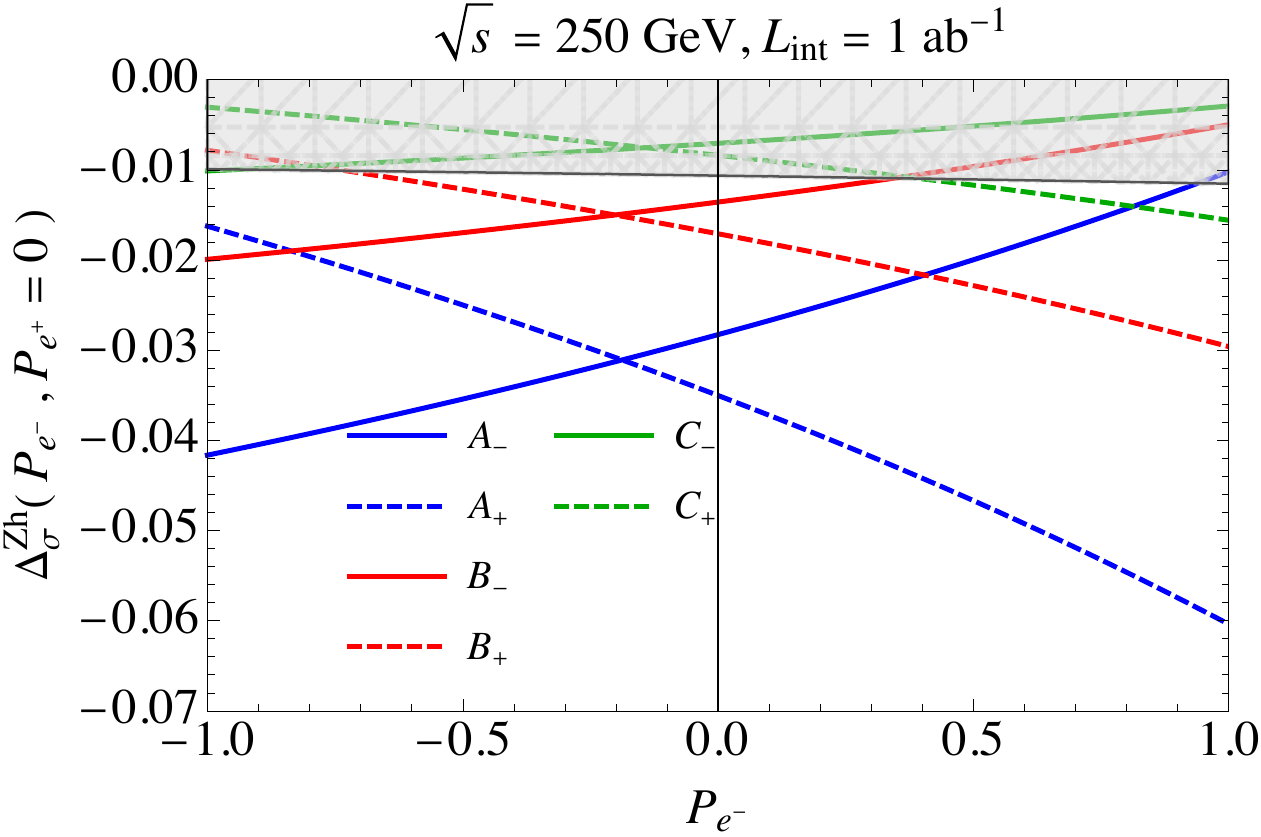}
\includegraphics[bb=0 0 363 241,height=5.25cm]{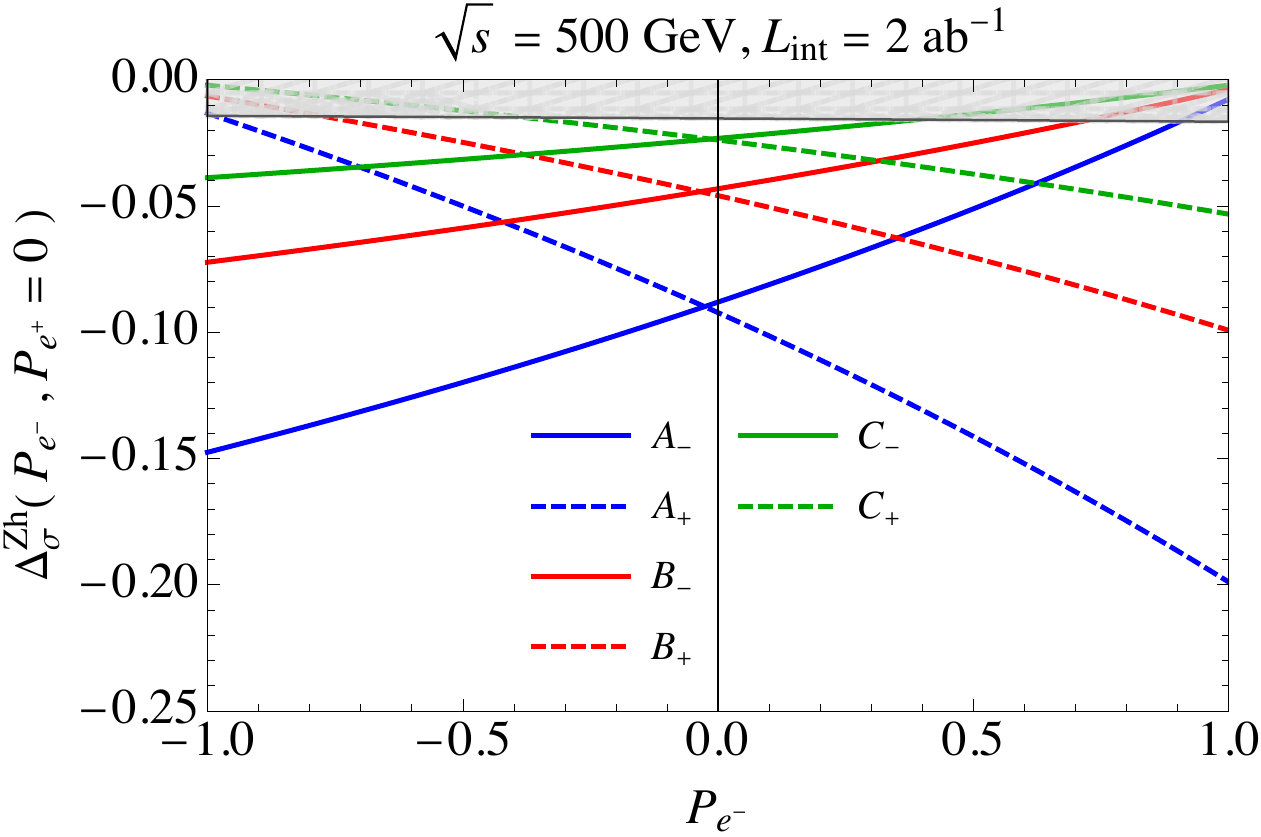}\\
 \caption{\small
 The electron polarization dependence of the deviation from the SM in
 the GHU models for the total cross sections,
 $\Delta_{\sigma}^{Zh}(P_{e^-},P_{e^+})$,
 is shown, where 
 we fix $P_{e^+}=0$.
 The gray region represents the $1\sigma$ statistical error
 estimated  by using the decay mode of $Z$ to $\mu^-\mu^+$, where 
 in the first and second figures,
 we used 
 the sets of the center-of-mass energies and the integrated luminosity
 $(\sqrt{s},L_{\rm int})=(250\,\mbox{GeV},1\,\mbox{ab}^{-1}),
 (500\,\mbox{GeV},2\,\mbox{ab}^{-1})$, respectively.
 Note that $\mbox{Br}(Z\to\mu^-\mu^+)=(3.3662\pm 0.0066)$\% 
 \cite{ParticleDataGroup:2022pth}.
 } 
 \label{Figure:Delta-sigma}
\end{center}
\end{figure}

In Figure~\ref{Figure:Delta-sigma}, the electron polarization dependence
of the deviation from the SM in  the GHU models for the total cross
sections, $\Delta_{\sigma}^{Zh}(P_{e^-},P_{e^+})$, is shown, where
$\Delta_\sigma^{Zh}(P_{e^-},P_{e^+})$ is given in
Eq.~(\ref{Eq:Delta_sigma}).
The $1\sigma$ statistical errors are estimated 
by using the decay mode of the $Z$ boson to $\mu^-$ and $\mu^+$, where 
$\mbox{Br}(Z\to\mu^-\mu^+)=(3.3662\pm 0.0066)$\% 
\cite{ParticleDataGroup:2022pth}.
We use the sets of the center-of-mass energies and the integrated
luminosity $(\sqrt{s},L_{\rm int})=(250\,\mbox{GeV},1\,\mbox{ab}^{-1}),
(500\,\mbox{GeV},2\,\mbox{ab}^{-1})$, respectively.
For reference, here we consider only the decay channel of the $Z$
boson to a pair of the muons \cite{Yan:2016xyx}. In the ILC experiment,
smaller statistical errors may be available by using the decay channel
of the $Z$ boson to hadron jets
\cite{Yan:2016xyx,ILCInternationalDevelopmentTeam:2022izu}.

\begin{figure}[htb]
\begin{center}
\includegraphics[bb=0 0 576 386,height=5.25cm]{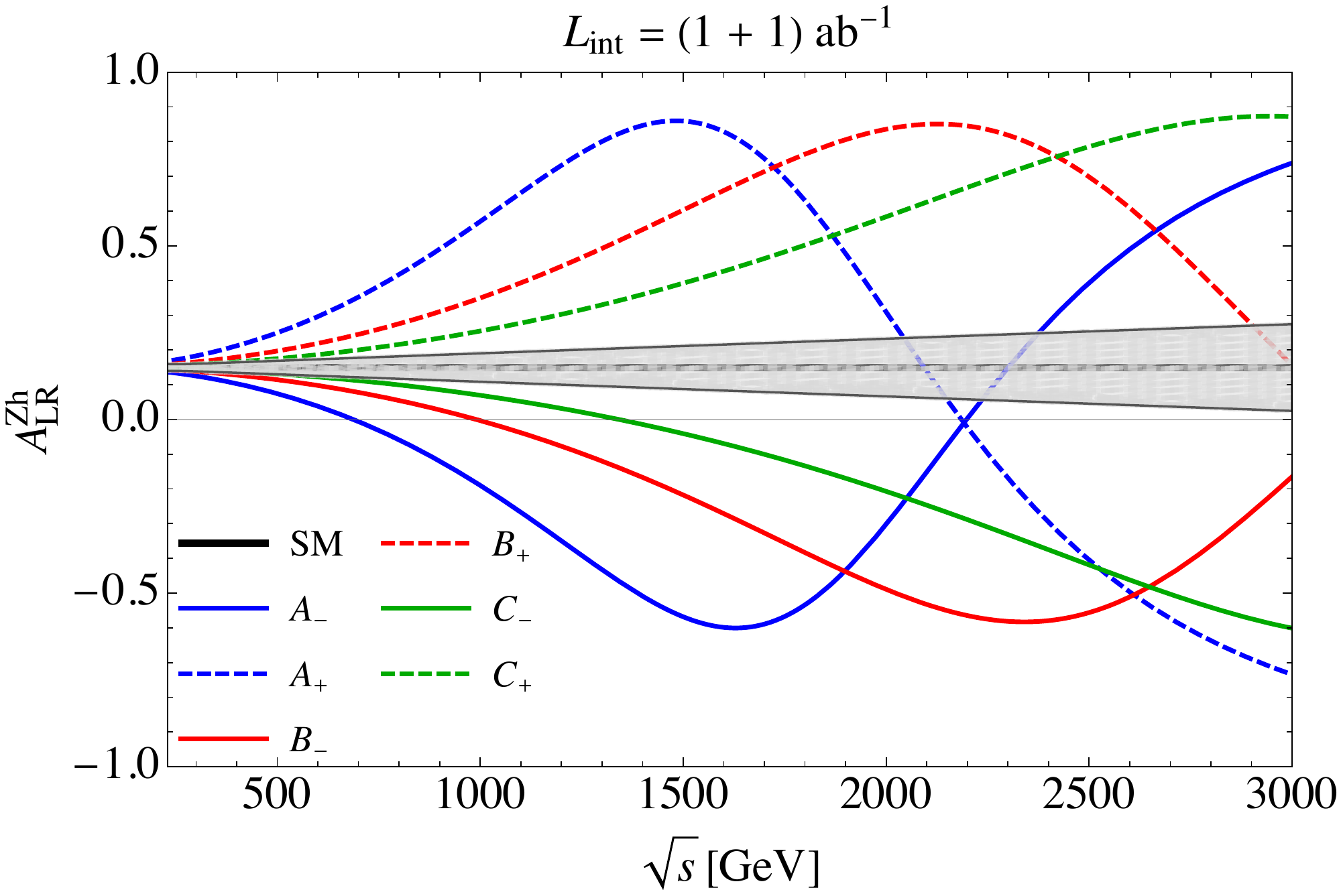}
\includegraphics[bb=0 0 504 341,height=5.25cm]{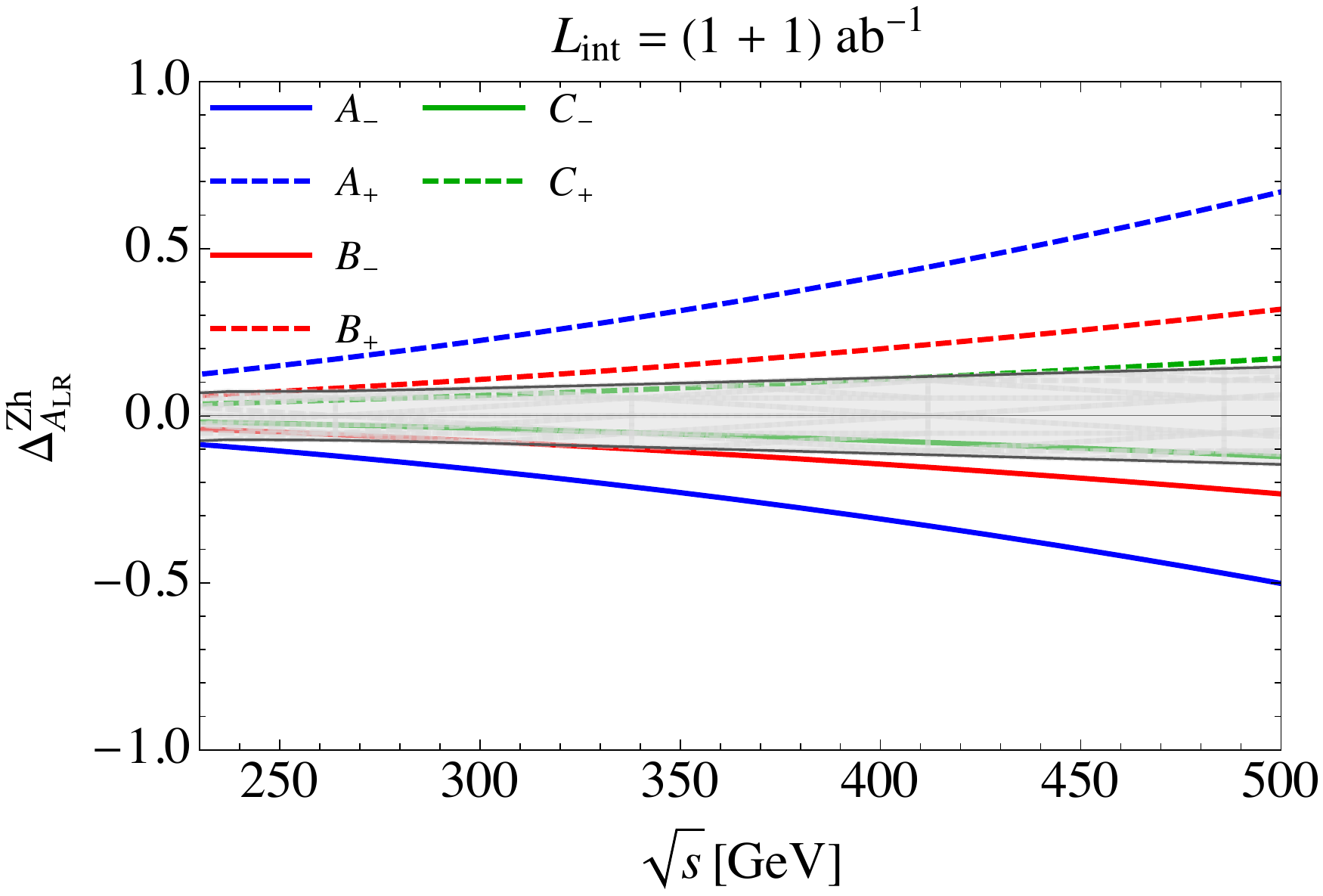}
 \caption{\small
 The $\sqrt{s}$ dependence of the left-right asymmetry of 
 the $e^-e^+\to Zh$ process and the deviation from the SM  are shown. 
 The left figure shows the $\sqrt{s}$ dependence of
 $A_{LR}^{Zh}$ in the SM and the GHU model.
 The right figure shows the $\sqrt{s}$ dependence of
 $\Delta_{A_{LR}}^{Zh}$.
 The energy ranges $\sqrt{s}$ in the first and second figures are
 $\sqrt{s}=[200,3000] \,$GeV, $\sqrt{s}=[200,1000] \,$GeV, respectively.
 The gray region represents the $1\sigma$ statistical error in the SM 
 at each $\sqrt{s}$ with $1$$\,$ab$^{-1}$ for 
 each polarized initial states
 $(P_{e^-},P_{e^+})=(\mp0.8,\pm0.3)$
 by using the decay mode of $Z$ to $\mu^-\mu^+$.
 Other information is the same as in 
 Figure~\ref{Figure:Delta-sigma}.
 } 
 \label{Figure:ALR-Zh}
\end{center}
\end{figure}

In Figure~\ref{Figure:ALR-Zh}, 
the $\sqrt{s}$ dependence of the left-right asymmetry $A_{LR}^{Zh}$
of the $e^-e^+\to Zh$ processes and the deviation from the SM
$\Delta_{A_{LR}}^{Zh}$ are shown,
where $A_{LR}^{Zh}$ and $\Delta_{A_{LR}}^{Zh}$ are 
given in Eqs.~(\ref{Eq:ALR-def}) and (\ref{Eq:Delta_A_LR}),
respectively.
The $1\sigma$ statistical error in the SM 
at each $\sqrt{s}$ with $1$$\,$ab$^{-1}$ for 
each polarized initial electron and positron
$(P_{e^-},P_{e^+})=(\mp0.8,\pm0.3)$
is estimated  by using the decay mode of $Z$ to $\mu^-\mu^+$.

From Figure~\ref{Figure:ALR-Zh}, we find that for
$\sqrt{s}\lesssim 2$\,TeV, the left-right asymmetry 
$A_{LR}^{Zh}$ in the GHU model is larger that
that in the SM when the bulk mass of the leptons is positive, while
$A_{LR}^{Zh}$ in the GHU model is smaller that in the SM when the bulk
mass of the leptons is negative.
As can be seen from Table~\ref{Table:Higgs-Gauge-Couplings}, there is
almost no bulk mass dependence for the coupling constants of the Higgs
boson to the $W$ and $Z$ bosons in the SM and the GHU model, and the
cubic coupling constants of the $Z$ boson, the
$Z'(=Z^{(1)},Z_R^{(1)})$ boson, and the Higgs boson,
$g_{ZZ^{(1)}h}$ and $g_{ZZ_R^{(1)}h}$,
are larger than the cubic coupling constant of 
the $Z$ boson, the $Z$ boson, and the Higgs boson, $g_{ZZh}$.
From Tables~\ref{Table:Gauge-Fermion-Couplings-W-lepton},
\ref{Table:Gauge-Neutrino-Couplings}, and
\ref{Table:Gauge-Charged-Lepton-Couplings},
there is almost no dependence on the sign of the bulk masses 
for the coupling constants of zero modes of the charged leptons to the
the zero mode of the $Z$ boson, while there is large difference between
the zero modes of the right-handed and left-handed 
leptons to the 1st KK modes of the $Z$ and $Z_R$ bosons.
Some gauge coupling constants of the zero modes of the right-handed or
left-handed leptons to the 1st KK modes of the $Z'$ bosons $Z^{(1)}$ and
$Z_R^{(1)}$ are larger than those of zero modes of the leptons to 
the zero mode of the $Z$ boson.

In summary, for the $e^-e^+\to Zh$ process,
some of the coupling constants of the $Z'$ bosons to the leptons and the
Higgs boson are much larger than the coupling constants of the $Z$
boson to the leptons and the Higgs boson, so that even a center-of-mass
energy $\sqrt{s}$ that is an order of magnitude smaller than the masses
of the $Z'$ bosons can yield experimentally verifiable predictions.

\subsection{$e^-e^+\to \nu\bar{\nu}h$}
\label{Sec:ee-to-nunuh}

Here we calculate the total cross sections of the 
$e^-e^+\to \nu\bar{\nu}h$
process given in Eq.~(\ref{Eq:Total-cross-section-ee-to-ffh}),
where if $\nu$ has no subscript, the contribution of all flavors shall
be considered.
Since neutrinos are not observable experimentally, we add up the
contributions
not only from the $e^-e^+\to \nu_e\bar{\nu}_e h$ process but also 
from the $e^-e^+\to \nu_\mu\bar{\nu}_\mu h$ and
$e^-e^+\to \nu_\tau\bar{\nu}_\tau h$ processes.
In the SM, the contributions to 
$\sigma(e^-e^+\to \nu_\mu\bar{\nu}_\mu h)$ and
$\sigma(e^-e^+\to \nu_\tau\bar{\nu}_\tau h)$
come only from the processes
$\sigma(e^-e^+\to Z\to Zh \to\nu_\mu\bar{\nu}_\mu h)$ and
$\sigma(e^-e^+\to Z\to Zh \to\nu_\tau\bar{\nu}_\tau h)$, respectively.
When we ignore the neutrino masses in the SM,
$\sum_{\ell=e,\mu,\tau}\sigma(e^-e^+\to Z\to Zh\to
\nu_\ell\bar{\nu}_\ell h)
=3\sigma(e^-e^+\to Z\to Zh\to
\nu_e\bar{\nu}_e h)$. 
In the GHU model, the coupling constants of the $Z$ boson to the
neutrinos given in Table~\ref{Table:Gauge-Neutrino-Couplings} are almost
the same regardless of neutrino flavor. Therefore, 
to a very good approximation, the following relationship holds:
$\sum_{\ell=e,\mu,\tau}
\sigma(e^-e^+\to Z\to Zh\to \nu_\ell\bar{\nu}_\ell h)
\simeq3\sigma(e^-e^+\to Z\to Zh\to \nu_e\bar{\nu}_e h)$.

\begin{figure}[htb]
\begin{center}
\includegraphics[bb=0 0 363 246,height=5.25cm]{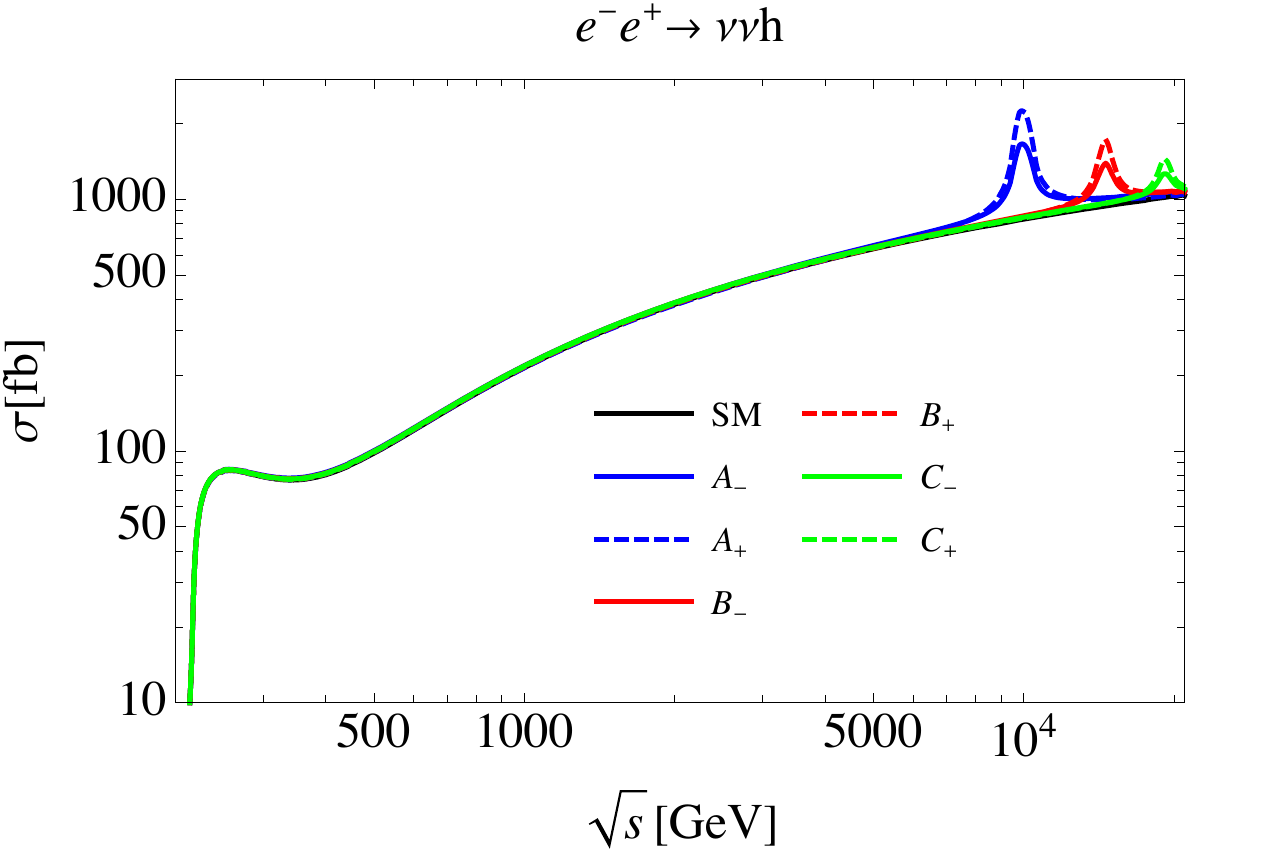}
\includegraphics[bb=0 0 504 341,height=5.25cm]{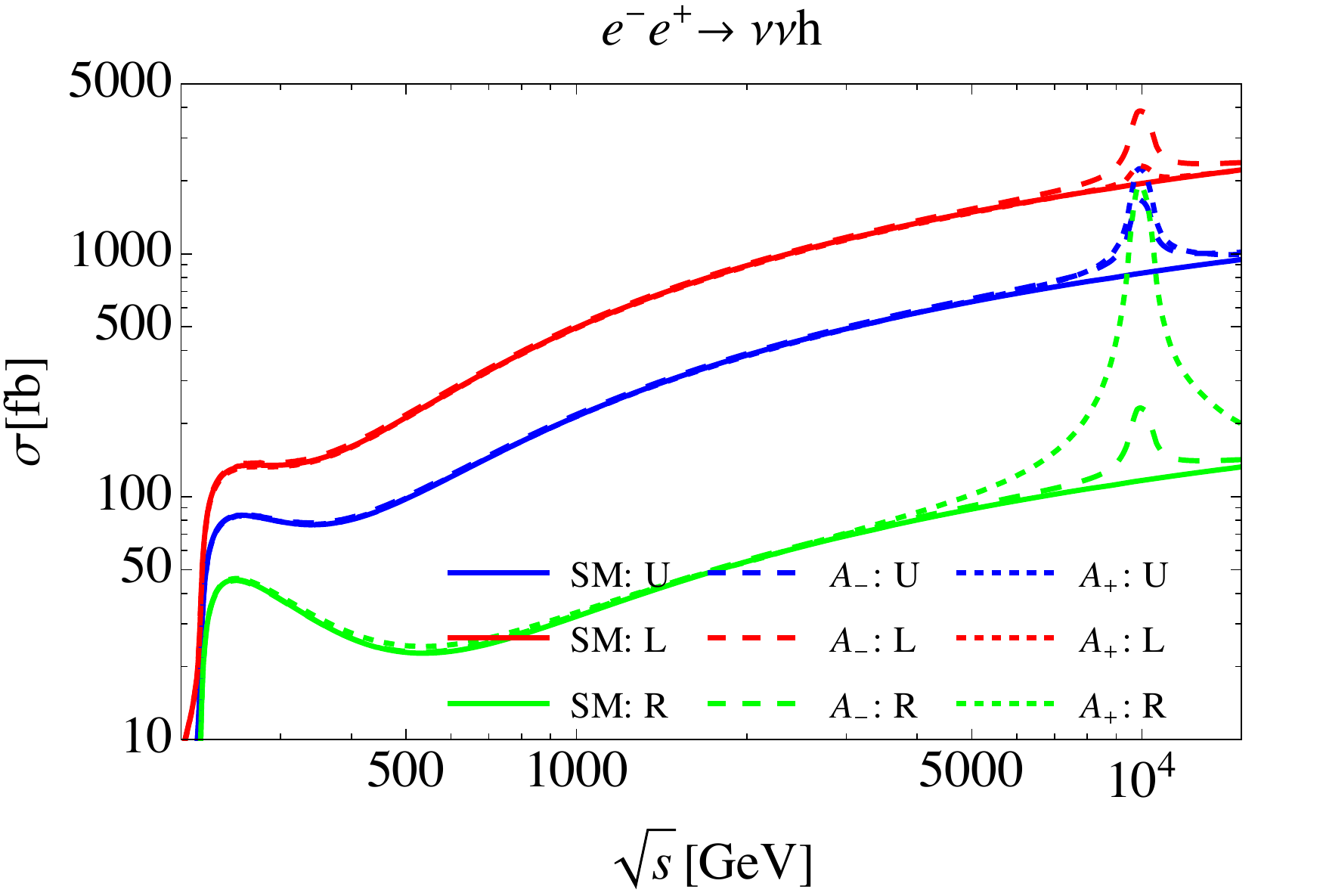}
\end{center}
 \caption{\small
 The $\sqrt{s}$ dependence of the total cross sections of 
 the $e^-e^+\to  \nu\bar{\nu}h$ process
 with unpolarized and polarized electron and positron  beams in the SM
 and the GHU  model is shown.
 The left figure shows the $\sqrt{s}$ dependence of the total cross
 sections with unpolarized electron and positron  beams in the SM and
 the GHU model  whose parameter sets are $A_\pm$, $B_\pm$, $C_\pm$.
 The right figure shows the $\sqrt{s}$ dependence of the total cross
 sections with unpolarized and polarized electron and positron  beams in
 the SM and  the GHU model whose parameter sets are $A_\pm$.
 The energy ranges $\sqrt{s}$ in the left and right figures are
 $\sqrt{s}=[200,21000]$\,GeV and  $\sqrt{s}=[200,15000]$\,GeV,
 respectively.
 U, L, R stand $(P_{e^-},P_{e^+})=(0,0),(-0.8,+0.3),(+0.8,-0.3)$,
 respectively.
 }
 \label{Figure:sigma-ee-to-nunuh}
\end{figure}

In Figure~\ref{Figure:sigma-ee-to-nunuh}, the total cross sections of
the $e^-e^+\to \nu\bar{\nu}h$ process
with unpolarized and polarized electron and positron beams 
in the SM and the GHU model 
whose parameter sets are $A_\pm$, $B_\pm$, $C_\pm$
are showed up to the energy scale $\sqrt{s}$
where resonances due to the KK gauge boson can be confirmed.
The left figure shows the $\sqrt{s}$ dependence of the total cross
sections with unpolarized electron and positron beams in the SM and the
GHU model whose parameter sets are $A_\pm$, $B_\pm$, $C_\pm$.
The right figure shows the $\sqrt{s}$ dependence of the total cross
sections with polarized electron and positron beams for the SM and the
GHU model whose parameter sets are $A_\pm$.
In both the figures, the main contribution to the peak around
$\sqrt{s}\simeq 250$\,GeV comes
from the $e^-e^+\to Z\to Zh\to \nu\bar{\nu}h$ process, and 
the main contributions to the peaks around $O(10)$\,TeV come from 
the $e^-e^+\to Z'(=Z^{(1)},Z_R^{(1)})\to Zh\to \nu\bar{\nu} h$
process for each parameter set.
In our calculation, we take into account the contribution to the cross 
section from the $W$ and $Z$ bosons and the 1st KK gauge bosons 
$W^{\pm(1)}$, $W_R^{\pm(1)}$, $Z^{(1)}$, and $Z_R^{(1)}$
but the contribution from higher KK modes cannot be ignored above
$O(m_{\rm KK})$.
We verify that for the $e^-e^+\to\nu\bar{\nu}h$ process the contribution
from the KK gauge bosons is 
small, independent of the KK masses and the initial polarization of
the electron and the positron, except around the $Z'$ boson mass scale.
Since the coupling constants of the $W$ and $Z$ bosons to the
left-handed leptons are different from that of the $W$ and $Z$ bosons to 
the right-handed leptons, the cross section depend on the initial
polarizations of the electron and the positron.
Furthermore, since the parity violation of the  $W$ boson to the leptons
are larger than that of the $Z$ boson to the leptons, 
the vector boson fusion (VBF) process, in which the $W$ boson gives the
main contribution, depends more strongly on the initial polarizations
than the $Zh$ process, in which the $Z$ boson gives the main
contribution.
The coupling constants of the $Z'$ bosons to the leptons depend on the
sign o the bulk masses. When the bulk mass is negative, the coupling
constants to the left-handed fermions is larger, and when the bulk mass
is positive, the coupling constant to the right-handed fermions is
larger. As a result, 
the magnitude of the cross section by polarization at resonance energies
strongly depends on the sign of the bulk mass.

\begin{figure}[htb]
\begin{center}
\includegraphics[bb=0 0 363 246,height=5.25cm]{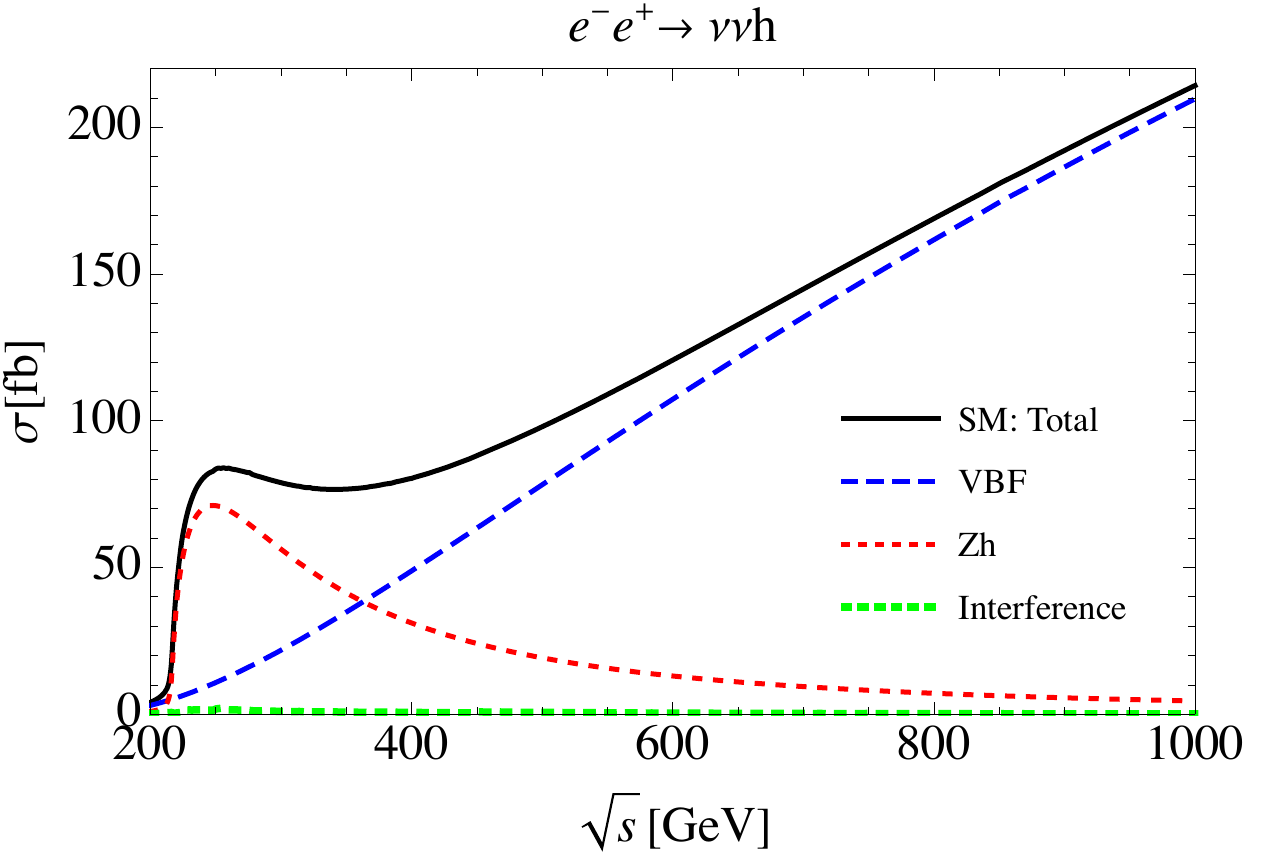}
\includegraphics[bb=0 0 363 246,height=5.25cm]{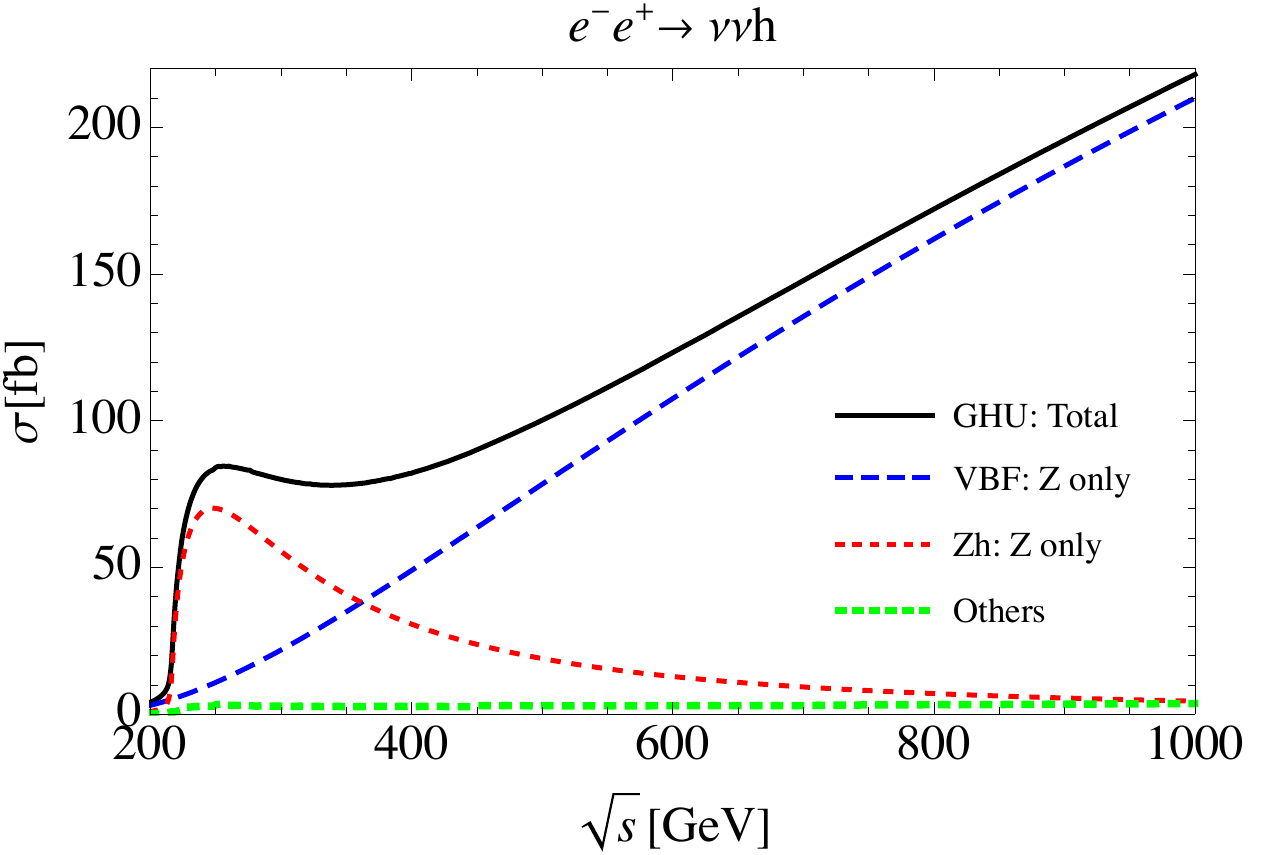}
\end{center}
 \caption{\small
 The $\sqrt{s}$ dependence of the total cross sections of 
 the $e^-e^+\to  \nu\bar{\nu}h$  process
 with unpolarized electron and positron beams is shown.
 The left figure shows the $\sqrt{s}$ dependence of the cross
 sections for total, vector boson fusion (VBF) only, $Zh$ only,
 and interference between VBF and $Zh$ processes only in the SM.
 The right figure shows the $\sqrt{s}$ dependence of the cross
 sections for total, VBF from $Z$ boson, $Zh$ from $Z$ boson, and 
 the others without VBF and $Zh$ only from $Z$ boson
 in the GHU model with the parameter set $A_-$.
 The others includes such as VBF and $Zh$ including at least one KK mode.
 The energy ranges $\sqrt{s}$ in the figures are
 $\sqrt{s}=[200,1000]$\,GeV.
 }
 \label{Figure:sigma-ee-to-nunuh-LS}
\end{figure}

In Figure~\ref{Figure:sigma-ee-to-nunuh-LS}, the $\sqrt{s}$  dependence
of the total cross sections of the $e^-e^+\to \nu\bar{\nu}h$ process
with unpolarized electron and positron beams 
in the SM and the GHU model whose parameter set is $A_-$ 
is shown up to $\sqrt{s}=1$\,TeV.
The contributions to the total cross sections 
from the VBF, $Zh$, and interference (other) in the SM and the GHU model
are decomposed and displayed in the left and right figures, respectively.
There is almost no difference between the cross sections in the SM and
the GHU model for $\sqrt{s}\leq 1$\,TeV.
In both the SM and the GHU model, 
the main contribution comes from the $Zh$ process below $\sqrt{s}\simeq
350$\,GeV, while 
the main contribution comes from the VBF process above 
$\sqrt{s}\simeq350$\,GeV, where 
the contributions from the $Zh$ and VBF processes depend on the initial
polarizations of the electron and the positron.

\begin{figure}[htb]
\begin{center}
\includegraphics[bb=0 0 363 234,height=5cm]{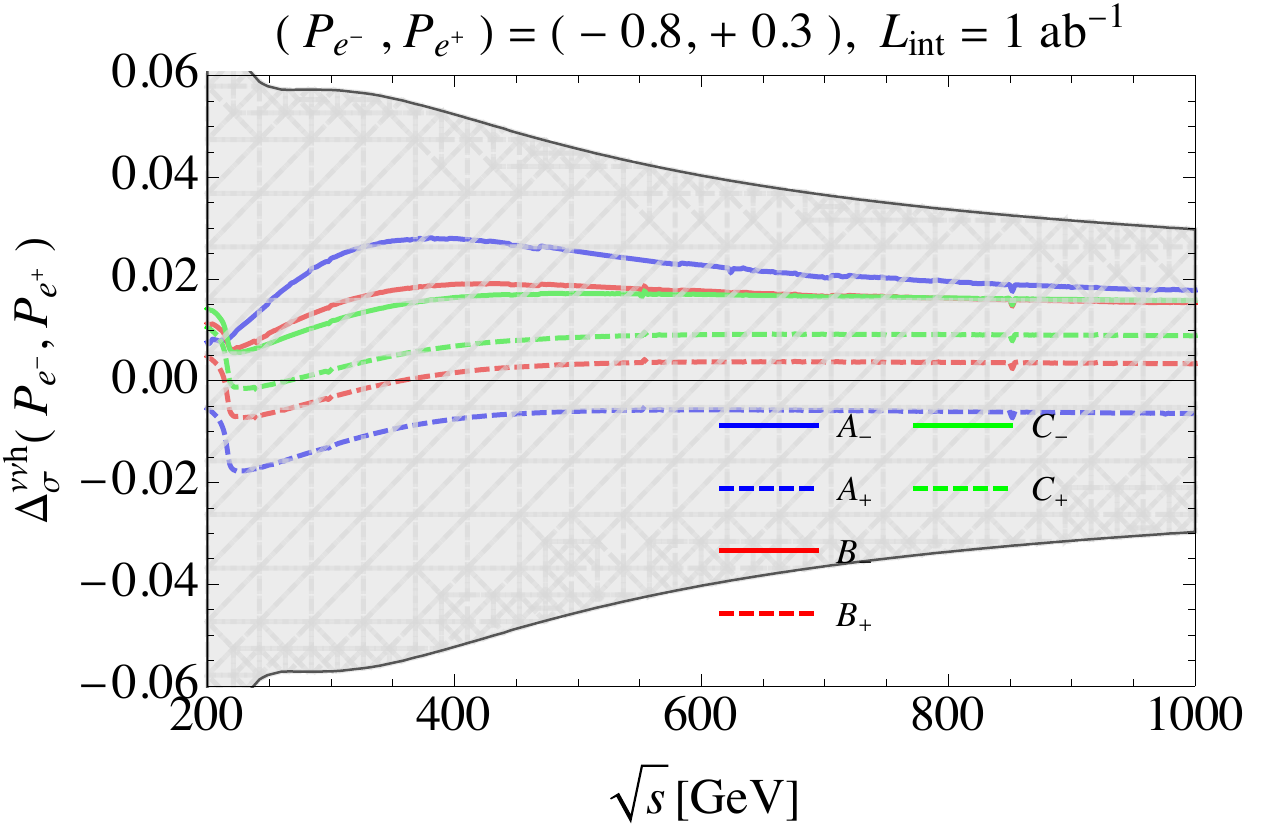}
\includegraphics[bb=0 0 363 234,height=5cm]{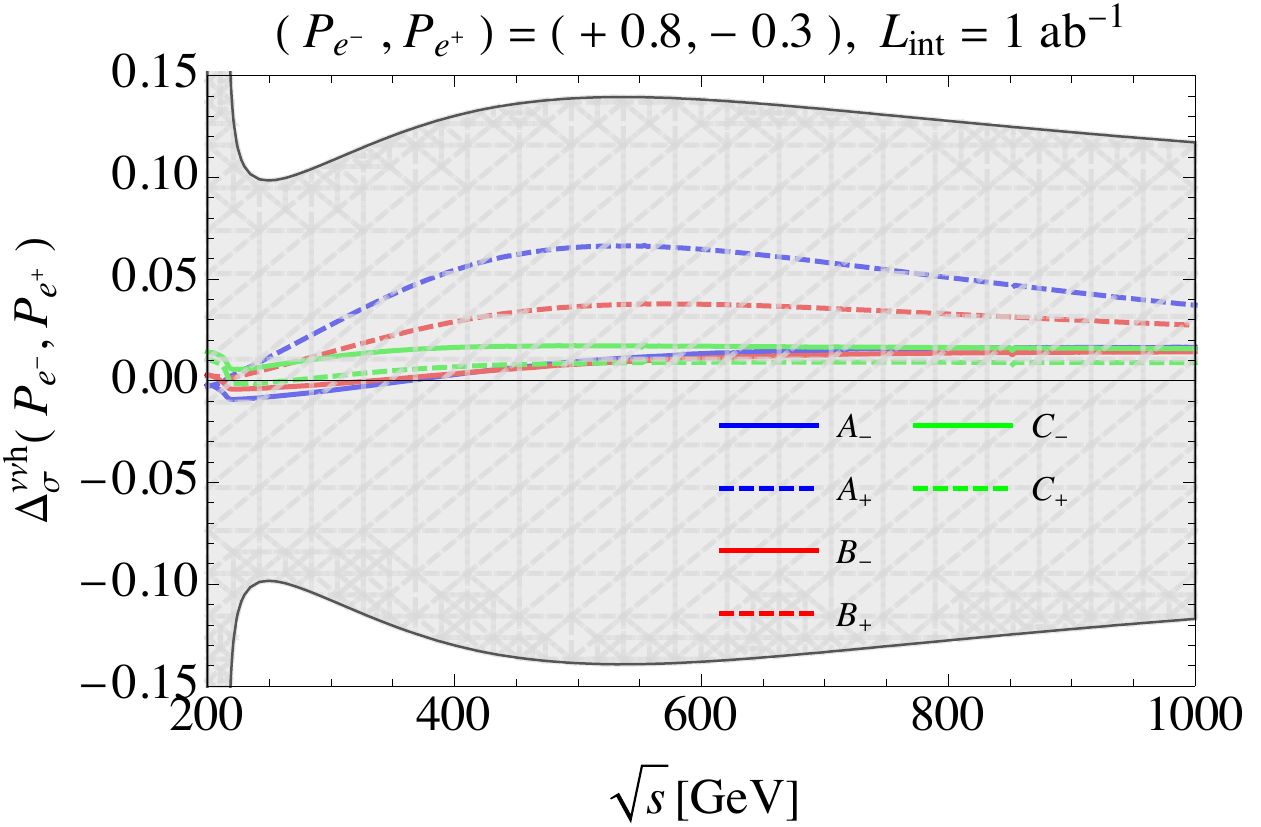}
\end{center}
 \caption{\small
 The $\sqrt{s}$ dependence of the deviation from the SM in 
 the GHU model with six parameter sets $A_\pm$, $B_\pm$, $C_\pm$ for
 total cross sections 
 $\Delta_\sigma^{\nu\nu h}(P_{e^-},P_{e^+})$
 is shown.
 The left and right figures show the center-of-mass energies 
 $\sqrt{s}=[200,1000]\,\mbox{GeV}$ and 
 $(P_{e^-},P_{e^+})=(-0.8,+0.3)$ and $(+0.8,-0.3)$, respectively.
 The gray regions represent the $1\sigma$ statistical errors
 in the SM at each $\sqrt{s}$ with $1$$\,$ab$^{-1}$ by using
 the decay mode of the Higgs boson $h$ to two photons $\gamma\gamma$,
 where the branching ratio for the SM Higgs boson with $m_H=125$\,GeV
 is given by $\mbox{Br}(h\to\gamma\gamma)=2.27(1\pm 0.021)\times 10^{-3}$
 \cite{ParticleDataGroup:2022pth}.
 } 
 \label{Figure:Delta-sigma-ee-to-nunuh}
\end{figure}

In Figure~\ref{Figure:Delta-sigma-ee-to-nunuh},
the $\sqrt{s}$ dependence of the deviation from the SM in  the GHU model
with six parameter sets $A_\pm$, $B_\pm$, $C_\pm$ for the total cross
sections
$\Delta_\sigma^{\nu\nu h}$
are shown,
where $\Delta_\sigma^{ffh}(P_{e^-},P_{e^+})$ is given in
Eq.~(\ref{Eq:Delta_sigma-ffh}).
The left and right figures show the center-of-mass energies 
$\sqrt{s}=[200,1000]\,\mbox{GeV}$ and 
$(P_{e^-},P_{e^+})=(-0.8,+0.3)$ and $(+0.8,-0.3)$, respectively.
The $1\sigma$ statistical errors are estimated in the SM at each
$\sqrt{s}$ with $1$$\,$ab$^{-1}$ by using  the decay mode of the Higgs
boson $h$ to two photons $\gamma\gamma$, where the branching ratio for
the SM Higgs boson with $m_H=125$\,GeV  is given by
$\mbox{Br}(h\to\gamma\gamma)=2.27(1\pm 0.021)\times 10^{-3}$
\cite{ParticleDataGroup:2022pth}.
The branching ratio of the Higgs boson to the final state of the
muon pair, whose energy and momentum can be well observed experimentally
with high precision, is very small, 
where  $\mbox{Br}(h\to\mu^-\mu^+)=2.18(1\pm 0.017)\times 10^{-4}$
\cite{ParticleDataGroup:2022pth}.
We refer to the decay of the Higgs boson to two photons, but the
analysis of the signal by hadron jets in the ILC experiment is currently
in progress. The mass reconstruction of the Higgs boson by using bottom
quark pairs and $\tau$ lepton pairs may be available, which have much
larger branching ratios of the Higgs boson, 
where $\mbox{Br}(h\to b\bar{b})=5.82(1\pm 0.013)\times 10^{-1}$ and
$\mbox{Br}(h\to\tau^-\tau^+)=6.27(1\pm 0.016)\times 10^{-2}$
\cite{ParticleDataGroup:2022pth}.
If so, the deviation from the SM in the GHU model can be explored by
using the $e^-e^+\to \nu\bar{\nu} h$ process.

\begin{figure}[htb]
\begin{center}
\includegraphics[bb=0 0 363 248,height=5.2cm]{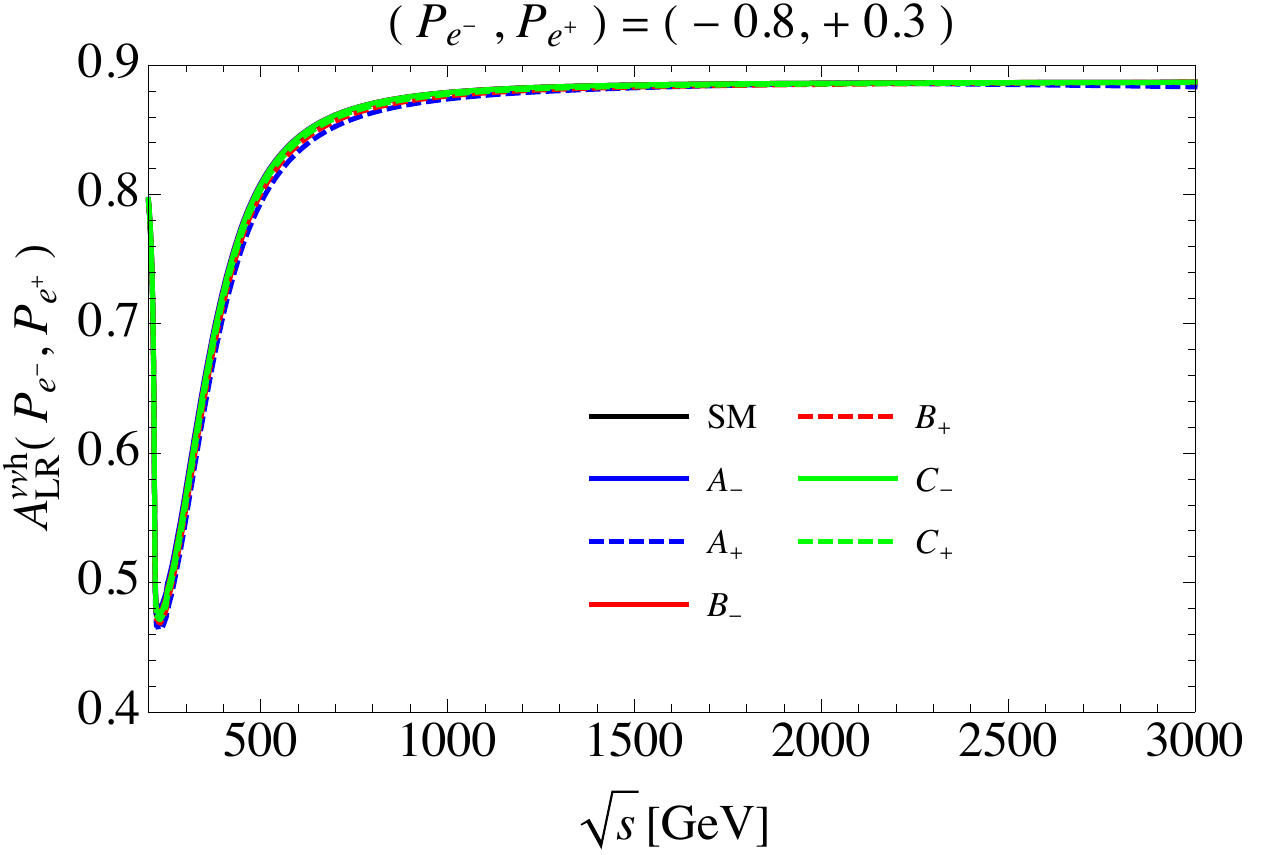}
\includegraphics[bb=0 0 363 237,height=5.2cm]{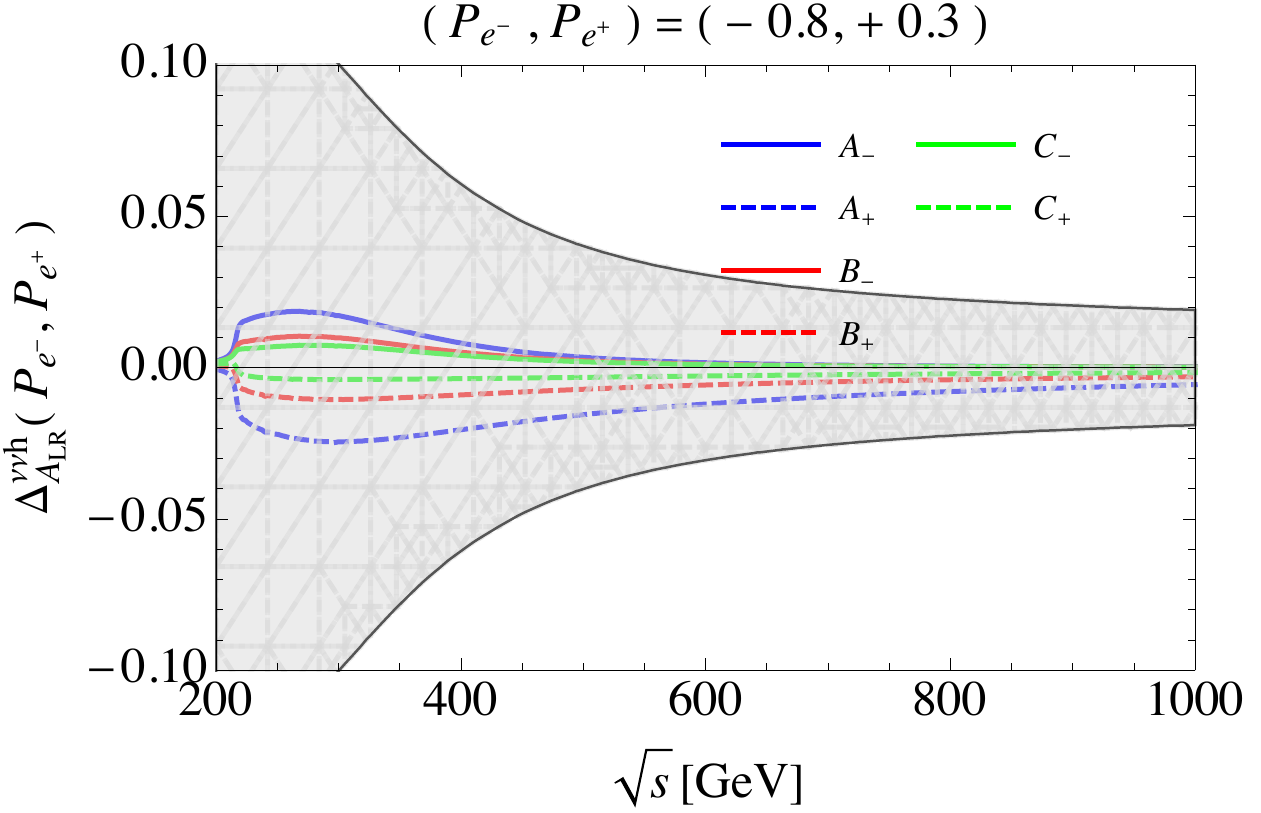}
 \caption{\small
 The left-right asymmetry of  the $e^-e^+\to \nu\bar{\nu} h$ process
 $A_{LR}^{\nu\nu h}(P_{e^-},P_{e^+})$ 
 in the SM and the GHU model
 and the deviation of the left-right asymmetry from the SM
 in the GHU model
 $\Delta_{A_{LR}}^{\nu\nu h}(P_{e^-},P_{e^+})$ 
 are shown.
 The left and right figures show the $\sqrt{s}$ dependence of
 $A_{LR}^{\nu\nu h}(P_{e^-},P_{e^+})$ 
 and 
 $\Delta_{A_{LR}}^{\nu\nu h}(P_{e^-},P_{e^+})$ 
 with the polarization $(P_{e^-},P_{e^+})=(-0.8,+0.3)$
 and the six parameter sets
 $A_\pm$, $B_\pm$, $C_\pm$.
 The energy ranges $\sqrt{s}$ in the left and right figures are
 $\sqrt{s}=[200,3000] \,$GeV, $\sqrt{s}=[200,1000]\,$GeV, respectively.
 The gray region represents the statistical error explained  
 in Table~\ref{Figure:Delta-sigma-ee-to-nunuh}.
 }
 \label{Figure:ALR-ee-to-nunuh}
\end{center}
\end{figure}

In Figure~\ref{Figure:ALR-ee-to-nunuh},
the $\sqrt{s}$ dependence of the left-right asymmetry of 
the $e^-e^-\to \nu\bar{\nu}h$ process
$A_{LR}^{\nu\nu h}(P_{e^-},P_{e^+})$
and the deviation of the left-right asymmetry from the SM 
$\Delta_{A_{LR}}^{\nu\nu h}(P_{e^-},P_{e^+})$
with $(P_{e^-},P_{e^+})=(-0.8,+0.3)$
are shown, where 
$A_{LR}^{\nu\nu h}(P_{e^-},P_{e^+})$ and 
$\Delta_{A_{LR}}^{\nu\nu h}(P_{e^-},P_{e^+})$ are 
given in Eqs.~(\ref{Eq:ALR-ffh-def}) and (\ref{Eq:Delta_A_LR-ffh}),
respectively.
From Eq.~(\ref{Eq:ALR-ffh-obs}), when 
$(P_{e^-},P_{e^+})=(\mp0.8,\pm0.3)$ and 
$\sigma_{LL}^{\nu\nu h}\gg \sigma_{RR}^{\nu\nu h},
\sigma_{LR}^{\nu\nu h}, \sigma_{RL}^{\nu\nu h}$, we find 
$A_{LR}^{\nu\nu h}(P_{e^-}=-0.8,P_{e^+}=+0.3)\simeq 0.887$.
Thus, we confirmed that when $\sqrt{s}$ is sufficiently large, the main 
contribution to the $e^-e^+\to \nu\bar{\nu}h$ process comes almost
exclusively from the VBF processes.
Furthermore, the deviation from the SM becomes smaller when 
$\sqrt{s}$ is larger. Therefore, it it very difficult to confirm the
deviation from the SM by using the left-right asymmetry of 
the $e^-e^+\to \nu\bar{\nu}h$ process.

\subsection{$e^-e^+\to e^-e^+h$}

Here we calculate the total cross section of the $e^-e^+\to e^-e^+h$
process given in Eq.~(\ref{Eq:Total-cross-section-ee-to-ffh}).
The cross section of the $e^-e^+\to e^-e^+h$ process is smaller than
that of the $e^-e^+\to\nu\bar{\nu}h$ process in all energy regions
unless the initial polarization of the electron (the positron) is
extremely right-handed (left-handed). Therefore, this process is usually
a subdominant Higgs boson production process. 

The following figures in this subsection are made by using exactly the
same formula and calculations as those in the previous
subsection~\ref{Sec:ee-to-nunuh}, except for the difference between the
final and intermediate state particles, and their coupling
constants. The explanation of the figures is basically the same as in
the previous section. We will omit duplicated explanations.

\begin{figure}[htb]
\begin{center}
\includegraphics[bb=0 0 363 246,height=5.25cm]{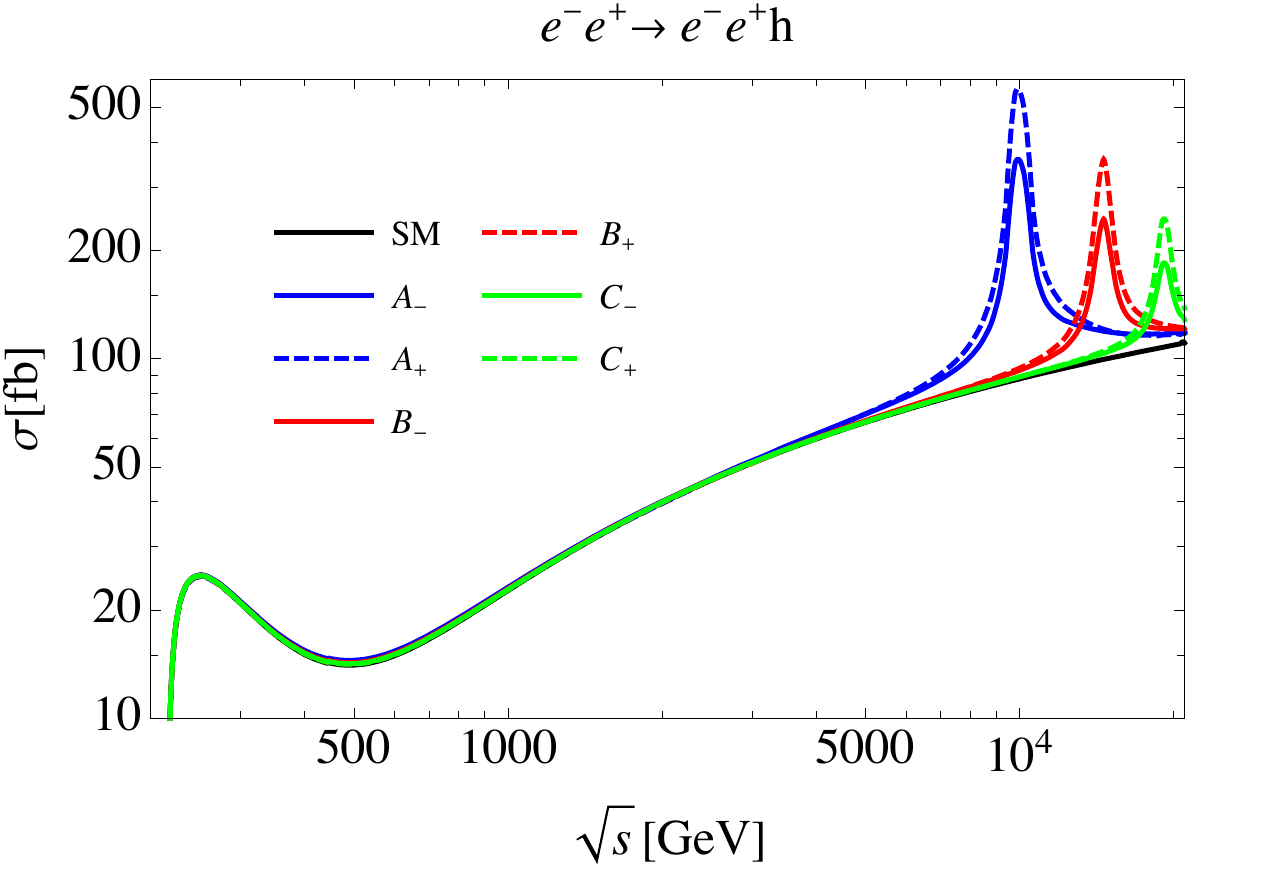}
\includegraphics[bb=0 0 363 246,height=5.25cm]{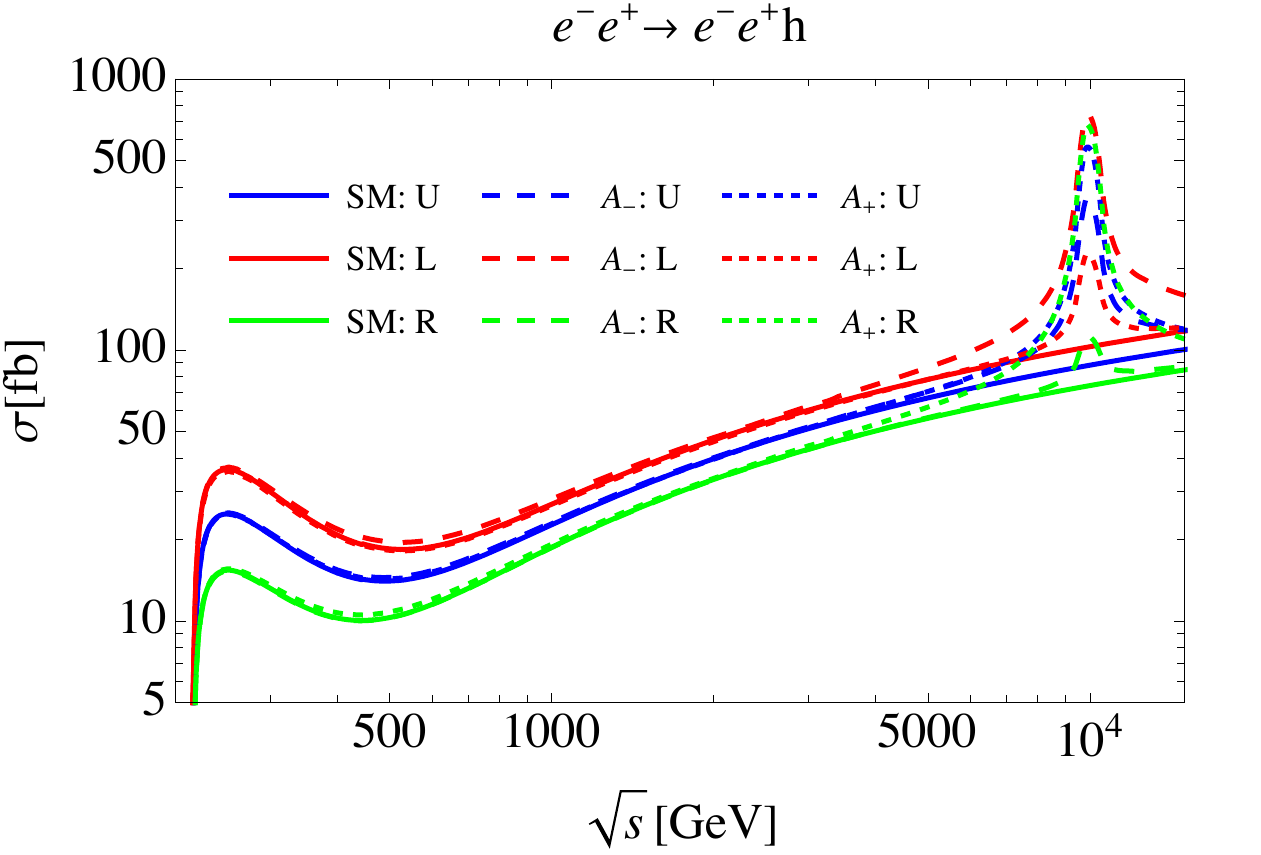}
\end{center}
 \caption{\small
 The $\sqrt{s}$ dependence of the total cross sections of
 the $e^-e^+\to e^-e^+h$ process
 with unpolarized and polarized electron and positron beams in the SM
 and the GHU  model are shown.
 Other information is the same as in 
 Figure~\ref{Figure:sigma-ee-to-nunuh}.
 }
 \label{Figure:sigma-ee-to-eeh}
\end{figure}

In Figure~\ref{Figure:sigma-ee-to-eeh}, 
the $\sqrt{s}$ dependence of the total cross sections of 
the $e^-e^+\to e^-e^+h$ process with unpolarized and polarized 
electron and positron beams in the SM and the GHU model are showed.
The left figure shows the $\sqrt{s}$ dependence of the total cross
sections with unpolarized electron and positron beams in the SM and the
GHU model whose parameter sets are $A_\pm$, $B_\pm$, $C_\pm$.
The right figure shows the $\sqrt{s}$ dependence of the total cross
sections with polarized electron and positron beams in the SM and the
GHU model whose parameter sets are $A_\pm$.
In both the figures, the main contribution to the peak around
$\sqrt{s}\simeq 250$\,GeV comes
from the $e^-e^+\to Z\to Zh\to e^-e^+h$ process, and 
the main contributions to the peaks around $O(10)$\,TeV come from 
the $e^-e^+\to Z'(=Z^{(1)},Z_R^{(1)})\to Zh\to e^-e^+ h$
process for each parameter set.
We verify that the contribution to the $e^-e^+\to e^-e^+h$ process
from the KK gauge bosons is small, independent of the KK masses and
initial polarization of the electron and the positrons, except around
the $Z'$ boson mass scale. 

From Figures~\ref{Figure:sigma-ee-to-nunuh} and
\ref{Figure:sigma-ee-to-eeh}, we confirm that the cross section of
the $e^-e^+\to\nu\bar{\nu}h$ process is larger than that of the
$e^-e^+\to e^-e^+h$ process, unless the initial polarization is
extremely right-handed polarization since the coupling constants of the
$W$ boson to the left-handed leptons are larger than the coupling
constants of the $Z$ boson to leptons.

\begin{figure}[htb]
\begin{center}
\includegraphics[bb=0 0 363 246,height=5.25cm]{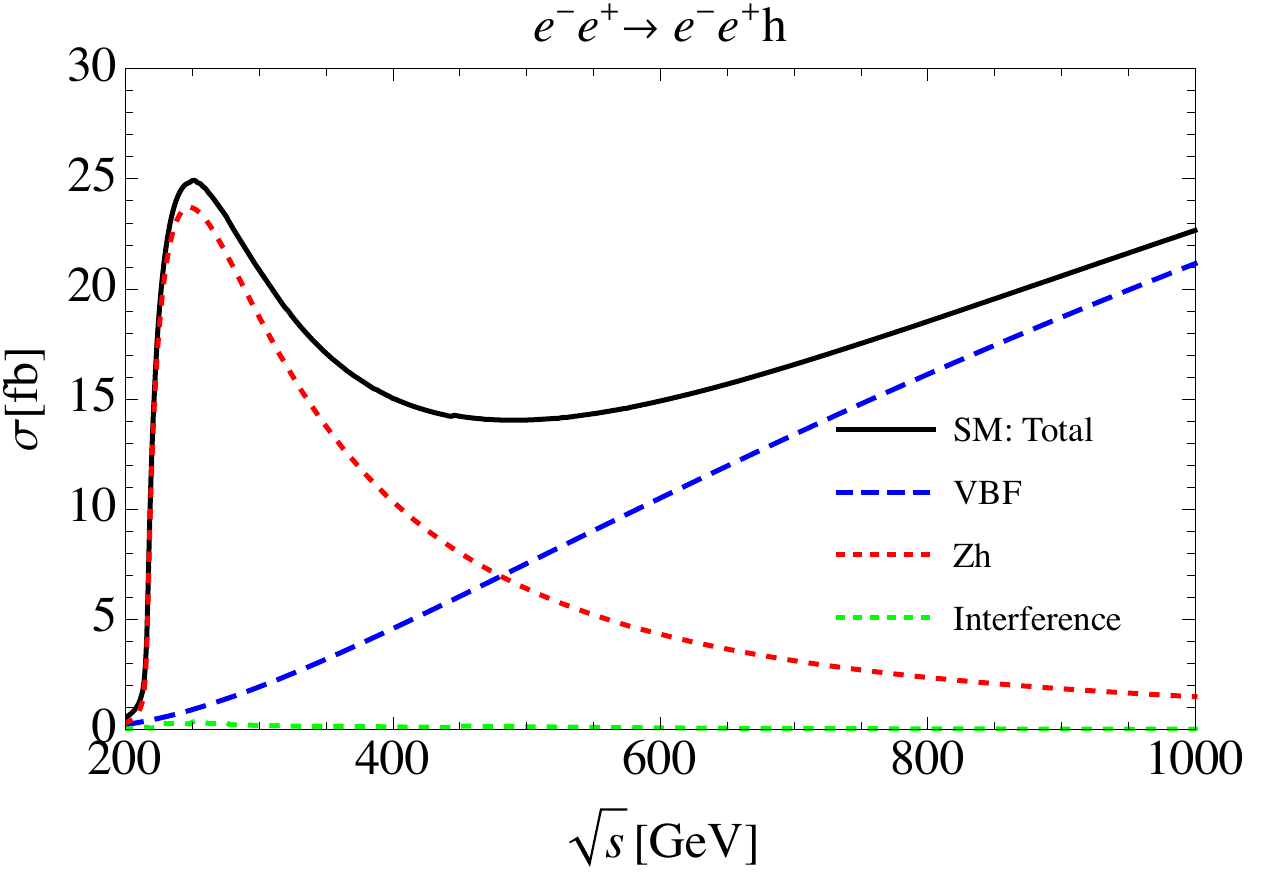}
\includegraphics[bb=0 0 363 246,height=5.25cm]{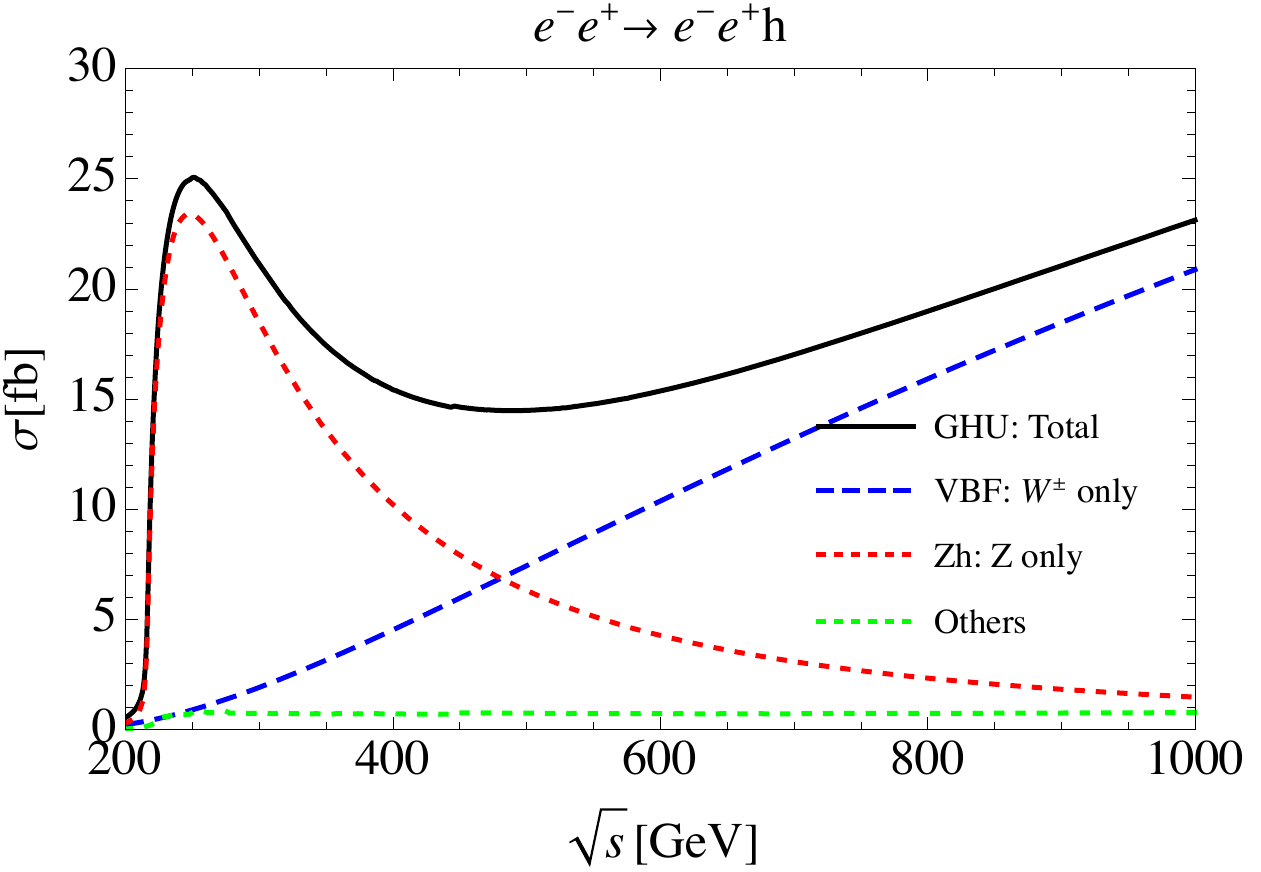}
\end{center}
 \caption{\small
 The $\sqrt{s}$ dependence of the total cross sections of 
 the $e^-e^+\to e^-e^+h$ process
 with unpolarized electron and positron beams in the SM and the GHU
 model are shown  for the left and right figures, respectively.
 Other information is the same as in 
 Figure~\ref{Figure:sigma-ee-to-nunuh-LS}.
 }
 \label{Figure:sigma-ee-to-eeh-LS}
\end{figure}

In Figure~\ref{Figure:sigma-ee-to-eeh-LS}, 
the $\sqrt{s}$ dependence of the total cross sections of 
the $e^-e^+\to e^-e^+h$ process
in the SM and the GHU model whose parameter set is $A_-$ with
unpolarized electron and positron beams are shown up to
$\sqrt{s}=1$\,TeV. There is almost no difference between the cross
sections in the SM and the GHU model.
In both the SM and the GHU model, 
the main contribution comes from the $Zh$ process below $\sqrt{s}\simeq
500$\,GeV, while 
the main contribution comes from the VBF process above 
$\sqrt{s}\simeq500$\,GeV, where 
the contributions from the $Zh$ and VBF processes depend on the initial
polarizations of the electron and the positron.

\begin{figure}[htb]
\begin{center}
\includegraphics[bb=0 0 363 234,height=5cm]{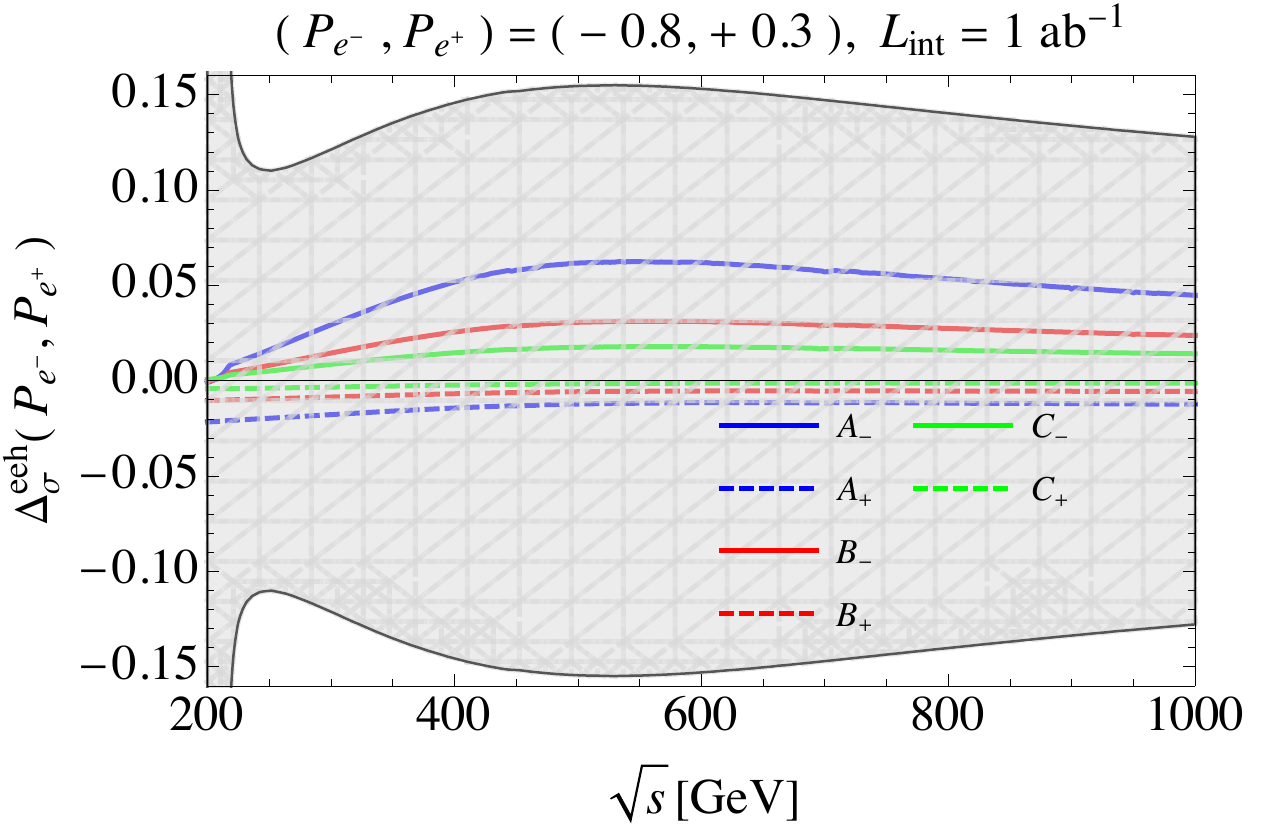}
\includegraphics[bb=0 0 363 234,height=5cm]{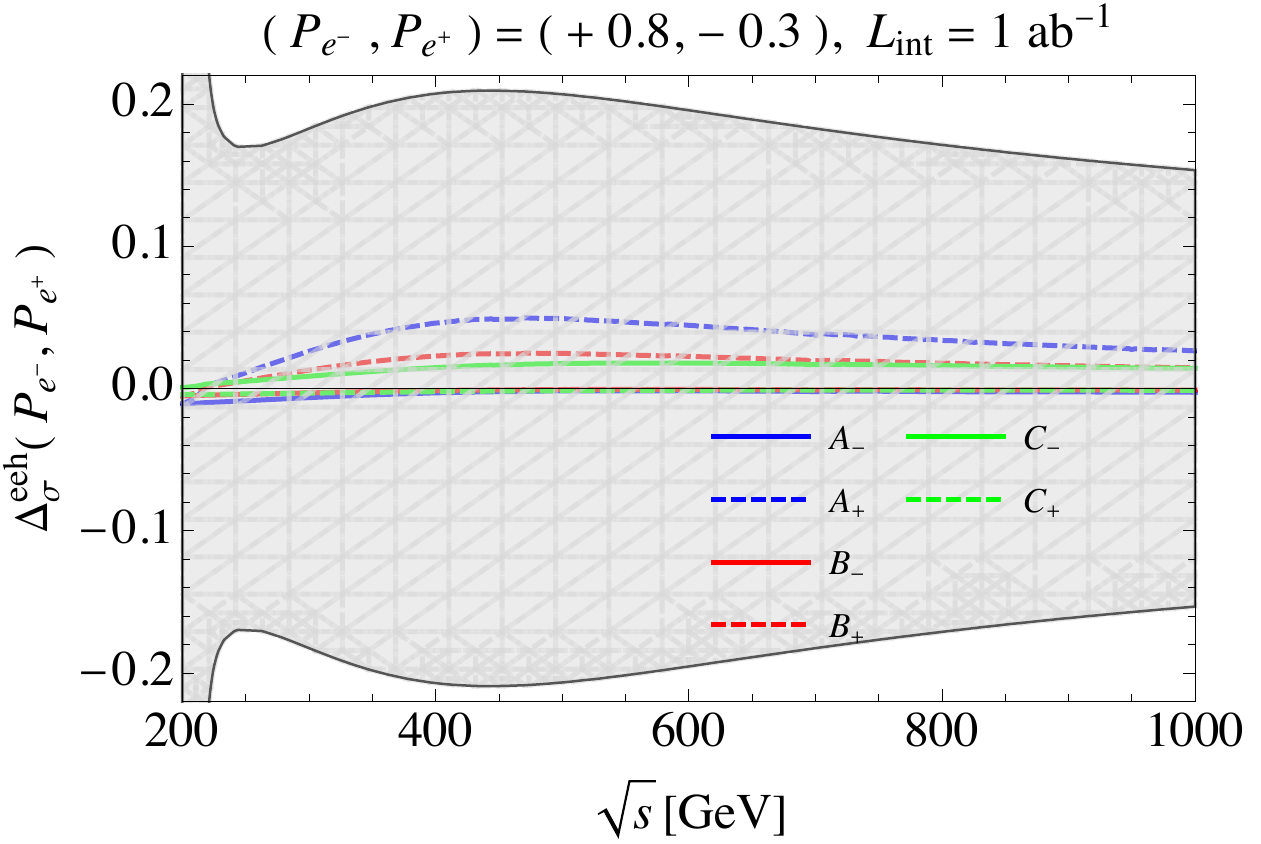}
\end{center}
 \caption{\small
 The deviations from the SM in 
 the GHU model with six parameter sets $A_\pm$, $B_\pm$, $C_\pm$ for
 total cross sections 
 $\Delta_\sigma^{eeh}(P_{e^-},P_{e^+})$ 
 are shown for 
 $\sqrt{s}=[200,1000]\,\mbox{GeV}$ and 
 $(P_{e^-},P_{e^+})=(-0.8,+0.3)$ and $(+0.8,-0.3)$, respectively.
 Other information is the same as in 
 Figure~\ref{Figure:Delta-sigma-ee-to-nunuh}.
 } 
 \label{Figure:Delta-sigma-ee-to-eeh}
\end{figure}

In Figure~\ref{Figure:Delta-sigma-ee-to-eeh},
the $\sqrt{s}$ dependence of the deviation from the SM in the GHU model
with six parameter sets $A_\pm$, $B_\pm$, $C_\pm$ for the total cross
sections $\Delta_\sigma^{eeh}(P_{e^-},P_{e^+})$ are shown.
As the same as in Figure~\ref{Figure:Delta-sigma-ee-to-nunuh},
the left and right figures show the center-of-mass energies 
$\sqrt{s}=[200,1000]\,\mbox{GeV}$ and 
$(P_{e^-},P_{e^+})=(-0.8,+0.3)$ and $(+0.8,-0.3)$, respectively.
The $1\sigma$ statistical errors are estimated in the SM at each
$\sqrt{s}$ with $1$$\,$ab$^{-1}$ by using  the decay mode of the Higgs
boson $h$ to two photons $\gamma\gamma$, where the branching ratio for
the SM Higgs boson with $m_H=125$\,GeV  is given by
$\mbox{Br}(h\to\gamma\gamma)=2.27(1\pm 0.021)\times 10^{-3}$
\cite{ParticleDataGroup:2022pth}.

\begin{figure}[htb]
\begin{center}
\includegraphics[bb=0 0 363 248,height=5.2cm]{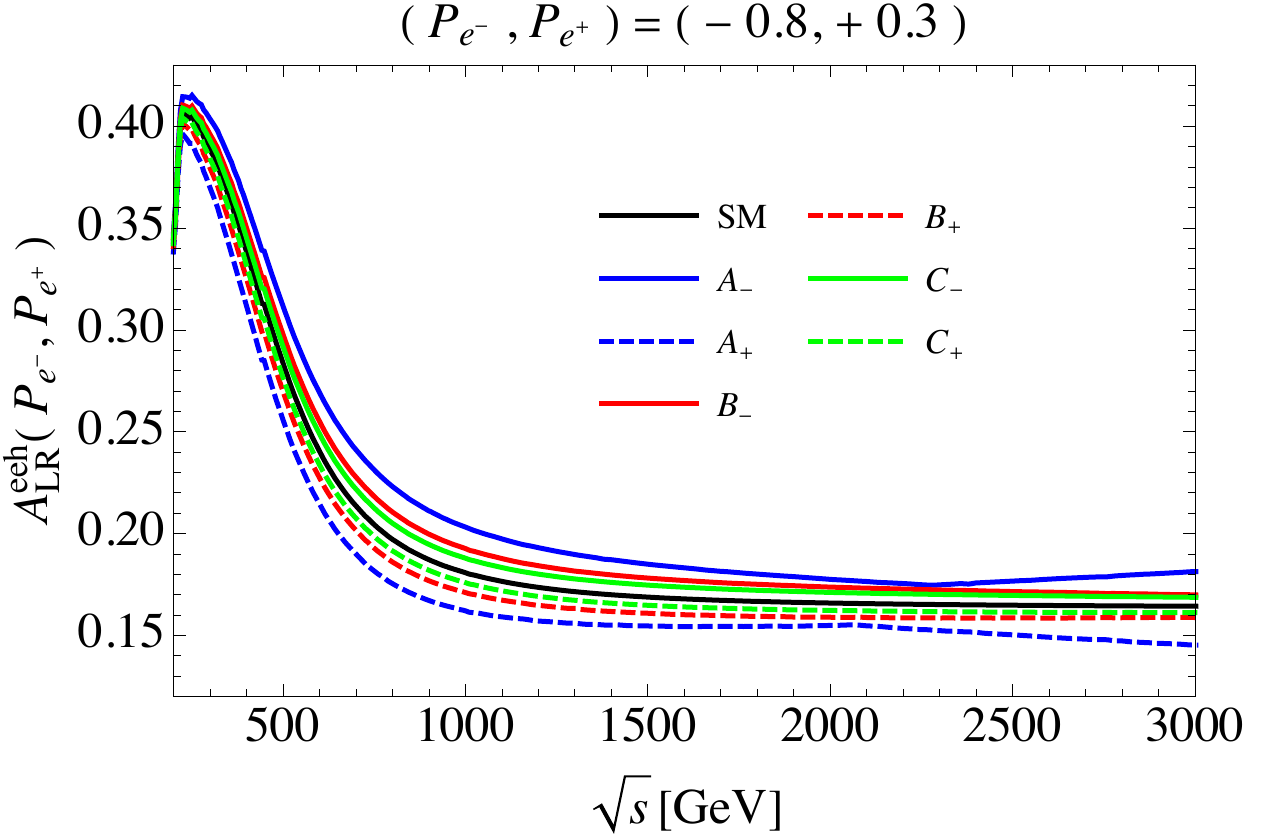}
\includegraphics[bb=0 0 363 237,height=5.2cm]{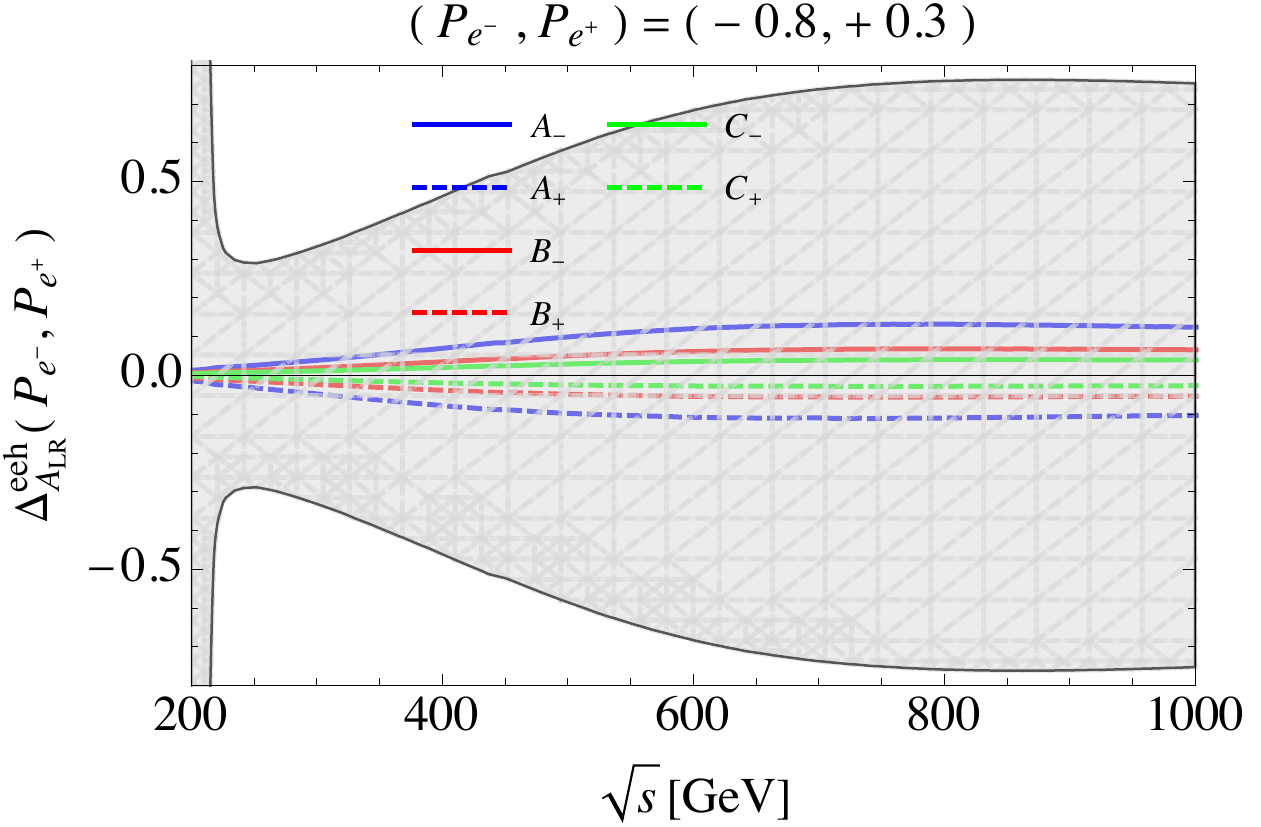}
 \caption{\small
 The left-right asymmetry of the $e^-e^-\to e^-e^+h$ process 
 and the deviation 
 of the left-right asymmetry $\Delta_{A_{LR}}^{eeh}(P_{e^-},P_{e^+})$
 from the SM are shown.
 Other information is the same as in 
 Figure~\ref{Figure:ALR-ee-to-nunuh}.
 }
 \label{Figure:ALR-ee-to-eeh}
\end{center}
\end{figure}

In Figure~\ref{Figure:ALR-ee-to-eeh},
the $\sqrt{s}$ dependence of the left-right asymmetry
$A_{LR}^{eeh}(P_{e^-},P_{e^+})$
and
the deviation from the SM of the left-right asymmetry 
$\Delta_{A_{LR}}^{eeh}(P_{e^-},P_{e^+})$
with $(P_{e^-},P_{e^+})=(-0.8,+0.3)$ are shown,
where $A_{LR}^{eeh}(P_{e^-},P_{e^+})$ and  
$\Delta_{A_{LR}}^{eeh}(P_{e^-},P_{e^+})$
are given in Eqs.~(\ref{Eq:ALR-ffh-def}) and (\ref{Eq:Delta_A_LR-ffh}).
Although the deviation from the SM of the left-right asymmetry is quite
large, but the number of events are small 
due to the small cross-section. Therefore, it it very difficult to
confirm the deviation from the SM by using the left-right asymmetry of 
the $e^-e^+\to e^-e^+h$ process.

\section{Summary and discussions}
\label{Sec:Summary}

In this paper, we analyzed the three main single Higgs boson production
processes $e^-e^+\to Zh$, $e^-e^+\to \nu\bar{\nu}h$, and 
$e^-e^+\to e^-e^+h$.
For the Higgs strahlung process $e^-e^+\to Zh$,
we showed that for a parameter region satisfying the current experimental
constraints ($m_{\rm KK}\geq 13$\,TeV, $\theta_H\leq 0.10$),
a maximum deviation from the SM is about up to 6\% for
$\sqrt{s}=250$\,GeV and that from the SM is about up to 20\% for
$\sqrt{s}=500$\,GeV,
depending on the initial polarizations of the electron and the positron.
The deviation from the SM in the GHU model is monotonically increasing
with respect to $\sqrt{s}\lesssim 1$\,TeV.
By observing the decay mode of the $Z$ boson to the muon pair for
$\sqrt{s}=250$\,GeV, $L_{\rm int}=1$\,ab$^{-1}$, and
$(P_{e^-},P_{e^+})=(\mp0.8,\pm0.3)$,
the deviation from the SM is larger than 3$\sigma$ for 
$m_{\rm KK}\gtrsim 15$\,TeV.
By observing the decay mode of the $Z$ boson to the muon pair for
$\sqrt{s}=500$\,GeV, $L_{\rm int}=2$\,ab$^{-1}$, and
$(P_{e^-},P_{e^+})=(\mp0.8,\pm0.3)$,
the deviation from the SM is larger than 3$\sigma$ for 
$m_{\rm KK}\gtrsim 20$\,TeV.
We also showed that the sign of the bulk mass of the electron multiplet
in the GHU model can be determined by examining the deviation of the
left-right symmetry of the $e^-e^+\to Zh$ process from the SM.
The $e^-e^+\to\nu\bar{\nu}h$ process gives the main contribution to the
single Higgs boson production at sufficiently high $\sqrt{s}$, but in
the GHU  model a sufficiently large deviation from the SM can only be
observed at energy scales close to the $Z'$ boson mass.

Although we focuses on single Higgs boson production processes in this
paper, double Higgs boson production processes are important to 
determine the self-coupling constant of the Higgs boson because we need
to confirm whether the currently observed Higgs boson fully satisfies the
properties of the Higgs boson predicted by the SM or not.
In the GHU model, it is shown in Ref.~\cite{Funatsu:2020znj} that
the coupling constants of the Higgs boson to the $W$ and $Z$ bosons are
very close to those in the SM, while the cubic and quartic self-coupling
constants of the Higgs boson in the GHU model are deviated from those in
the SM for about 8\% and 30\%, respectively.
According to Ref.~\cite{Abramowicz:2016zbo}, the total cross section of
the main double Higgs boson production $e^-e^+\to Zhh$ at
$\sqrt{s}=500$\,GeV is about $0.1$\,fb, 
where the main contribution comes from the $e^-e^+\to Zh\to Zhh$
process.
Even if almost all the decay modes of the Higgs boson could
be used in the analysis, the 1\,$\sigma$ statistical error of the 
$e^-e^+\to Zhh$ process is about $5$\% by the ILC experiment with 
$\sqrt{s}=500$\,GeV and $L_{\rm int}=4$\,ab${}^{-1}$.
Therefore, it is difficult to check the deviation from the SM in the
GHU model at the ILC experiment.
The planned experiments at higher $\sqrt{s}$, such as the CLIC
experiment, may confirm the deviation of 
the self-coupling constant of the Higgs boson from the SM by using 
a double Higgs boson production process $e^-e^+\to\nu_e\bar{\nu}_ehh$
because the cross section of the process becomes larger when 
$\sqrt{s}$ becomes larger, where the main contribution comes from the 
$e^-e^+\to\nu_e\bar{\nu}_eW^{+}W^{-}
\to\nu_e\bar{\nu}_eh\to\nu_e\bar{\nu}_ehh$
process.

{

We comment on electroweak precision measurements. In this paper, we focus
on the search for the GHU model by using the single Higgs boson
production processes.
By comparing the exploration region of the KK mass in the GHU model by
using the single Higgs boson production processes with
that by using the fermion pair production processes analyzed in 
Ref.~\cite{Funatsu:2020haj}, we find that fermion pair production
processes can be explored to higher KK mass scale.
Therefore, the first sign of the new physics in the
ILC experiment is expected to be found not in the Higgs boson production
process but in the fermion pair production processes 
because the cross section in the electron-positron collider experiments
is larger for the fermion pair production processes than for the single
Higgs boson production processes.
This analysis of the Higgs boson production process provides one
important piece of information to distinguish the GHU model from other
new physics models.

}

\section*{Acknowledgments}

The authors would like to thank Yutaka Hosotani for useful discussions
and comments. One of the authors (N.Y.) would like to thank Cheng-Wei
Chiang for valuable comments. 
This work was supported in part 
by European Regional Development Fund-Project Engineering Applications
of Microworld Physics (No.\ CZ.02.1.01/0.0/0.0/16\_019/0000766) (Y.O.),
and the Ministry of Science and Technology of Taiwan under
Grant No. MOST-111-2811-M-002-047-MY2 (N.Y.).

\appendix

\section{Cross section}
\label{Sec:Cross-section}

In this section, we give the formulas necessary to calculate
observables of the $e^-e^+\to Zh$ and $e^-e^+\to f\bar{f}h$ processes,
where $f\bar{f}$ stands for a pair of fermions such as the electron and
the positron ($e^-e^+$).

\subsection{Cross section of $e^-e^+\to Zh$}

In the SM, the contribution to the cross section of $e^-e^+\to Zh$
comes from the $e^-e^+\to Z{}\to Zh$ process at tree level. In the GHU
model, additional contributions come from not only 
the $e^-e^+\to Z'\to Zh$ process but also the interference between
the $e^-e^+\to Z\to Zh$ process and 
the $e^-e^+\to Z'\to Zh$ process.

\subsubsection{Amplitude of $e^-e^+\to Zh$}

The s-channel production process of $Zh$ via vector bosons $V_a=Z,Z'$ is
given by 
\begin{align}
e_X^-(p_{e^-}) e_{\overline{X}}^+(p_{e^+}) 
\longrightarrow
V_a(q=p_{e^-}+p_{e^+}=p_Z+p_h)
\longrightarrow
 Z(p_Z) h(p_h),
\end{align}
where $X=L,R$, $\overline{L}=R$, $\overline{R}=L$.
The amplitude of the $e^-e^+\to Zh$ process is given by
\begin{align}
{\cal M}^{X} 
&= J_{\mu}^{X} \epsilon^{\mu*}(p_Z)Q_{X},
\end{align}
where $\epsilon^\mu$ is a polarization vector of the finial state of the
$Z$ boson, 
$J^{X}_{\mu}$ is the fermion current defined as
\begin{align}
J^{X}_{\mu} := \bar{v}(p_{e^+})\gamma_{\mu}P_{X}u(p_{e^-}),
\end{align}
and 
\begin{align}
Q_{X}:=
\sum_{a} \frac{g_{V_{a}e}^{X} g_{V_{a}Zh} }
{(q^2-m_{V_{a}}^{2}) + i m_{V_{a}}\Gamma_{V_{a}}},
\label{Eq:QL-QR}
\end{align}
where  $X=L,R$;
$P_{R/L} = (1 \pm \gamma_{5})/2$;
$g_{V_{a}e}^{L}$  and $g_{V_{a}e}^{R}$ 
are the coupling constants of the left- and right-handed 
electron (the right- and left-handed positron) to the gauge boson
$V_{a}$, respectively;
$g_{V_{a}Zh}$ is the coupling constant of the gauge bosons $V_a$ and $Z$
to the Higgs boson $h$;
$m_{V_{a}}$ and $\Gamma_{V_{a}}$ are the mass and the total decay width
of $V_{a}$.

\subsubsection{Kinematics of $e^-e^+\to Zh$}

We use the following four moments for the initial state of the
electron ($e^{-}$) and the positron ($e^+$), and the final state of the
$Z$ boson ($Z$) and the Higgs boson ($h$) in the center-of-mass system:
\begin{align}
&p_{e^-} = \frac{\sqrt{s}}{2}(1,0,0,+1),\ \ 
p_{e^+} = \frac{\sqrt{s}}{2}(1,0,0,-1),\ \ 
\nonumber\\
&p_Z = 
\left(E_Z, +\kappa\sin\theta,0, +\kappa\cos\theta\right),\ \ 
p_h = 
\left(E_h, -\kappa\sin\theta,0, -\kappa\cos\theta\right),
\nonumber\\
&q = (\sqrt{s},0,0,0),
\label{Eq:kinematics}
\end{align}
where we neglect the mass of electron; $m_Z$ and $m_h$ stand for the
masses of the $Z$ boson and the Higgs boson, respectively;
\begin{align}
&\kappa:=\frac{\sqrt{[s-(m_Z+m_h)^2][s-(m_Z-m_h)^2]}}{2\sqrt{s}}
=\frac{\lambda^{1/2}(s,m_Z^2,m_h^2)}{2\sqrt{s}},
\nonumber\\
&\lambda(\alpha,\beta,\gamma):=
\alpha^2+\beta^2+\gamma^2-2\alpha\beta-2\alpha\gamma-2\beta\gamma
=(\alpha-\beta-\gamma)^2-4\beta\gamma
\end{align}
From the above, we find
\begin{align}
(p_{e^-} \cdot p_Z)=
\frac{s}{2}\left(E_Z-p\cos\theta\right),\ \
(p_{e^+} \cdot p_Z)=
\frac{s}{2}\left(E_Z+p\cos\theta\right).
\end{align}
The Mandelstam variables are given by
\begin{align}
&s=(p_{e^-}+p_{e^+}) ^2=(p_Z+p_h)^2=q^2,\ \
t=(p_{e^-}-p_Z)^2=(p_{e^+}-p_h)^2
=m_Z^2-\sqrt{s}(E_Z-\kappa\cos\theta),\nonumber\\
&u=(p_{e^-}-p_h)^2=(p_{e^+}-p_Z)^2
=m_Z^2-\sqrt{s}(E_Z+\kappa\cos\theta),\ \ 
s+t+u=m_Z^2+m_h^2.
\end{align}

\subsubsection{Cross section of $e^-e^+\to Zh$}

For the $e_X^- e_{\overline{X}}^+ \to Zh$ process, when only vector
bosons are involved in the intermediate states, the cross section 
of the initial states of of the polarized electron and positron is given
by 
\begin{align}
\frac{d\sigma^{Zh}}{d\cos\theta}(P_{e^-},P_{e^+},\cos\theta)
&= \frac{1}{4}\biggl\{
 (1 - P_{e^-})(1 + P_{e^+}) \frac{d\sigma_{L}^{Zh}}{d\cos\theta}(\cos\theta)
+(1 + P_{e^-})(1 - P_{e^+}) \frac{d\sigma_{R}^{Zh}}{d\cos\theta}(\cos\theta)
\biggr\},
\label{Eq:dsigma}
\end{align}
where $P_{e^-}$ and $P_{e^+}$ are the initial polarizations of the
electron and the positron, and $\sigma_L^{Zh}$ and $\sigma_{R}^{Zh}$ are 
cross sections of $e^-_L e^+_R \to Zh$ and $e^-_R e^+_L \to Zh$,
respectively:
\begin{align}
\frac{d\sigma_{X}^{Zh}}{d\cos\theta}(\cos\theta) :=
\frac{d\sigma}{d\cos\theta}(e^{-}_{X} e^{+}_{\overline{X}}\to Zh)
&= \frac{\kappa}{16\pi s\sqrt{s}}
|{\cal M}^{X}|^{2},
\label{Eq:dsigma-LR}
\end{align}
where the formula for the differential cross section in the SM is given
in e.g., Refs.~\cite{Peskin:1995ev,ParticleDataGroup:2022pth}.
The amplitude of the $e_X^-e_{\overline{X}}^+\to Zh$ process can be
obtained by using the standard method of calculation given in e.g.,
Ref.~\cite{Peskin:1995ev} as
\begin{align}
\left|{\cal M}^X\right|^{2}
&=|Q_{X}|^2s\frac{2m_Z^2+\kappa^2\sin^2\theta}{m_{Z}^2},
\label{Eq:Amplitude-LR}
\end{align}
where $Q_{X=L,R}$ are defined in Eq.~(\ref{Eq:QL-QR}).
Therefore, we find the differential cross section of 
the $e^-(P_{e^-})e^+(P_{e^+})\to Zh$ process with the initial
polarizations $P_{e^{\pm}}$ as
\begin{align}
&\frac{d\sigma^{Zh}}{d\cos\theta}(P_{e^-},P_{e^+},\cos\theta)
\nonumber\\
&=
\frac{1}{4}
\left(
(1-P_{e^-})(1+P_{e^+})
|Q_L|^2
+(1+P_{e^-})(1-P_{e^+})
|Q_R|^2
\right)
\frac{\kappa}{16\pi \sqrt{s}}
\frac{2m_Z^2+\kappa^2\sin^2\theta}{m_{Z}^2}.
\label{Eq:differentical-cross-section-LR}
\end{align}

The total cross section of the $e^-e^+\to Zh$ process with the initial
polarizations can be defined by integrating the
differential cross section in Eq.~(\ref{Eq:dsigma}) with the angle
$\theta$
\begin{align}
 \sigma^{Zh}(P_{e^-},P_{e^+}):=
 \int_{-1}^{1}
 \frac{d\sigma^{Zh}}{d\cos\theta}(P_{e^-},P_{e^+},\cos\theta)
 d\cos\theta,
\end{align}
where the minimum and maximal values of $\cos\theta$ are determined
by each detector and we cannot use date near $\cos\theta\simeq \pm1$.
By using Eq.~(\ref{Eq:differentical-cross-section-LR}),
the total cross section of $e^-e^+\to Zh$ is given by
\begin{align}
\sigma^{Zh}(P_{e^-},P_{e^+})
=\frac{1}{4}
(1-P_{e^-})(1+P_{e^+})
\sigma_L^{Zh}
+\frac{1}{4}
(1+P_{e^-})(1-P_{e^+})
\sigma_R^{Zh},
\label{Eq:sigma-total-Zh}
\end{align}
where 
\begin{align}
\sigma_X^{Zh}&:=
 \int_{-1}^{1}
 \frac{d\sigma_X^{Zh}}{d\cos\theta}(\cos\theta)
 d\cos\theta
=|Q_X|^2
\frac{\kappa}{12\pi \sqrt{s}}
\frac{3m_Z^2+\kappa^2}{m_{Z}^2},
\label{Eq:sigma-total-Zh-LR}
\end{align}
where $X=L,R$; $Q_{X}$ is given in Eq.~(\ref{Eq:QL-QR}).

The statistical error of the cross section 
$\sigma^{Zh}(P_{e^-},P_{e^+})$ is given by
\begin{align}
\Delta \sigma^{Zh}(P_{e^-},P_{e^+})=
 \frac{\sigma^{Zh}(P_{e^-},P_{e^+})}
 {\sqrt{N^{Zh}}},\ \ 
 N^{Zh} =  L_{\rm int} \cdot
 \sigma^{Zh}(P_{e^-},P_{e^+}),
\label{Eq:stat-error-sigma}
\end{align}
where $L_{\rm int}$ is an integrated luminosity, and 
$N^{Zh}$ is the number of events for the $e^-e^+\to Zh$ process.
Note that the $Z$ boson and the Higgs boson cannot observed directly, so
we need to choose the decay modes of the $Z$ boson and/or the Higgs
boson, and then the available number of events must be $N^{Zh}$
multiplied by the branching ratio of each selected decay mode.
The amount of the deviation of the cross section of the $e^-e^+\to Zh$
process from the SM in the GHU model is given by
\begin{align}
\Delta_\sigma^{Zh}(P_{e^-},P_{e^+}):=
\frac{
\left[\sigma^{Zh}(P_{e^-},P_{e^+})\right]_{\rm GHU}}
{\left[\sigma^{Zh}(P_{e^-},P_{e^+})\right]_{\rm SM}}-1,
\label{Eq:Delta_sigma}
\end{align}
where $\left[\sigma^{Zh}(P_{e^-},P_{e^+})\right]_{\rm GHU}$ 
and $\left[\sigma^{Zh}(P_{e^-},P_{e^+})\right]_{\rm SM}$ 
stand for the cross sections of the $e^-e^+\to Zh$ process in
the SM and the GHU model, respectively.
The same notation is used for other cases in the followings.

\subsubsection{Left-right asymmetry of $e^-e^+\to Zh$}

As the same in fermion pair productions, we can define a left-right
asymmetry 
\cite{MoortgatPick:2005cw,Abe:1994wx,Abe:1996nj,Funatsu:2020haj,Funatsu:2022spb}
of the $e^-e^+\to Zh$ process as
\begin{align}
A_{LR} ^{Zh}
:=\frac{\sigma_L^{Zh}-\sigma_R^{Zh}}
{\sigma_L^{Zh}+\sigma_R^{Zh}}
=\frac{|Q_L|^2-|Q_R|^2}{|Q_L|^2+|Q_R|^2},
\label{Eq:ALR-def}
\end{align}
where $\sigma_{X}^{Zh}$ and $Q_X$ ($X=L,R$) are given in 
Eq.~(\ref{Eq:sigma-total-Zh-LR}) and Eq.~(\ref{Eq:QL-QR}), respectively.

An observable left-right asymmetry is defined as 
\begin{align}
A_{LR}^{Zh}(P_{e^-},P_{e^+})
:=\frac{\sigma^{Zh}(P_{e^-},P_{e^+})-\sigma^{Zh}(-P_{e^-},-P_{e^+})}
{\sigma^{Zh}(P_{e^-},P_{e^+})+\sigma^{Zh}(-P_{e^-},-P_{e^+})}
\end{align}
for $P_{e^-}<0$ and $|P_{e^-}|>|P_{e^+}|$.
From Eq.~(\ref{Eq:differentical-cross-section-LR}), we find
\begin{align}
A_{LR} ^{Zh}(P_{e^-},P_{e^+},\cos\theta)
&=-\frac{P_{e^-}-P_{e^+}}{1-P_{e^-}P_{e^+}}
\frac{|Q_L|^2-|Q_R|^2}
{|Q_L|^2+|Q_R|^2}.
\label{Eq:ALR-obs}
\end{align}

The relation
between the theoretically defined left-right asymmetry 
$A_{LR}^{Zh}$ given in Eq.~(\ref{Eq:ALR-def})
and the observable left-right asymmetry 
$A_{LR}^{Zh}(P_{e^-},P_{e^+})$
given in Eq.~(\ref{Eq:ALR-obs}) is given by
\begin{align}
 A_{LR}^{Zh}
 =\frac{1}{-P_{\rm eff}}A_{LR}^{Zh}(P_{e^-},P_{e^+}),\ \ 
P_{\rm eff}:=\frac{P_{e^-}-P_{e^+}}{1-P_{e^-}P_{e^+}}.
\end{align}

The statistical error of the left-right asymmetry is given by
\begin{align}
 \Delta A_{LR}^{Zh} &=
 2\frac{\sqrt{N_{L}^{Zh}N_{R}^{Zh}}
 \left(\sqrt{N_{L}^{Zh}}+\sqrt{N_{R}^{Zh}}\right)}
 {(N_{L}^{Zh}+N_{R}^{Zh})^2},
\label{Eq:Error-A_LR}
\end{align}
where $N_{L}^{Zh}=L_{\rm int} \, \sigma^{Zh}(P_{e^-},P_{e^+})$
and $N_{R}^{Zh}=L_{\rm int} \, \sigma^{Zh}(-P_{e^-},-P_{e^+})$
are the numbers of the events 
for $P_{e^-}<0$ and $|P_{e^-}|>|P_{e^+}|$.
The amount of the deviation from the SM in Eq.~(\ref{Eq:ALR-def})
is characterized by
\begin{align}
\Delta_{A_{LR}}^{Zh} &:=
 \frac{\left[A_{LR}^{Zh}\right]_{\rm GHU}}
{\left[A_{LR}^{Zh}\right]_{\rm SM}}-1.
\label{Eq:Delta_A_LR}
\end{align}

\subsection{Cross section of $e^-e^+\to f\bar{f}h$}

The cross sections of the $e^-e^+\to \nu\bar{\nu}h$ process via the
$W^\pm$ bosons and of the $e^-e^+\to e^-e^+ h$ process via the $Z$
bosons are estimated in
Refs.~\cite{Jones:1979bq,Cahn:1984tx,Altarelli:1987ue}. 
When we consider the cross section of the $e^-e^+\to f\bar{f}h$ processes
$(f\bar{f}=\nu\bar{\nu},e^-e^+)$ in the SM,
we need to take into account not only the vector boson fusion process
$e^-e^+\to f\bar{f}VV\to f\bar{f}h$ 
but also the $Zh$ process
$e^-e^+\to Z\to Zh\to f\bar{f}h$.
In the GHU model, additional contributions come from the $W'$ and $Z'$
bosons.

\subsubsection{Amplitude of $e^-e^+\to f\bar{f}VV\to f\bar{f}h$}

The Higgs boson production processes via vector gauge bosons
$V_{a}=W^{(\prime)\pm}$ or $Z^{(\prime)}$ are given by 
\begin{align}
&e_X^-(p_1) e_{\overline{Y}}^+(p_2) \longrightarrow
\nu_e(p_3)\bar\nu_e(p_4)
W^{(\prime)-}(q_1=p_1-p_3)W^{(\prime)+}(q_2=p_2-p_4)
\longrightarrow
\nu_e(p_3)\bar\nu_e(p_4)h(p_h)\,
\nonumber\\
&e_X^-(p_1) e_{\overline{Y}}^+(p_2) \longrightarrow
e^-(p_3) e^+(p_4)
Z^{(\prime)}(q_1=p_1-p_3)Z^{(\prime)}(q_2=p_2-p_4)
\longrightarrow
e^-(p_3) e^+(p_4)h(p_h),
\label{Eq:ee-to-ffVV-ffh}
\end{align}
where $X,Y=L,R$; $\overline{L}=R$, $\overline{R}=L$;
$p_1$ and $p_2$ are the momenta of the initial states of electron and
positron; $p_3$ and $p_4$ are the momenta of the final states of
fermions; and $p_h$ is the momenta of the Higgs boson.
The amplitude is given by
\begin{align}
{\cal M}_{VVh}^{XY}
&= \sum_{V_a,V_b=\{W^{(\prime)}\}\ \mbox{or}\ \{Z^{(\prime)}\}}
g_{V_{a}V_{b}h}
g_{V_{a}e}^{X} g_{V_{b}e}^{Y}
g^{\mu\nu}
{J_{\mu\nu}^{XY}}
\frac{1}{(q_1^2-m_{V_{a}}^{2}) + i m_{V_{a}} \Gamma_{V_{a}}}
\frac{1}{(q_2^2-m_{V_{b}}^{2}) + i m_{V_{b}} \Gamma_{V_{b}}},
\label{Eq:Amplitude-ee-to-ffVV-ffh}
\end{align}
where we defined
\begin{align}
J_{\mu\nu}^{XY}:=
\bar{u}(p_3)\gamma_\mu P_{X} u(p_1)
\bar{v}(p_2)\gamma_{\nu} P_{Y}v(p_4),
\label{Eq:Current-ee-to-ffVV-ffh}
\end{align}
and $P_{R/L} = (1 \pm \gamma_{5})/2$.
We neglected electron and neutrino masses.

After some tedious calculations, we obtain
\begin{align}
&\sum_{\rm spin}
\left|{\cal M}_{VVh}^{XY} \right|^2
\nonumber\\
&= 
16\sum_{V_a,V_b,V_c,V_d=\{W^{(\prime)}\}\ \mbox{or}\ \{Z^{(\prime)}\}}
g_{V_{a}V_{b}h}g_{V_{c}V_{d}h}^*
\times 
\left\{
\begin{array}{cc}
g_{V_{a}e}^{X} g_{V_{b}e}^{X} 
g_{V_{c}e}^{X*} g_{V_{d}e}^{X*}
(p_{1}\cdot p_{2})
(p_{3}\cdot p_{4})
&\mbox{for}\ X=Y
\\
g_{V_{a}e}^{X} g_{V_{b}e}^{Y} 
g_{V_{c}e}^{X*} g_{V_{d}e}^{Y*} 
(p_{1}\cdot p_{4})
(p_{2}\cdot p_{3})
&\mbox{for}\ X\not=Y\\
\end{array}
\right\}
\nonumber\\
&\ \ \times 
\frac{1}{(q_1^2-m_{V_{a}}^{2}) + i m_{V_{a}} \Gamma_{V_{a}}}
\frac{1}{(q_2^2-m_{V_{b}}^{2}) + i m_{V_{b}} \Gamma_{V_{b}}}
\frac{1}{(q_1^2-m_{V_{c}}^{2}) - i m_{V_{c}} \Gamma_{V_{c}}}
\frac{1}{(q_2^2-m_{V_{d}}^{2}) - i m_{V_{d}} \Gamma_{V_{d}}}.
\label{Eq:M2-general-1}
\end{align}

\subsubsection{Amplitude of  $e^-e^+\to Z^{(\prime)}
\to Z^{(\prime)}h \to f\bar{f}h$}

The Higgs boson production processes via neutral gauge bosons $V_a=Z,Z'$
are given by 
\begin{align}
&e_X^-(p_1) e_{\overline{Y}}^+(p_2) 
\longrightarrow
Z^{(\prime)}(q_1^{\prime}=p_1+p_2)
\longrightarrow
Z^{(\prime)}(q_2^{\prime}=p_3+p_4)h(p_h)
\longrightarrow
\nu(p_3)\bar\nu(p_4)h(p_h),
\nonumber\\
&e_X^-(p_1) e_{\overline{Y}}^+(p_2) \longrightarrow
Z^{(\prime)}(q_1^{\prime}=p_1+p_2)
\longrightarrow
Z^{(\prime)}(q_2^{\prime}=p_3+p_4)h(p_h)
\longrightarrow
e^-(p_3) e^+(p_4)h(p_h),
\label{Eq:ee-to-Z-to-Zh-to-ffh}
\end{align}
The amplitude is given by
\begin{align}
{\cal M}_{Zh}^{XY}
&= \sum_{V_a,V_b=\{Z^{(\prime)}\}}
g_{V_{a}V_{b}h}
g_{V_{a}e}^{X} g_{V_{b}e}^{Y} 
g^{\mu\nu}{J_{\mu\nu}^{\prime XY}}
\frac{1}{(q_1^{\prime 2}-m_{V_{a}}^{2}) + i m_{V_{a}} \Gamma_{V_{a}}}
\frac{1}{(q_2^{\prime 2}-m_{V_{b}}^{2}) + i m_{V_{b}} \Gamma_{V_{b}}},
\label{Eq:Amplitude-ee-to-Zh-ffh}
\end{align}
where we define $J_{\mu\nu}^{\prime XY}$ $(X,Y=L,R)$ as
\begin{align}
&J_{\mu\nu}^{\prime XY}=
\bar{v}(p_2)\gamma_\mu P_{X} u(p_1)
\bar{u}(p_3)\gamma_{\nu} P_{Y}v(p_4).
\label{Eq:Current-ee-to-Zh-ffh}
\end{align}

By the same calculation as for the square of the amplitude
given in  Eq.~(\ref{Eq:M2-general-1}), we find
\begin{align}
&\sum_{\rm spin}
\left|{\cal M}_{Zh}^{XY} \right|^2
\nonumber\\
&= 
16
\sum_{V_a,V_b,V_c,V_d=\{Z^{(\prime)}\}}
g_{V_{a}V_{b}h}g_{V_{c}V_{d}h}^*
\times
\left\{
\begin{array}{cc}
g_{V_{a}e}^{X} g_{V_{b}e}^{X} 
g_{V_{c}e}^{X*} g_{V_{d}e}^{X*} 
(p_{1}\cdot p_{3})
(p_{2}\cdot p_{4}) 
&\mbox{for}\ X=Y \\
g_{V_{a}e}^{X} g_{V_{b}e}^{Y} 
g_{V_{c}e}^{X*} g_{V_{d}e}^{Y*} 
(p_{1}\cdot p_{4})
(p_{2}\cdot p_{3})
&\mbox{for}\ X\not=Y \\
\end{array}
\right\}
\nonumber\\
&\ \ \times 
\frac{1}{(q_1^{\prime 2}-m_{V_{a}}^{2}) + i m_{V_{a}} \Gamma_{V_{a}}}
\frac{1}{(q_2^{\prime 2}-m_{V_{b}}^{2}) + i m_{V_{b}} \Gamma_{V_{b}}}
\frac{1}{(q_1^{\prime 2}-m_{V_{c}}^{2}) - i m_{V_{c}} \Gamma_{V_{c}}}
\frac{1}{(q_2^{\prime 2}-m_{V_{d}}^{2}) - i m_{V_{d}} \Gamma_{V_{d}}},
\label{Eq:M2-general-2}
\end{align}
where when the fermion pair in the final state is $e^-e^+$,
$g_{V_{b}e}^{L/R}$
and $g_{V_{d}e}^{L/R}$ are gauge coupling constants of the left- or
right-handed electron, while when the fermion pair in the final state is
$\nu_e\bar{\nu}_e$, $g_{V_{b}e}^{L/R}$ and $g_{V_{d }e}^{L/R}$ are 
gauge coupling constants of the left-handed or right-handed neutrinos
The $g_{V_{a}e}^{L/R}$ and $g_{V_{c}e}^{L/R}$ are always the gauge
coupling constants of the left-handed or right-handed electron.

\subsubsection{Interference of
$e^-e^+\to f\bar{f}VV\to f\bar{f}h$ and 
$e^-e^+\to Z^{(\prime)} \to Z^{(\prime)}h \to f\bar{f}h$}

The square of the amplitude of the $e_X^+e_Y^-\to f\bar{f}h$ process is
given by
\begin{align}
\left|{\cal M}^{XY}\right|^2=
\left|{\cal M}_{VVh}^{XY}+{\cal M}_{Zh}^{XY}\right|^2=
\left|{\cal M}_{VVh}^{XY}\right|^2
+\left|{\cal M}_{Zh}^{XY}\right|^2
+2\mbox{Re}\left({\cal M}_{VVh}^{XY}{\cal M}_{Zh}^{XY*}\right).
\label{Eq:M2-total}
\end{align}
The last term stands for the interference term between 
$e^-e^+\to f\bar{f}VV\to f\bar{f}h$ and 
$e^-e^+\to Z^{(\prime)} \to Z^{(\prime)}h \to f\bar{f}h$.
This term also need to be calculated.

As the same as Eqs.~(\ref{Eq:Amplitude-ee-to-ffVV-ffh}) and 
(\ref{Eq:Amplitude-ee-to-Zh-ffh}), after some tedious calculations,
we find
\begin{align}
&
2\mbox{Re}\left[
\sum_{\rm spin}
{\cal M}_{VVh}{\cal M}_{Zh}^*
\right]=
\nonumber\\
&-16
\mbox{Re}\Bigg[
\sum_{V_a,V_b=\{W^{(\prime)}\}\ \mbox{or}\ \{Z^{(\prime)}\}}
\sum_{V_c,V_d=\{Z^{(\prime)}\}}
g_{V_{a}V_{b}h}g_{V_{c}^*V_{d}h}^*
\nonumber\\
&\times
\left\{
\begin{array}{cl}
g_{V_{a}e}^{X} g_{V_{b}e}^{X} 
g_{V_{c}e}^{X*} g_{V_{d}e}^{X*} 
A
+ ig_{V_{a}e}^{X} g_{V_{b}e}^{X} 
g_{V_{c}e}^{X*} g_{V_{d}e}^{X*} 
B
&\mbox{for}\ X=Y=L \\
g_{V_{a}e}^{X} g_{V_{b}e}^{X} 
g_{V_{c}e}^{X*} g_{V_{d}e}^{X*} 
A
- ig_{V_{a}e}^{X} g_{V_{b}e}^{X} 
g_{V_{c}e}^{X*} g_{V_{d}e}^{X*} 
B
&\mbox{for}\ X=Y=R \\
0 &\mbox{for}\ X\not=Y \\
\end{array}
\right\}
\nonumber\\
&\times 
\frac{1}{(q_1^2-m_{V_{a}}^{2}) + i m_{V_{a}} \Gamma_{V_{a}}}
\frac{1}{(q_2^2-m_{V_{b}}^{2}) + i m_{V_{b}} \Gamma_{V_{b}}}
\frac{1}{(q_1^{\prime 2}-m_{V_{c}}^{2}) - i m_{V_{c}} \Gamma_{V_{c}}}
\frac{1}{(q_2^{\prime 2}-m_{V_{d}}^{2}) - i m_{V_{d}} \Gamma_{V_{d}}}
\Bigg],
\label{Eq:M2-general-3}
\end{align}
where 
\begin{align}
A:=
\left(p_1\cdot p_4\right)\left(p_2\cdot p_3\right)
+\left(p_1\cdot p_3\right)\left(p_2\cdot p_4\right)
-\left(p_1\cdot p_2\right)\left(p_3\cdot p_4\right),\ \
B:=
p_1^\alpha p_2^\beta p_3^\gamma p_4^\delta
\epsilon_{\gamma\beta\alpha\delta}.
\end{align}

\subsubsection{Kinematics of $e^-e^+\to f\bar{f}h$}

In the center-of-mass system of the initial electron and
positron, the initial energies are $E_1=E_2=\sqrt{\hat{s}}/2$,
where $\hat{s}$ is the center-of-mass energy for the initial 
electron and positron.
When we take the same convention in Ref.~\cite{Cahn:1984tx}, we define
$\eta$ and $\zeta$ as 
\begin{align}
 E_3=(1-\eta)\frac{1}{2}\sqrt{\hat{s}},\ \ 
 E_4=(1-\zeta)\frac{1}{2}\sqrt{\hat{s}},
\end{align}
so that the energy of the Higgs boson is
\begin{align}
E_h=(\eta+\zeta)\frac{1}{2}\sqrt{\hat{s}}.
\end{align}
The final state momenta lie in a plane and 
\begin{align}
({\bf p_h})^2=({\bf p}_3+{\bf p}_4)^2=E_h^2-m_h^2.
\end{align}
We define the angle of $\theta$ by
\begin{align}
\cos\theta=-\hat{\bf p}_3\cdot\hat{\bf p}_4=
1-2\frac{\zeta\eta-\frac{m_h^2}{\hat{s}}}{(1-\zeta)(1-\eta)}.
\label{Eq:theta-ee-to-VVh}
\end{align}

The kinematic boundaries occur for $\cos\theta=\pm 1$, and the allowed
region must be satisfied as
\begin{align}
\zeta\eta\geq \frac{m_h^2}{\hat{s}},\ \ 
\zeta+\eta\leq 1+\frac{m_h^2}{\hat{s}}.
\end{align}

To specify the final state in the center-of-mass system of the initial
electron and positron,
we take the following basis:
\begin{align}
&\hat{\bf p}_1=\hat{\bf x},\ \
\hat{\bf p}_2=-\hat{\bf x},\ \ 
\hat{\bf p}_3=
\hat{\bf x}\cos\alpha\cos\beta
+\hat{\bf y}\sin\alpha
-\hat{\bf z}\sin\beta\cos\alpha,\nonumber\\
&\hat{\bf p}_4=
-\hat{\bf x}\cos(\alpha-\theta)\cos\beta
-\hat{\bf y}\sin(\alpha-\theta)
+\hat{\bf z}\sin\beta\cos(\alpha-\theta).
\end{align}
The Lorentz scalars are given as
\begin{align}
&(p_1\cdot p_2)=\frac{1}{2}\hat{s},\ \
(p_2\cdot p_3) =\frac{1}{4}(1-\eta)(1-\cos\alpha\sin\beta)\hat{s},
\nonumber\\
&(p_1\cdot p_3) =\frac{1}{4}(1-\eta)(1+\cos\alpha\sin\beta)\hat{s},\ \
(p_3\cdot p_4)=\frac{1}{4}(1-\zeta)(1-\eta)(1+\cos\theta)\hat{s},
\nonumber\\
&(p_1\cdot p_4)=
\frac{1}{4}(1-\zeta)(1-\cos(\alpha-\theta)\sin\beta)\hat{s},\ \ 
(p_2\cdot p_4)
=\frac{1}{4}(1-\zeta)(1+\cos(\alpha-\theta)\sin\beta)\hat{s},
\nonumber\\
&(q_1)^2=-\frac{1}{2}(1-\eta)(1+\cos\alpha\sin\beta)\hat{s},\ \
(q_2)^2 =-\frac{1}{2}(1-\zeta)(1+\cos(\alpha-\theta)\sin\beta)\hat{s},
\nonumber\\
&(q_1')^2=\hat{s},\ \
(q_2')^2=\frac{1}{2}(1-\zeta)(1-\eta)(1+\cos\theta)\hat{s}.
\label{Eq:Kinematics-ee-to-VVh}
\end{align}

\subsubsection{Cross section of $e^-e^+\to f\bar{f}h$}

The differential cross section of the $e^-e^+\to f\bar{f}h$ process 
is given by
\begin{align}
d\sigma^{ffh}&=\frac{1}{16(2\pi)^4\hat{s}}
|{\cal M}|^2dE_3dE_4d\alpha d\cos\beta
=\frac{1}{64(2\pi)^4}
|{\cal M}|^2
d\eta d\zeta d\alpha d\cos\beta,
\end{align}
by using the formula in e.g.,
Refs.~\cite{Cahn:1984tx,ParticleDataGroup:2022pth}.
The total cross section with the initial polarization of the electron
and the positron is give by 
\begin{align}
\sigma^{ffh}(P_{e^-},P_{e^+})&=
\frac{1}{4}
\bigg\{
(1-P_{e^-})(1+P_{e^+})
\sigma_{LL}^{ffh}
+(1+P_{e^-})(1-P_{e^+})
\sigma_{RR}^{ffh}
\nonumber\\
&\hspace{2em}
 +(1-P_{e^-})(1-P_{e^+})
\sigma_{LR}^{ffh}
+(1+P_{e^-})(1+P_{e^+})
\sigma_{RL}^{ffh}
\bigg\},
\label{Eq:Total-cross-section-ee-to-ffh}
\end{align}
where 
\begin{align}
\sigma_{XY}^{ffh}&:=\frac{1}{64(2\pi)^4}
\int_{\frac{m_h^2}{\hat{s}}}^1d\eta
\int_{\frac{m_h^2}{\hat{s}\eta}}^{1+\frac{m_h^2}{\hat{s}}-\eta}d\zeta
\int_0^{2\pi} d\alpha
\int_{-1}^1d\cos\beta
|{\cal M}^{XY}|^2,
\end{align}
where $|{\cal M}^{XY}|^2(X,Y=L,R)$ are given by substituting
Eqs.~(\ref{Eq:M2-general-1}), (\ref{Eq:M2-general-2}),
(\ref{Eq:M2-general-3}) into Eq.~(\ref{Eq:M2-total}).
and we omitted $\sum_{\rm spin}$ to simplify our notation. 

The amount of the deviation of the cross section of 
the $e^-e^+\to f\bar{f}h$ process
from the SM in the GHU model is given by
\begin{align}
\Delta_\sigma^{ffh}(P_{e^-},P_{e^+}):=
\frac{
\left[\sigma^{ffh}(P_{e^-},P_{e^+})\right]_{\rm GHU}}
{\left[\sigma^{ffh}(P_{e^-},P_{e^+})\right]_{\rm SM}}-1,
\label{Eq:Delta_sigma-ffh}
\end{align}
where $\left[\sigma^{ffh}(P_{e^-},P_{e^+})\right]_{\rm GHU}$ 
and $\left[\sigma^{ffh}(P_{e^-},P_{e^+})\right]_{\rm SM}$ 
stand for the cross sections of the $e^-e^+\to f\bar{f}h$ process in 
the SM and the GHU model, respectively.

\subsubsection{Left-right asymmetry of $e^-e^+\to f\bar{f}h$}

Unlike the left-right asymmetry of the $e^-e^+\to Zh$ process,
we cannot define a left-right asymmetry of the $e^-e^+\to f\bar{f}h$ 
process that does not depend on the initial polarization of electron and
positron even if the off-shell particles are limited to vector bosons
only. Still we can define a left-right asymmetry that depends on the
initial polarization of the electron and the positron as
\begin{align}
A_{LR} ^{ffh}(P_{e^-},P_{e^+})
:=\frac{\sigma^{ffh}(P_{e^-},P_{e^+})-\sigma^{ffh}(-P_{e^-},-P_{e^+})}
{\sigma^{ffh}(P_{e^-},P_{e^+})+\sigma^{ffh}(-P_{e^-},-P_{e^+})}
\label{Eq:ALR-ffh-def}
\end{align}
for $P_{e^-}<0$ and $|P_{e^-}|>|P_{e^+}|$.
By using Eq.~(\ref{Eq:Total-cross-section-ee-to-ffh}), we find
\begin{align}
A_{LR} ^{ffh}(P_{e^-},P_{e^+})
&=-\frac{(P_{e^-}-P_{e^+})(\sigma_{LL}^{ffh}-\sigma_{RR}^{ffh})
-(P_{e^-}+P_{e^+})(\sigma_{LR}^{ffh}-\sigma_{RL}^{ffh})}
{(1-P_{e^-}P_{e^+})(\sigma_{LL}^{ffh}+\sigma_{RR}^{ffh})
+(1+P_{e^-}P_{e^+})(\sigma_{LR}^{ffh}+\sigma_{RL}^{ffh})}.
\label{Eq:ALR-ffh-obs}
\end{align}

We can define the amount of the deviation from the SM in
Eq.~(\ref{Eq:ALR-ffh-def}) as
\begin{align}
\Delta_{A_{LR}}^{ffh}(P_{e^-},P_{e^+})
 &:=
 \frac{
\left[A_{LR}^{ffh}(P_{e^-},P_{e^+})\right]_{\rm GHU}}
{\left[A_{LR}^{ffh}(P_{e^-},P_{e^+})\right]_{\rm SM}}-1,
\label{Eq:Delta_A_LR-ffh}
\end{align}
where $\left[A_{LR}^{ffh}(P_{e^-},P_{e^+})\right]_{\rm GHU}$
and ${\left[A_{LR}^{ffh}(P_{e^-},P_{e^+})\right]_{\rm SM}}$
are $A_{LR}^{ffh}(P_{e^-},P_{e^+})$ in Eq.~(\ref{Eq:ALR-ffh-obs})
calculated with the parameter sets of the SM and the GHU model.

\bibliographystyle{utphys}
\bibliography{../../../arxiv/reference}

\end{document}